\documentclass[12pt]{book}
\usepackage[utf8]{inputenc}
\usepackage[english]{babel}
\usepackage{axodraw4j}
\usepackage{latexsym}
\usepackage{cite}
\usepackage{amsmath}
\usepackage{amssymb}
\usepackage{float}
\usepackage{afterpage}
\usepackage{gensymb}
\usepackage{enumerate}
\usepackage{wrapfig}
\usepackage{lipsum}
\usepackage{color}
\usepackage{multirow}
\usepackage{subcaption}
\usepackage{doi}
\usepackage{graphicx}
\usepackage{hyperref}
\usepackage[font=small,labelfont=bf]{caption}
\hypersetup{
    unicode=false,          
    pdftex=true,
    pdftoolbar=true,        
    pdfmenubar=true,        
    pdffitwindow=false,     
    pdfstartview={FitH},    
    pdftitle={Non-linear response in periodic systems: a real-time approach},
    pdfauthor={Claudio Attaccalite},     
    pdfnewwindow=true,      
    colorlinks=false,       
    linkcolor=red,          
    citecolor=green,        
    filecolor=magenta,      
    urlcolor=cyan           
}

\renewcommand{\baselinestretch}{1.3}

\newcommand{\clearemptydoublepage}{\newpage{\pagestyle{empty}\cleardoublepage}}

\def\pp		{{\bf p}}
\def\xx		{{\bf x}}
\def\bxi      	{\boldsymbol{\xi}}
\def\rr		{{\bf r}}

\def\bA		{{\bf A}}
\def\BB		{{\bf B}}
\def\GG		{{\bf G}}
\def\RR		{{\bf R}}
\def\TT		{{\bf T}}

\def\EE		{{\bf E}}
\def\qq		{{\bf q}}
\def\kk		{{\bf k}}
\def\redk	{{\color{red} {\bf k}}}
\def\jj		{{\bf j}}

\newcommand{\ra}{\rightarrow}

\newcommand{\chitwo}{\chi^{(2)}}

\renewcommand{\[}{\left[}
\renewcommand{\]}{\right]}
\renewcommand{\(}{\left(}
\renewcommand{\)}{\right)}

\newcommand{\JJ}{{\boldsymbol J}}
\newcommand{\lv}{{\bf a}}
\newcommand{\bb}{{\bf b}}

\newcommand{\hh}{{\bf h}}
\newcommand{\PP}{{\bf P}}
\newcommand{\HH}{{\bf H}}
\newcommand{\SiS}{{\bf \Sigma}}
\newcommand{\VV}{{\bf V}}
\newcommand{\UU}{{\bf U}}
\newcommand{\w}{\omega}

\newcommand{\be}{\begin{equation}}
\newcommand{\ee}{\end{equation}}
\newcommand{\ben}{\begin{equation*}}
\newcommand{\een}{\end{equation*}}
\newcommand{\bea}{\begin{eqnarray}}
\newcommand{\eea}{\end{eqnarray}}
\newcommand{\bean}{\begin{eqnarray*}}
\newcommand{\eean}{\end{eqnarray*}}

\renewcommand{\[}{\left[}
\renewcommand{\]}{\right]}
\renewcommand{\(}{\left(}
\renewcommand{\)}{\right)}
\def\efield{\boldsymbol{\cal E}}
\def\ket#1{\vert#1\rangle}

\def\susc#1{\chi^{(#1)}}
\def\ket#1{\vert#1\rangle}

\def\ai{\emph{ab-initio} }

\newcommand{\newtensor}[1] {\underline{\underline{#1}}}

\def\sss{\scriptscriptstyle\rm}
\def\Efield{\boldsymbol{\cal E}}
\def\efield{\boldsymbol{\scriptstyle \cal E}}
\def\ket#1{\vert#1\rangle}

\def\susc#1{\chi^{(#1)}}
\def\PPo#1{{\boldsymbol {\cal P}}^{(#1)}}
\def\tsusc#1{\tilde{\chi}^{(#1)}}
\def\ket#1{\vert#1\rangle}

\newcommand{\Av}{{\bf A}}

\newcommand{\chirr}{\chi^{\rho\rho}}
\newcommand{\bchirr}{\bar\chi^{\rho\rho}}
\newcommand{\tchirr}{\tilde\chi^{\rho\rho}}
\newcommand{\zero}{{\bf 0}}
\def\tensor#1{\overset\leftrightarrow{#1}}

\newcommand{\PPc}{{\boldsymbol {\cal P}}}
\newcommand{\DDc}{{\boldsymbol {\cal D}}}
\newcommand{\EEc}{{\boldsymbol {\cal E}}}

\def\bverb      {\renewcommand{\baselinestretch}{1}\begin{verbatim}}
\def\everb  	{\end{verbatim}\renewcommand{\baselinestretch}{1.3}}

\restylefloat{figure}
\restylefloat{table}

\begin{document}

\pagestyle{empty}

\begin{center}

\begin{figure}
\hspace{2cm}
\includegraphics[width=10cm]{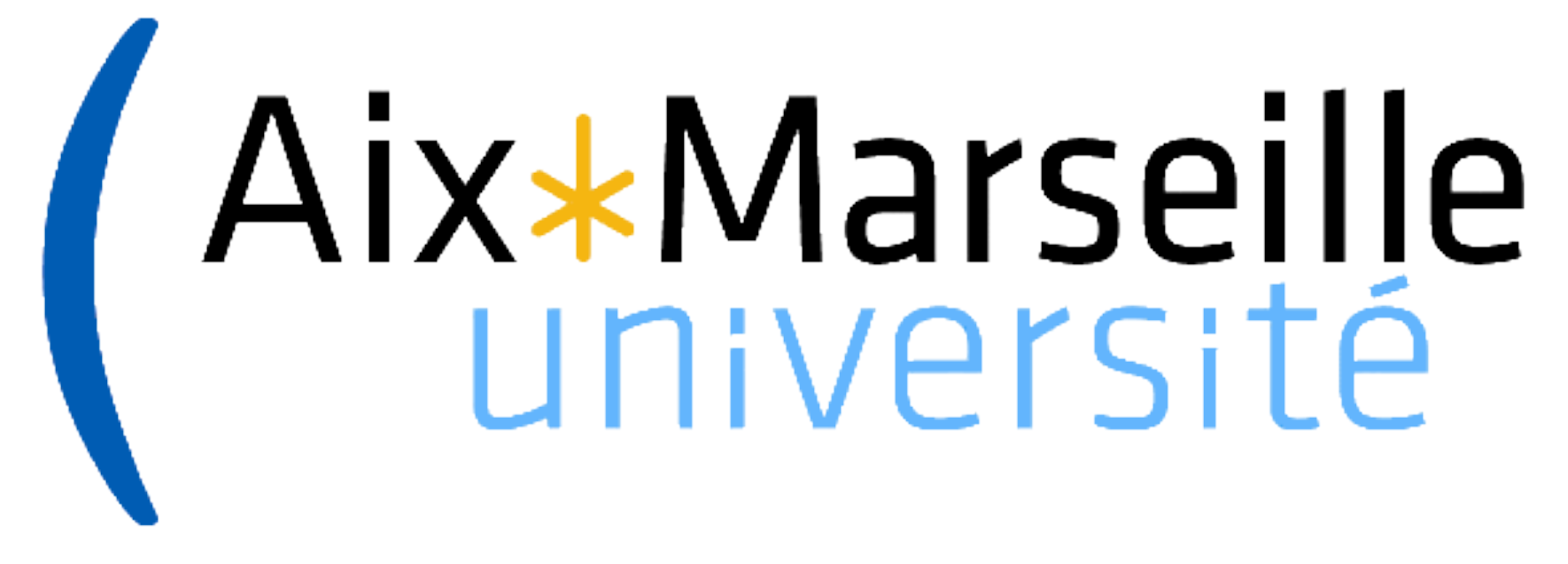}
\end{figure}
\vspace{-2cm}
\Large{\underline{Aix Marseille Universit\'e}}
\vspace{0.5cm}
\Large{\underline{Habilitation \'a Diriger des Recherches (HDR)}}\\
\vspace{1cm}
\LARGE{
{\bf 
Non-linear response in extended systems: a real-time approach}}\\
\vspace{1cm}
\Large{\underline{\it Claudio Attaccalite}}\\[1cm]
\vspace{0.2cm}
\Large{2016}\\[1cm]

\end{center}

\clearemptydoublepage

\pagestyle{plain}
\pagenumbering{roman}

\setcounter{tocdepth}{4}
\renewcommand{\baselinestretch}{1.5}
\tableofcontents
\clearemptydoublepage
\clearemptydoublepage

\pagenumbering{arabic}

\chapter*{Abstract}
In this thesis we present a new formalism to study linear and non-linear response in extended systems. Our approach is based on real-time solution of an effective Schr\"odinger equation. The coupling between electrons and external field is described by means of Modern Theory of Polarization. Correlation effects are derived  from Green's function theory. We show that the inclusion of local-field effects and electron-hole interaction is crucial to predict and reproduce second and third harmonic generation in low dimensional structures, where strong bound excitons are present. Finally in the last part we introduce a real-time density functional approach suitable for infinite periodic crystals in which we work within the so-called length gauge and calculate the polarisation as a dynamical Berry's phase. This approach, in addition to the electron density considers also the macroscopic polarisation as a main variable to correctly treat periodic crystals in presence of electric fields within a density functional framework.

\chapter*{List of abbreviations}
The following table describes the significance of various abbreviations and acronyms used throughout the thesis.  
\begin{table}[h]
\footnotesize
\begin{tabular}{c|c}
\hline
\textbf{Abbreviation} & \textbf{Meaning} \\
\hline
BBGKY & Bogoliubov-Born-Green-Kirkwood-Yvon hierarchy \\
BSE & Bethe-Salpeter Equation\\
BZ & Brilluoin Zone\\
COH & Coulomb-hole  \\
COHSEX & Coulomb-hole plus Screened Exchange self-energy \\
DFT & Density Functional Theory   \\
DFTP & Density Functional Polarisation Theory   \\
EDA & Electric-dipole Approximation \\
EOM & Equation of Motion   \\
GW & GW approximation for self-energy of a many-body system\\
G$_0$W$_0$ & Zero order of the GW self-energy \\
HF & Hartree-Fock\\
HHG & High Harmonic Generation   \\
IPA & Indepent particle approximation  \\
IPC & Infinite Periodic Crystal  \\
KBE & Kadanoff-Baym Equations\\
KS & Kohn and Sham\\
KSV & King-Smith Vanderbilt Polarisation\\
IP & Independent Particle\\
IPA & Independent Particle Approximation\\
JGM & Jellium with Gap Model\\
    & \emph{continued to the next page}   \\
\end{tabular}
\end{table}

\begin{center}
\begin{table}
\footnotesize
\begin{tabular}{c|c}
\hline
\textbf{Abbreviation} & \textbf{Meaning} \\
\hline
JGM-PF & Jellium with Gap Model Polarisation Functional\\
LDA & Local Density Approximation\\
LRC & Long-range corrected approximation\\
MBPT & Many-Body Perturbation Theory\\
opt-PF & optimal Polarisation Functional\\
PBC & Periodic Boundary Conditions\\
QPA & Quasi-particle approximation\\
RPA & Random Phase Approximation (including local field effects)\\
SEX & Screened Exchange self-energy \\
SHG & Second Harmonic Generation\\
THG & Third Harmonic Generation   \\
RT-BSE & Real-Time Bethe-Salpeter Equation   \\
TDDFT & Time-Dependent Density Functional Theory   \\
TDCDFT & Time-Dependent Current-Density Functional Theory   \\
TD-LDA & Time-Dependent Local Density Approximation\\
TDH & Time-Dependent Hartree   \\
TDHF & Time-Dependent Hartree-Fock   \\
TD-KS & Time-Dependent Kohn-Sham   \\
\end{tabular}
\end{table}
\end{center}

\chapter{Introduction to the non-linear optics} 

\section{What is non-linear optics?}
\begin{wrapfigure}{r}{0.5\textwidth}
    \vspace{-0.7cm}
  \begin{center}
    \includegraphics[width=0.4\textwidth]{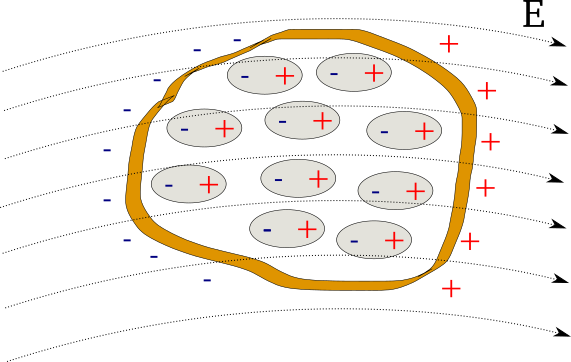}
  \end{center}
  \caption{A solid immersed in an electric field. \label{immerse}}
\end{wrapfigure}

When you immerse a solid, either an insulator or a semiconductor, in an electric field (see Fig. \ref{immerse}), the dipoles inside the material get orientated along the field lines and create an internal field, 
the polarisation $\PPc$, opposite to the field that generates it. This naive picture, even if not correct for extended systems, gives us an idea of the effect of an external electric field on a material.
The total electric field inside the solid $\EEc(\rr,t)$ is the sum of the external plus the polarisation one:
\be
\EEc(\rr,t) = \DDc(\rr,t) - \PPc(\rr,t).
\label{materialeq}
\ee
This equation is one of the so-called  "Materials equations", namely the Maxwell equations for electric and magnetic fields in bulk materials, where $\DDc(\rr,t)$ is the Electric Displacement and $ \EEc(\rr,t)$ is the Total Electric Field. In order to understand the origin of these two fields, it is possible to write down their corresponding Gauss's equations:
\ben
\nabla \DDc(\rr,t)  = \frac{\rho_{ext}(\rr,t)}{\epsilon_0} \mbox{ , } \nabla \EEc(\rr,t)  = \frac{\rho_{tot}(\rr,t)}{\epsilon_0}. 
\een
From the above equations one can see that the Electric Displacement is generated by the external charges while the Electric Field is due to the sum of external plus the internal ones, namely the total charge. This explains the structure of Eq.~\ref{materialeq} being the total field equal to the external one minus the polarisation that is the field generated by the internal charges and  opposed to the external one. In general we can expand the polarisation  $\PPc$ in a power series of the total electric field $\EEc$: 
\be
\PPc = \PPc_0 + \chi^{(1)} \EEc + \chi^{(2)} \EEc^2 + \chi^{(3)} \EEc^3 + ....
\label{pexpansion}
\ee
where $\PPc_0$ is the intrinsic polarisation of the material and  $\chi^{(1)}, \chi^{(2)},...$ are the response functions of increasing order. Equation~\ref{pexpansion} is valid for a wide range of situations. However there are cases where this expansion is not valid: 1) for very strong fields, beyond the convergence radius of the expansion\cite{lee2014first}; 2) when there is an hysteresis and therefore there is not a univocal relation between polarisation and electric field; 3) close to phase transitions where a small external field can completely change the material properties. In this thesis we will not consider any of these cases but we will concentrate on the "simpler" one, when the relation between polarisation and electric field can be written in a power series. Moreover in the present work we will always assume $\PPc_0=0$ because we are not interested in materials with intrinsic polarisation as for instance ferroeletrics.\\
\begin{figure}[ht]
  \begin{center}
    \includegraphics[width=0.9\textwidth]{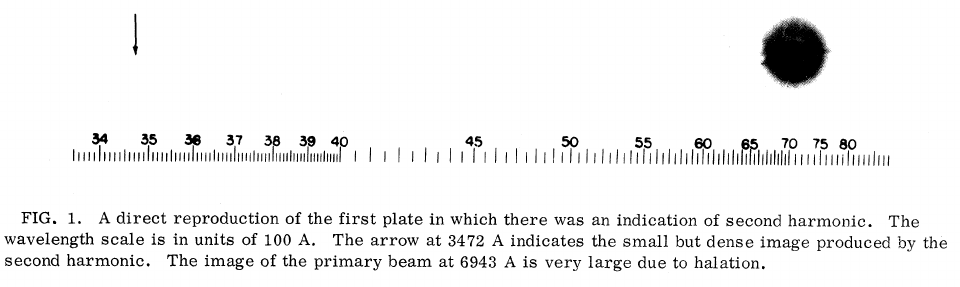}
  \end{center}        
  \caption{The original photographic image of the SHG with the corresponding caption from the Franken's paper\cite{franken1961generation}. \label{frankenfig}} 
\end{figure}
After all these elucidations it is time to introduce non-linear optics. If in Eq.~\ref{pexpansion} we limit the power series to the first term $\chi^{(1)}$ we can describe all the phenomena which belong to the linear optics regime. All the other terms  $\chi^{(2)}, \chi^{(3)},.... $  describe the non-linear response. What does non-linear response mean in practice?  The simplest non-linear phenomena, we obtain from the additional terms,  can be easily understood if we rewrite Eq.~\ref{pexpansion} in frequency domain for an homogeneous material:
\be
\PPc(\omega) = \chi^{(1)} (\omega) \EEc(\omega)  + \chi^{(2)} (\omega = \omega_1 + \omega_2) \EEc(\omega_1) \EEc(\omega_2) + ....
\label{pexpomega}
\ee
From the first term of the RHS we immediately realise that the outgoing light [i.e. the polarisation $\PPc (\omega)$] has the same frequency $\omega$ of the incoming one [i.e. the electric filed $\EEc (\omega)$]. On the contrary terms beyond the linear one contain sum or difference of many electric fields and therefore the outgoing light could have a different frequency, namely a colour different from the incoming one.  For example in the second harmonic generation(SHG), the outgoing light has a frequency that is two times larger than the incoming one. \\
This simple effect, even it is easy to understand and visualise, it is not something evident in our everyday lives. In fact the non-linear coefficients of the polarisation expansion are very small, and therefore in order to obtain a detectable non-linear response one needs a strong light source. This is the reason why the first experimental measurement of second-harmonic generation (SHG) dates 1961\cite{franken1961generation}, just few years after the laser invention.\cite{maiman1960stimulated} In this first experiment of non-linear optics Franken and his collaborators were able to obtain a SHG signal from a ruby crystal employing a monochromatic laser beam with an intensity of $10^5$ volts/cm, see Fig.~\ref{frankenfig}.\\
Nowadays lasers with an intensity equivalent to the one used in the Franken's experiment are commercial available in shops, and SHG became a common technique to double the laser frequency.\\
Of course non-linear optics is not limited to the SHG but the term covers a large spectra of phenomena spanning saturation, sum frequency generation, high harmonic and so on. In the next section we will show some usages and applications of non-linear response.
\section{What can be done with non-linear optics?} 
\begin{wrapfigure}{r}{0.5\textwidth}
    \vspace{-0.7cm}
  \begin{center}
    \includegraphics[width=0.4\textwidth]{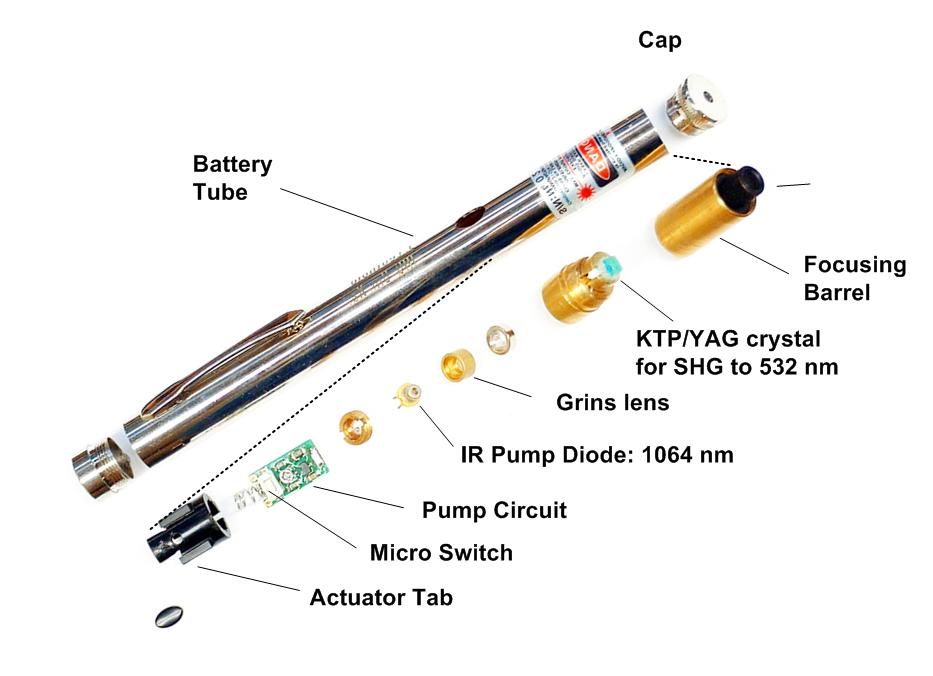}
  \end{center}
  \caption{Schematic of the green laser pointer. \label{greenlaser}}
\end{wrapfigure}
In the last thirty years the field of non-linear spectroscopy\cite{bloembergen1982nonlinear} made progresses in leaps and bounds. The simplest commercial application that everybody knows, is the green laser pointer used in conferences. In this device the green light is obtained combining a red laser with a non-linear crystal that doubles the frequency, see Fig.~\ref{greenlaser}. Nowadays non-linear crystals are routinely used in laboratories to change shape, length and intensity of laser beams. \\
        Applications of non-linear optics are not limited to physics, they range from optoelectronics to medicine. \\
        For example nanocrystals with non-linear properties can be bounded to proteins and then inserted in living systems.
        They become a tool to probe protein dynamics. In fact under intense illumination, such as the focus of a laser-scanning microscope, these SHG nanocrystals modify the light colour and thus they can be imaged by means of the two-photon microscopy. Since biological tissue do not present a particular non-linear response, scientists can visualise the dynamics of the proteins thanks to the nanocrystals.        
        Unlike commonly used fluorescent probes, SHG nanoprobes neither bleach nor blink. The resulting contrast and detectability of SHG nanoprobes provided therefore unique advantages for molecular imaging of living cells and tissues. \cite{pantazis2010second}

\begin{wrapfigure}{r}{0.4\textwidth}
    \vspace{-0.7cm}
  \begin{center}
    \includegraphics[width=0.4\textwidth]{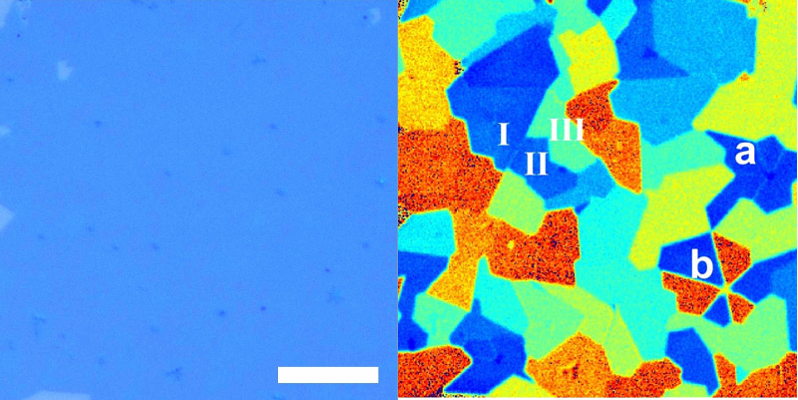}
  \end{center}
  \caption{On the left a linear optics image of a  MoS$_2$ single layer. On the right SHG image of the same layer. \label{mos2} [Figure from Ref.\cite{yin2014edge}]}
\end{wrapfigure}

\begin{wrapfigure}{r}{0.4\textwidth}
 \vspace{-0.8cm}
\begin{center}
\includegraphics[width=0.4\textwidth]{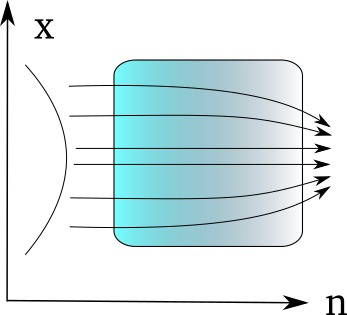}
\end{center}
\vspace{-0.5cm}
\caption{A schematic representation of the self-focusing phenomena in optical fibers. \label{selffocusing}}
\end{wrapfigure}

In quantum optics, non-linear crystals are used to create entangled photons. A photon at high energy is transformed in two or more photons with lower energy by means of reverse second or third harmonic generation. These news outgoing photons can be used in quantum information studies, quantum cryptography or for quantum computation, due to their entangled states.\cite{PhysRevLett.75.4337}\\ 
\begin{wrapfigure}{l}{0.4\textwidth}
    \vspace{-0.7cm}
  \begin{center}
    \includegraphics[width=0.4\textwidth]{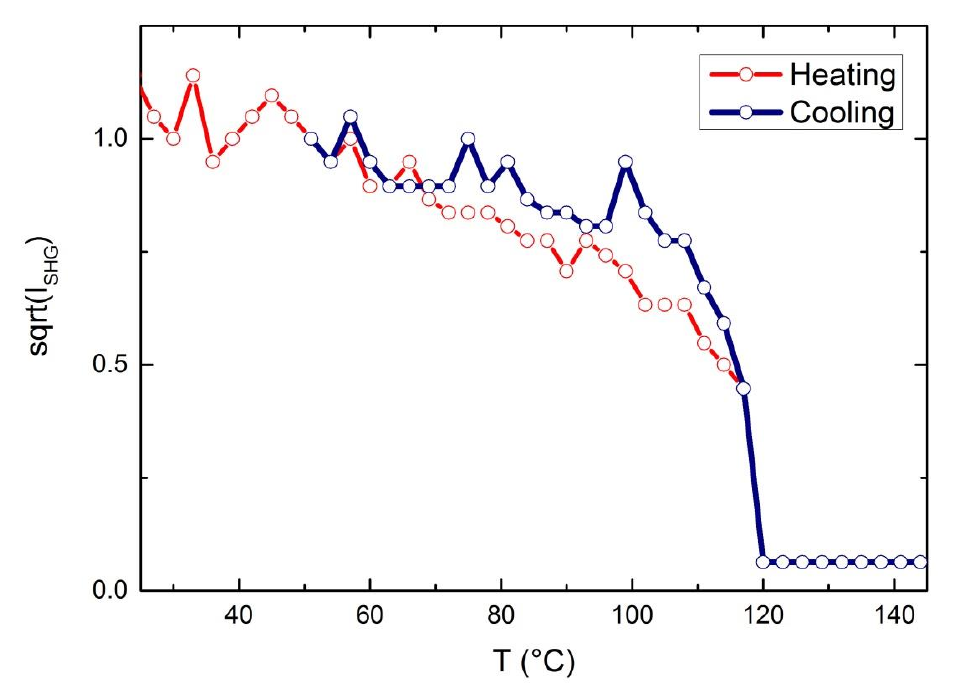}
  \end{center}
  \caption{Square root of SHG signal changing due to temperature variation, in a BaTiO$_3$ crystal. \label{ferroelectric}}
\end{wrapfigure}
In condensed matter the non-linear response remains an essential tool to characterise and explore electronic and structural properties of materials. 
For example, since second harmonic generation can be produced only in materials that lack of inversion symmetry, it became a tool to probe phase transitions and phenomena the break this symmetry. \\
In presence of a macroscopic electric field as the one of piezoelectrics, pyroelectrics, and ferroelectrics or a bulk magnetizations as in ferromagnets the inversion symmetry is broken and a simple SHG measurement as function of the temperature can be used to discriminate between the different phases of these materials. In Fig.~\ref{ferroelectric} it is shown how the SHG signal changes with the temperature in the ferroelectric BaTiO$_3$. At 120\degree~C there is no more signal and the change is very abrupt. 
This confirms that 120\degree~C is the temperature where inversion symmetry is restored in BaTiO$_3$ crystal as it loses its ferroelectric properties.\\
Another import application of non-linear response is the characterisation of surfaces and interfaces. Since SHG is much more sensitive to the lattice orientation, compared with linear optics, it can be used to scan a layer deposited on a surface and to identify  dimension and orientation of the different flakes. 
In a recent experiment~\cite{yin2014edge}  X. Yin et al. used this idea to develop a nonlinear optical imaging technique that allows a rapid and all-optical determination of the crystal orientations in 2D materials at a large scale. In Fig.~\ref{mos2} one sees the main results of  X. Yin et al., on the left there is an image of a single-layer of MoS$_2$ in linear optics and on the right a SHG image of the same layer, where the different colours represent different intensity of the SHG response. The flakes and their orientation are clearly visible in the SHG image.
Instead, Y. Li at al. used SHG to probe the number of layers deposited on a surface, using the fact that an even number of layers posses inversion symmetry while an odd one does not.\cite{doi:10.1021/nl401561r}\\
Another domain where SHG plays a major role is surface spectroscopy. One of the problem occurring to experimentalists is to disentangle bulk from surface contribution in their measurements. SHG is one of the few techniques that can probe the surface without the contributions from the bulk. The reason lies in the fact that in solids with inversion symmetries the bulk contribution is zero and the only source of SHG is the surface one. This is true not only for bulk materials but also for liquids that are in average symmetric but not at the liquid-liquid or gas-liquid interfaces. In  this cases SHG provides  great insights on the surface structure that sometime are difficult to probe with other techniques.\cite{eisenthal1996liquid} \\ 
The importance of non-linear response for solids characterisation is not limited to the SHG, but also other response functions find applications in condensed matter physics. For example two-photon absorption that is proportional to the imaginary part of the $\chi^{(3)}$ can be used to probe excited states that are dark in linear optics.\cite{wang2005optical,cassabois2015hexagonal}\\ 
Finally we want to conclude this section showing some negative sides of the non-linear response. While in many applications the non-linear response is the desired effect, there are cases where one tries to avoid any non-linear phenomena. 
For example one of the limiting factor of the light power that can be transported by optical fibers is the self-focusing phenomena. Self-focusing is a non-linear optical process generated by the third harmonic response in materials exposed to intense electromagnetic radiation. 
A medium whose refractive index is modified by the $\chi^{(3)}$ response acts as a focusing lens for an electromagnetic wave characterised by an initial transverse intensity gradient, as the one generated by a laser beam (see Fig.~\ref{selffocusing}).

The peak intensity of the self-focused region keeps increasing as the wave travels through the medium until medium damage interrupts this process. At present no method is known for increasing the self-focusing limit in optical fibers\cite{encylaser}.\\
In this section we covered a minimal part of the positive and negative non-linear phenomena in research and applications, for a general overview different books and reviews are available in literature.\cite{boyd,bloembergen1982nonlinear}
\section{How to calculate non-linear response}
The first attempt to calculate non-linear optical response in solids from a quantum mechanics was done by means of density matrix formalism.\cite{bloembergen1964quantum} This formalism was already used in the past to derive local field effects in linear optics\cite{PhysRev.126.413,wiser1963dielectric}, to investigate  a non-linear phenomena like saturation of microwave resonances\cite{karplus1948note}, and to describe nuclear magnetic relaxation\cite{kubo1954general,RevModPhys.33.249,PhysRev.102.104}.              
One particular advantage of this formalism is that it allows to include damping in an easy way.\\  
The Hamiltonian that enters in the equations of motion(EOM) [Eq.~\ref{eomsimple}] for the density matrix\cite{neumann} can be decomposed in three parts:   $\HH_A$ that determines the unperturbed energy levels of the system, $\HH_{coh}$ that describes the coupling  with the external perturbation, in our case a monochromatic electro-magnetic field, and finally $\HH_{random}$ that includes all relaxation processes.\\
We can thus write down the EOM for the density matrix as:
\be
i \hbar \frac{\partial \rho}{\partial t} = [ \HH_A, \rho] + [ \HH_{coh}, \rho] + i\hbar \left ( \frac{\partial \rho}{\partial t} \right )_{damping}.
\label{eomsimple}
\ee
The last term of this equation is the one generated by the random processes  $\HH_{random}$ and it is due to the coupling with phonon modes, external environments or generated by the electronic correlation. In the literature different phenomenological models have been proposed for the damping therm, the simplest one is:
\be
\left ( \frac{\partial \rho}{\partial t} \right )_{damping} = - \left (\Gamma \rho + \rho \Gamma \right ).
\label{dmeq}
\ee
Where the anti-commutator on the RHS is generated by the non-Hermitian part of the Hamiltonian.\cite{tokman}\\
A steady-state solution for EOM [Eq.~\ref{eomsimple}] in ascending powers of the coupling term may be found from the following hierarchy equations:
\bea
i \hbar \frac{\partial \rho^{(0)}}{\partial t} &=& [ \HH_A, \rho^{(0)}] +  i\hbar \left ( \frac{\partial \rho^{(0)}}{\partial t} \right )_{damping}\\ 
i \hbar \frac{\partial \rho^{(1)}}{\partial t} &=& [ \HH_A, \rho^{(1)}] + [ \HH_{coh}, \rho^{(0)}] + i\hbar \left ( \frac{\partial \rho^{(1)}}{\partial t} \right )_{damping} \\
i \hbar \frac{\partial \rho^{(2)}}{\partial t} &=& [ \HH_A, \rho^{(2)}] + [ \HH_{coh}, \rho^{(1)}] + i\hbar \left ( \frac{\partial \rho^{(2)}}{\partial t} \right )_{damping}. 
\eea
The first equation gives the density matrix at equilibrium. The second equation describes the linear response. By Fourier analysis it is easy to show that $\rho^{(1)}$ must contain the same frequencies as $\HH_{coh}$. The $\rho^{(2)}$  is the first non-linear term. Differently from $\rho^{(1)}$,  $\rho^{(2)}$ oscillates at a frequency that can be the sum or difference of the incoming fields. This term describes second harmonic generation and optical rectification and the dc term. The other terms   $\rho^{(n)}$, describes higher harmonic generations, saturation phenomena and so on.
From these hierarchy equations it is possible to derive the corresponding equations for the response functions $\chi^{(1)}$, $\chi^{(2)}$ ... by differentiating the density matrix respect to the external perturbation. \\
Density matrix formalism is not the only possibility to calculate non-linear response. Expressions for the second order response functions can be also derived directly from perturbation theory.\cite{PhysRevB.56.1787,PhysRevB.42.3567,PhysRevB.82.235201} \\
In the both the methods mentioned above, the response functions and their corresponding Dyson equations are generally expressed by means of sum over states, i. e. valence and conduction bands.\\
Expressing response functions in terms of valence and conduction bands has the advantage to make easy the interpretation of the different peaks appearing in the  $\chi^{(1)}$, $\chi^{(2)}$ ...., however calculations can become prohibitive as the number of bands and k-points increases. For this reason, some groups took a different road to calculate non-linear response functions. Dal Corso and Mauri used the "2n+1" theorem in the time-dependent density functional theory (TDDFT) framework to calculate static nonlinear susceptibilities avoiding the sum over states.\cite{PhysRevB.50.5756}
Other groups used a frequency dependent Sternheimer equation to obtain dynamic polarisabilities and hyperpolarizabilities in molecular systems.\cite{andrade2007time} \\
Finally there is the possibility to follow in real-time the excitation of the system, by solving Eq.~\ref{dmeq} and then analysing the outgoing polarisation or current. Although the real-time solution has a better scaling  with the system size  than  previous mentioned approaches, it is not so common in the scientific literature. The reasons are twofold: first the real-time solution has a large prefactor in the computational time  and therefore only for large systems it starts to become more convenient, and second results analysis is more involved that in the other methods. \\ 
However in last years different works appeared in the literature that use real-time propagation to calculate non-linear response both for molecular\cite{takimoto:154114,ding2013efficient} and periodic systems.\cite{goncharov2013nonlinear}\\
In the next chapters I will present a new real-time approach to study  non-linear response in extended systems that offers different advantages respect the previous methods.\cite{nloptics2013}

\subsection{Correlation effects and non-linear response}
Until now we discussed how to calculate non-linear response but we did not say anything about correlation effects neither on the Hamiltonian that appears in Eq.~\ref{dmeq}. In this section we  briefly outline the different approaches used in the past to take into account these effects in the non-linear response.\\
The first calculations of non-linear response were based on empirical pseudo-potentials, often underestimating or overestimating the experimental values by one or two order of magnitudes.\cite{PhysRevB.12.2325,PhysRevB.36.9708}
In the nineties Levine\cite{PhysRevB.42.3567} presented for the first time an \emph{ab-initio} formalism for the calculation of the second-harmonic generation. Sipe and coworkers extended the calculation of non-linear response to the third harmonic generation eliminating  unphysical divergences that are present in the velocity gauge.\cite{PhysRevB.61.5337,PhysRevB.48.11705}\\ 
Calculations based on ab-initio band structures already improved results respect to previous approaches, however correlation effects were not taken in account yet. Few years later, again Levine and coworkers presented the first calculations of the second harmonic generation including local-field effects and self-energy effects by means of a scissor operator.\cite{PhysRevLett.63.1719,PhysRevB.56.1787} \\
Beyond these effects only a few works included electron-hole interaction in the non-linear response. In particular excitonic effects have been derived by Green's function theory and included by means of generalisations of Bethe-Salpeter equation (BSE)\cite{strinati} to higher order response functions. Following this idea Chang et al.\cite{Chang2002}  and Leitsman et al.\cite{Leitsmann2005} presented an \emph{ab-initio} many body framework for computing the frequency dependent second-harmonic generation that includes local fields and excitonic effects through an effective two-particle Hamiltonian derived from the BSE and found a good agreement with the experimental results.\\
More recently Hubener\cite{PhysRevA.83.062122} presented a full Bethe-Salpeter equation for the second order response functions, while Virk and Sipe derived a similar equation for the third harmonic generation.\cite{PhysRevB.56.1787}\\
Another possible way to include correlation effects, alternative to the Green's function theory, is Time-Dependent Density Functional Theory (TDDFT)\cite{PhysRevLett.52.997}. TDDFT is in principle an exact theory to calculate response functions in finite systems. However the exchange-correlation functional that enters in the equations is unknown and has to be approximated. Standard approximations that rely on local or semi-local functionials miss long range contributions that are responsible of excitonic effects\cite{botti2007time}. Long range contribution can be included in reciprocal or real-space or obtained by means of hybrid functionals\cite{botti2007time,faber2014excited}. For extended systems the situation is more complicated since TDDFT is not an exact theory for the optical response (see Chapter~\ref{chaptertddft} for a discussion).\\
Despite of these problems, TDDFT has been used to calculate linear and non-linear response functions in both finite and extended systems, often with very good results, and also in its real-time formulation.\cite{takimoto:154114,andrade2007time} 

\clearemptydoublepage

\chapter{Dynamical Berry's phase and non-linear response} 
\label{chapterberry}
\section{Why do we need Berry's phase?}
\begin{wrapfigure}{l}{0.5\textwidth}
    \vspace{-0.7cm}
  \begin{center}
    \includegraphics[width=0.4\textwidth]{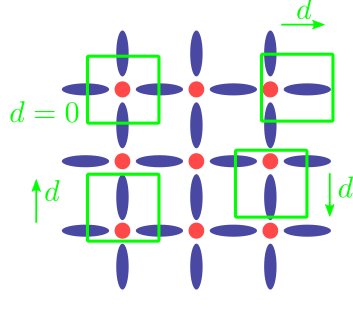}
  \end{center}
  \caption{Polarisation vector versus the choice of the unit cell. \label{cellpol} }
\end{wrapfigure}

This section presents a simple introduction to the problem of polarisation definition in extended systems and how it can be solved by means of Berry's phase concept.
\begin{wrapfigure}{r}{0.5\textwidth}
  \begin{center}
    \includegraphics[width=0.4\textwidth]{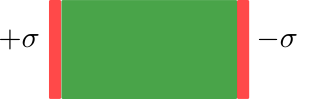}
  \end{center}
  \caption{Surface contribution to the polarisation in a solid. \label{surfacepol}}
\end{wrapfigure}
For many years an unsolved problem in solid state physics was the correct definition of polarisation in periodic systems.
This definition is intrinsically related to the one the dipole operator, that is a problematic object for extended systems.
In the literature different wrong definitions of bulk polarisation have been proposed that we will not cite here\cite{restanotes}.
In order to understand the problem, let's start the discussion from the polarisation in isolated systems.
In a system with open boundary conditions, the dipole operator is well defined and therefore one can write down the polarisation as:
\be
\PP = \frac{e \langle \vec \rr \rangle}{V} = \frac{e}{V}\int \vec \rr n (\rr) d \rr,
\label{polisolated}
\ee
where $n(\rr)$ is the electronic density.
The simplest idea for the definition of the polarization in periodic systems would be to generalise the previous formula. The integral in Eq.~\ref{polisolated} can be redefined in different possible ways in periodic systems. We can average the dipole operator on the whole material or consider its unit cell. In the first case we obtain $ \PP = \langle \vec \rr \rangle_{sample}/V_{sample}$. In an insulator the contributions from the dipoles inside the material cancel each other (as one can see from  Fig.~\ref{immerse}) and only the surfaces contribute to the total polarisation (see Fig.~\ref{surfacepol}):                              
\be
\Delta \PP = \frac{(\Delta \sigma L^2) L }{L ^3},
\label{polsurface}
\ee
where $\Delta \sigma$ is related to the charges accumulated on the surfaces.\cite{vanderbilt1993electric} 
This definition [Eq.~\ref{polsurface}] is not suitable for numerical calculations because it requires the simulation of the entire sample and moreover the above defined polarisation is not a bulk property but it depends from the surfaces.\\ 
The second possibility is to define the polarisation as  $ \PP = \langle \vec \rr \rangle_{cell}/V_{cell}$. But this definition is completely arbitrary. In fact different choices of the unit cell give completely different polarisations for the same material, see Fig.~\ref{cellpol}. A last possibility exists, the use of the dipole matrix elements in terms of Bloch orbitals, but also in this case there is problem since the dipole operator is unbounded in periodic systems.\\
Finally let mention that also the well know Clausius-Mossotti formula for the polarisability\cite{Mossotti} cannot be used in real solids because wave-functions are not localised objects.\\
Two reasons make polarisation definition so difficult in solids. First the dipole operator is ill defined in periodic systems, because $\vec \rr$ is not periodic while wave-functions are. Second, differently from finite systems, the polarisation cannot be expressed as an integral on the charge density\cite{Martin1998}.
This second aspect can be better understood if we write down the general relation between polarisation and density:
\be
\nabla \cdot \PP(\rr) = - n(\rr).
\ee                              
In finite systems we impose the condition  $\PP(\rr) \rightarrow 0 $ outside the sample (Dirichlet boundary condition) and $\int n(\rr) d\rr=0$.
In periodic system, it is most useful to resolve the previous equation into Fourier components $\qq+ \GG$,
where $\GG$ denotes a reciprocal lattice vector and $\qq$ belongs to the first Brillouin zone(BZ):
\be
(\qq + \GG) \cdot \PP(\qq + \GG) = i n (\qq + \GG).
\label{poldens}
\ee
It follows from Eq.~\ref{poldens} that each Fourier component can be treated separately. Now let us consider the limit $\GG=0$ and $\qq \rightarrow 0$. In this limit  the macroscopic polarisation $\PP$ is not determined anymore by the zero Fourier component of the density, which must vanish by charge neutrality.  Thus in the limit $\qq = 0$ for an infinite crystal, the polarisation contains additional information not included in the density.\cite{Martin1998} \\
The problem of a correct definition of polarisation in periodic systems was solved in 1993 by  King-Smith and Vanderbilt.\cite{KSV1} In their seminal paper they shown that  bulk polarisation can be expressed as a closed integral on the wave-function phase in the Brillouin zone, a particular case of the Berry's phase. Their formulation solved all problems with the previous attempts to define the polarisation. In fact the King-Smith and Vanderbilt(KSV) polarisation is a bulk quantity, its time derivative gives the current and its derivatives respect to the external field reproduce the polarisabilities at all orders.\\
In the next section we will introduce the Berry's phase concept and will present the KSV formula.
\section{A simple introduction to the Berry's phase}
\label{berrysection}
In this section we will introduce the Berry's phase concept\cite{berry1984quantal} and show by means of simple arguments which is its relation with the bulk polarisation.
 
We will not present the full derivation of the King-Smith and Vanderbilt formula for the polarisation  but we will explain the physical meaning of the different terms appearing in the formula and how they are related to the Berry's phase.\\                                         
Mathematical derivation of the KSV polarisation can be found in their original paper\cite{KSV1} or in its generalisation to the many-body case\cite{PhysRevLett.80.1800}. \\ 
Suppose you have an Hamiltonian $\HH(\bxi)$ that depends from an external parameter $\bxi$. For each value of $\bxi$ it is possible to diagolalise the Hamiltonian and obtain:
\be
H(\bxi) | \psi(\bxi) \rangle = E(\bxi) | \psi(\bxi) \rangle.
\ee
where  $E(\bxi)$ and $\psi(\bxi)$ are respectively the eigenvalue and eigenstate of $H(\bxi)$ for a fixed value of $\bxi$.
Now we can define the phase difference between two ground states with different $\bxi$ values as: 
\bea
e^{-i \Delta \phi_{12}} &=& \frac{\langle \psi( \bxi_1) | \psi(\bxi_2) \rangle}{|\langle \psi(\bxi_1) | \psi (\bxi_2) \rangle |},\\
\Delta \phi_{12} &=& -Im \log{ \langle \psi (\bxi_1) | \psi (\bxi_2)} \rangle.
\eea
This definition is similar to the one used in geometry to define the angle between two vectors. 
Unfortunately $\Delta \phi_{12}$ cannot have a physical meaning, because the phase of the wave-function is arbitrary and so the phase difference. However if we construct a closed-path in the space spanned by the $\bxi$ parameter we get something new:
\bea  
\gamma &=&\Delta \phi_{12} + \Delta \phi_{23} + \Delta \phi_{34}  + \Delta \phi_{41} \nonumber \\
       &=& - Im \log \langle \psi (\bxi_1) | \psi (\bxi_2) \rangle  \langle \psi (\bxi_2) | \psi (\bxi_3) \rangle \langle \psi (\bxi_3) | \psi (\bxi_4) \rangle  \langle \psi (\bxi_4) | \psi (\bxi_1) \rangle.  \label{discretberry}
\eea
Now the total phase change $\gamma$ is gauge invariant because each wave-function appears both as ket and bra in the previous formula.  In physics a gauge-invariant object is a potential physical observable, as for instance the eigenvalues of an Hermitian operator. However  $\gamma$  is an "exotic" observable because it cannot be expressed in term of any Hermitian operator. The reason for the existence of this strange kind of observables lies in the fact that the Hamiltonian is not isolated and the parameter $\bxi$ represents the coupling with "the rest of the universe" (to use a sentence from Berry's paper). In a truly isolated system there cannot be any manifestation of the Berry's phase and all observables are eigenvalues of an Hermitian operator.\\
\begin{figure}[ht]
  \begin{center}
    \includegraphics[width=0.5\textwidth]{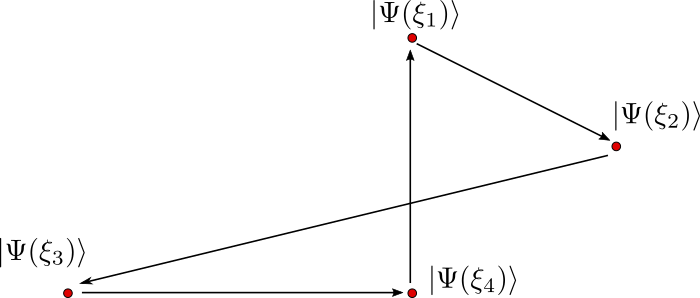}
  \end{center}
  \caption{Closed path in the space of $\bxi$ parameter. \label{closedpath}}
\end{figure}
Different phenomena can be described in terms of Berry's phase, as for instance the Aharonov-Bohm effect\cite{wilczek1989geometric}, the Wannier-Stack ladder spectra of semi-classical electrons\cite{zak1989berry}, and the Molecular Aharonov-Bohm Effect.\footnote{The interpretation of this phenomena in terms of Berry's phase has been questioned in recent years. For a discussion see Ref.~\cite{min2014molecular} and references there in.}\\
Now we will show that also the polarisation in periodic systems can be expressed as a Berry's phase integral. We start from the simple case of non-interacting electrons.
The solution of a single particle  Schr\"{o}dinger equation in an infinite crystal reduces to the one of the primitive cell with Born-von-Karman boundary conditions:
\be
\phi_{n,\kk} (\rr + \RR) = e^{i\kk \rr} \phi_{n,\kk}(\rr)
\ee
where $\phi_{n,\kk} (\rr)$ are solution of:
\be
\left [ \frac{1}{2m}  \pp^2 + V(\rr)\right] \phi_{n,\kk} (\rr)= \epsilon_n(\kk)\phi_{n,\kk} (\rr).
\label{schonebody}
\ee

The Bloch theorem\cite{bloch1929quantenmechanik} guarantees that these wave-functions can be expressed as:
\be
\phi_{n,\kk} (\rr) = e^{i\kk \rr} u_{n,\kk}(\rr)
\label{blochwf}
\ee
where $u_{n,\kk}(\rr)$ obeys to periodic boundary conditions. If we now insert the Bloch wave-functions Eq.~\ref{blochwf} in Eq.~\ref{schonebody} we get a new equation for the periodic orbitals $u_{n,\kk}(\rr)$:
\be
\left [ \frac{1}{2m} \left ( \pp + \hbar \redk \right ) + V(\rr)\right] u_{n,\redk} (\rr)= \epsilon_n(\redk) u_{n,\redk} (\rr).
\label{schequ}
\ee
In this new equation we mapped a problem with $\kk$-dependent boundary conditions in a new problem with periodic boundary condition plus an Hamiltonian that depends parametrically on $\kk$.
\begin{wrapfigure}{l}{0.5\textwidth}
    \vspace{-0.7cm}
  \begin{center}
    \includegraphics[width=0.4\textwidth]{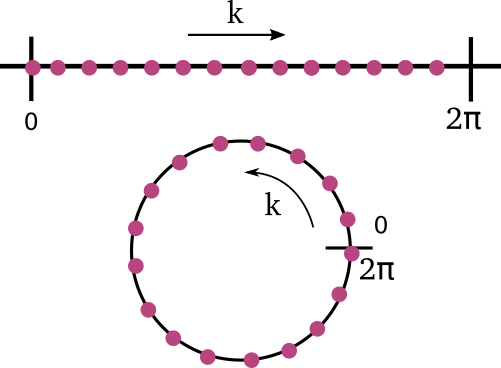}
  \end{center}
  \caption{Schematic representation of one-dimensional Brillouin zone, and the corresponding integration path of Eq.~\ref{berry1D}.\label{1dbz}}
    \vspace{-0.7cm}
\end{wrapfigure} 
A careful reader will now recognise that we are back to a situation similar to the one presented in section~\ref{berrysection}, an Hamiltonian that depends from an external parameter $\kk$.  The question arises as which physical observable represents the phase change generated by a closed path in the $\kk$ space. This question was answered for the first time in 1993 by King-Smith and Vanderbilt.\cite{KSV1}
They shown that the observable associated to the Berry's phase of the $\kk$ vector is the polarisation. Their formula reads:
\be
\PP_\alpha = \frac{2ie}{(2\pi)^3} \int_{BZ} d\kk \sum_{n=1}^{M} \langle  u_{n,\kk} | \frac{\partial}{\partial \kk_\alpha}| u_{n,\kk}\rangle
\label{berry1D}
\ee
where M is the number of valence bands, and the integral is performed along  the Brillouin zone, as shown in Fig.~\ref{1dbz}.

The KSV polarisation [Eq. \ref{berry1D}] is the natural extention to periodic systems of the well known formula [Eq. \ref{polisolated}] for isolated ones. In order to see this let's rewrite explicitly all terms appearing in both formula:
\bea
\PP^{KVS}_\alpha &=& {\color{magenta} \frac{2e}{v }} \frac{1}{V_{BZ}} \int_{BZ} \sum_{n=1}^{M} \langle  u_{n,\kk} | {\color{blue} i\frac{\partial}{\partial \kk_\alpha}}| u_{n,\kk}\rangle\\
\PP_\alpha &=& {\color{magenta} \frac{2e}{V} }\sum_{n=1}^{M} \langle  u_{n} | {\color{blue} \vec \rr_\alpha}| u_{n}\rangle.
\eea
We can see here that in periodic systems the dipole operator is replaced by $i\frac{\partial}{\partial \kk_\alpha}$, the volume of the system by the one of the cell  $v$ and the integral on the wave-functions is split in a sum on k-points plus an integral for each k-point.  Notice that the dipole operator as derivative respect to the $\kk$-point will appear again in the equation of motion when we define the Hamiltoanian in presence of an external electric field (see Sec.~\ref{ss:fldcpl})\\ 
The previous formula has been also extended to the case of a finite k-point sampling and more dimensions. The integral in dimensions larger than one is performed along periodic lines in the BZ  and summed up along the perpendicular directions (see Section~\ref{ss:fldcpl}).  King-Smith and Vanderbilt demonstrated the validity of Eq.~\eqref{berry1D} by means of Wannier functions. They supposed that is possible to map Bloch orbitals in maximal localised Wannier functions then wrote the polarisation in terms of the last ones, and finally show that this is equivalent to the Eq.~\eqref{berry1D}. Another proof was proposed some years later by R. Resta that generalised the previous formula to the many-body case and show that it reproduces the integral of the total current.\cite{PhysRevLett.80.1800} \\
Now that we introduced the relation between Berry's phase and polarisation we are ready to use it for the non-linear response. In the next sections we will introduce the non-linear response in solids and present a new real-time formalism to calculate it. 
\section{General introduction}
\emph{Ab-initio} approaches based on Green's function theory became a standard tool for quantitative and predictive calculations of linear response optical properties in Condensed Matter. In particular, the state-of-the-art approach combines the $G_0W_0$ approximation for the quasi-particle band structure~\cite{aryasetiawan1998gw} with the Bethe-Salpeter equation in static ladder approximation for the response function.~\cite{strinati} This approach proved to effectively and accurately account for the essential effects beyond independent particle approximation (IPA) in a wide range of electronic systems, including extended systems with strong excitonic effects.~\cite{Onida}

In contrast, for nonlinear optics \ai calculations of extended systems rely in large part on the IPA\cite{PhysRevB.48.11705} with correlation effects entering at most as a rigid shift of the conduction energy levels\cite{PhysRevB.80.155205}.  Within time-dependent density-functional theory (TDDFT), it has been recently proposed~\cite{PhysRevB.82.235201} an approach to calculate the second-harmonic generation (SHG) in semiconductors that takes into account as well crystal local-field and excitonic effects. However, this promising approach~\cite{Cazzanelli2012} is limited by the treatment of the electron correlation to systems with weakly bound excitons.~\cite{LRC} 

Within Green's function theory the inclusion of many-body effects into the expression for the nonlinear optical susceptibilities is extremely difficult. 
Furthermore the complexity of these expressions grows with the perturbation order. Therefore it is not surprising that there have been only few isolated attempts to calculate second-order optical susceptibility using the Bethe-Salpeter equation~\cite{Leitsmann2005,Chang2002} and no attempt to calculate higher-order optical susceptibilities.~\cite{PhysRevB.80.165318} 

Alternatively to the frequency-domain response-based approach, one can obtain the nonlinear optical susceptibility in time-domain from the dynamical polarisation $\PP$ of the system by using the expansion of $\PP$ in power of the applied field
\be
\label{eq:peopbf}
\PP= \chi^{(1)} \efield + \chi^{(2)} \efield^2 + \chi^{(3)} \efield^3 + \dots 
\ee 
This strategy is followed in several real-time implementations of TD-DFT\cite{PhysRevB.54.4484}. In these approaches the dynamical polarisation is obtained by numerical integration of the equations of motion (EOMs) for the Kohn-Sham system.\cite{takimoto:154114,castro:3425,meng:054110} So far applications regard mostly nonlinear optical properties in molecules.

The time-domain approach presents three major advantages with respect to frequency-domain response-based approaches. First, many-body effects are included easily by adding the corresponding operator to the effective Hamiltonian. Second, it is not perturbative in the external fields and therefore it treats optical susceptibilities at any order without increasing the computational cost and with the only limitation dictated by the machine precision. Third, several non-linear phenomena and thus spectroscopic techniques are described by the same EOMs. For instance, by the superposition of several laser fields one can simulate sum- and difference-frequency harmonic generation, or four-waves mixing.\cite{boyd}

Although this approach shows very promising results for molecular systems, its extention to periodic system remains still limited.
In fact, due to the problems in defining the position operator and thus $\PP$, it is not trivial to apply Eq.~\eqref{eq:peopbf} to systems in which periodic boundary conditions (PBC) are imposed. As it was recognised for example in Ref.~\cite{PhysRevB.52.14636}, the same problem appears in the direct evaluation of the nonlinear optical susceptibility in frequency-response based approaches. In particular the dipole matrix elements between the periodic part of the Bloch functions are ill-defined when using the standard definition of the  position operator. In that case, it is possible to obtain correct expressions for the dipole matrix elements from perturbation theory~\cite{PhysRevB.52.14636,PhysRevB.48.11705,PhysRevB.82.235201,korbel2015optical} at a given order in the external field. Instead, in the real-time approach one needs an expression valid at each order of the perturbation.

A correct definition of the polarisation operator in systems with PBC has been introduced by means of the geometric Berry phase in the Modern theory of polarisation.\cite{RevModPhys.66.899} 
To our knowledge different schemes for calculating the electron-field coupling consistently with PBC have been proposed in Refs.~\cite{springborg, PhysRevB.76.035213, souza_prb, korbel2015optical}. In those works the dipole matrix elements are evaluated numerically from the derivative in the crystal-momentum ($\kk$) space. The latter cannot be carried out trivially because of the freedom in the gauge of the periodic part of the Bloch functions. In fact, the gauge freedom leads to spurious phase differences in the Bloch functions at two neighbouring $\kk$ points and ultimately to spurious contributions to the numerical derivative.
Then, basically the four schemes~\cite{springborg, PhysRevB.76.035213, souza_prb, korbel2015optical} differ in how the gauge is fixed to eliminate the spurious phase.

This chapter presents a real-time \ai approach to nonlinear optical properties for extended systems with PBC in which the nonlinear optical susceptibility are obtained through Eq.~\eqref{eq:peopbf}. To derive the EOMs we follow the scheme of Souza et al.\cite{souza_prb} based on the generalisation of Berry's phase to the dynamical polarisation (Sec.~\ref{ss:fldcpl}). Originally applied to a simple tight-binding Hamiltonian, this approach is valid for any single-particle Hamiltonian and, as we discuss in Sec.~\ref{ss:correff}, it can be applied in an \ai context with inclusion of the relevant many-body effects. After detailing on how nonlinear optical susceptibility is extracted from the dynamical polarisation (Sec.~\ref{sc:compdet}), we show results for the second-harmonic generation (SHG) in semiconductors (Sec.~\ref{sc:results}) and successfully validate them against existing results from the literature obtained by response theory in frequency domain.   
   


\section{Theoretical background}\label{sc:theory}
We consider a system of $N$ electrons in a crystalline solid of volume $V=Mv$ (where $M$ is the number of the equivalent cells and $v$ the cell volume) coupled with a time-dependent electric field $\efield$
\be
H(t)=H^0 + H^{\efield}(t), \label{eq:startH}
\ee
where $H^0$ is the zero-field Hamiltonian, and $H^{\efield}(t)$ describes the coupling with the electric field. 
Here, we consider a generic single-particle Hamiltonian $H^0$. In Sec.~\ref{ss:correff} we specify the form of $H^0$ and show how many-body effects are included by means of effective single-particle operators. Of course, the choice of a single-particle Hamiltonian prevents applications to systems with strong static correlation such as Mott insulators or frustrated magnetic materials.
We assume the ground state of $H^0$ to be non-degenerate and a spin-singlet so that the ground-state wave-function can be expressed as a single Slater determinant. 
We also assume, as usual in treating cell-periodic systems, Born-von K\'arm\'an PBC and define a regular grid of  $N_\kk=M$  $\kk$-points in the Brillouin zone. With such assumptions, the single-particle solutions of $H^0$ are Bloch-functions.

Regarding the electron-field coupling we assume classic fields and use the dipole approximation, $H^{\efield}(t)=e\efield(t)\hat r$ ($-e$ is the electronic charge).  However, because of the PBC the position operator is ill-defined. In order to obtain a form for the field coupling operator compatible with  Born-von K\'arm\'an PBC, in this chapter we use the Berry's phase formulation of the position operator and  consequently the polarisation. As proved in Ref.~\cite{souza_prb}, in this formulation the solutions of  $H(t)$ are also in a Bloch function form: $\phi_{\kk,n}(\rr,t) = \mathrm{exp}(i\kk\cdot\rr) v_{\kk,n} (\rr,t)$, with  $v_{\kk,n}$ being the periodic part and $n$ being the band index. Notice that, even in the Berry's phase formulation, for very strong fields and with the number of $\kk$-points that goes to infinity the Hamiltonian Eq.~\ref{eq:startH} is unbounded from below due to the Zener tunnelling.\cite{springborg} Nevertheless the strength of the fields used in non-linear optics is well below this limit.\cite{springborg,souza_prb}\\
In Sec.~\ref{ss:fldcpl} we detail how, by starting from the Berry's phase formulation of polarisation, we  obtain the EOMs in presence of an external electric field within PBC.                 
\subsection{Treatment of the field coupling term}\label{ss:fldcpl}
\subsubsection{Berry's phase polarisation}
In this section we take a different path to obtain the KSV polarisation. We start from the many-body polarisation operator proposed by R. Resta and then we derive the single particle one, i.e.  the KSV polarisation.\\
Developed in the mid-90s the Modern Theory of Polarisation\cite{RevModPhys.66.899} provides a correct definition for the macroscopic bulk polarisation, not limited to the perturbative regime, in terms of the many-body geometric phase 
\be 
\PP_\alpha =  \frac{e N_{\kk_\alpha} \lv_\alpha}{2\pi V} \mbox{Im ln }  \langle \Psi_0 | {\rm e}^{i \qq_\alpha \cdot \hat{\mathbf X}} | \Psi_0 \rangle . \label{limit} 
\ee
In Eq.~\eqref{limit} $\PP_\alpha$ is the macroscopic polarisation along the primitive lattice vector $\lv_\alpha$, $\hat{\mathbf X} = \sum_{i=1}^{N} \hat{\mathbf x}_i$, $\qq_\alpha = \frac{\bb_\alpha}{N_{\kk_\alpha}}$ with $\bb_\alpha$ the primitive reciprocal lattice vector such that $\bb_\alpha\cdot\lv_\alpha=2\pi$, and $N_{\kk_\alpha}$ the number of $\kk$-points along $\alpha$, corresponding to the number of equivalent cells in that direction, $\qq_\alpha$ is the smallest distance between two k-points along the $\alpha$ direction.
Note that in this formulation the polarisation operator is a genuine many-body operator that cannot be split as a sum of single-particle operators. \\
The polarization defined by the Eq.~\ref{limit} is valid for any many-body wavefunction on lattice or continuum\cite{PhysRevLett.80.1800,resta1999electron}, now we will see how this expression gets simplified in case of a single Slater determinant.



By using the assumption that the wave-function can be written as a single Slater determinant,
the expectation value of the many-body geometric phase in Eq.~\eqref{limit} can be seen as the overlap between two single Slater determinants. 
The latter is equal to the determinant of the overlap $\cal S$ matrix built out of $\phi_{\kk_j,m}$, the occupied Bloch functions
\be
{\cal S}_{\kk m, \mathbf{k'} m'} = \langle \phi_{\kk,m} | e^{-i \mathbf q_\alpha \hat x}|\phi_{\mathbf {k'},m'} \rangle \label{sint}.
\ee 

Then we can rewrite Eq.~\eqref{limit} as
\be
\mathbf P_\alpha = -\frac{ef \lv_\alpha}{2 \pi N_{\kk_\alpha^\perp} v} \mbox{Im ln det } {\cal S}, \label{eq:Pipa}
\ee
where $f$ is the spin degeneracy, equal to $2$ since we consider here only spin-unpolarized systems, and $N_{\kk_\alpha^\perp}$ is the number of $\kk$-points in the plane perpendicular to reciprocal lattice vector $\bb_\alpha$, with $N_{\kk} = N_{\kk_\alpha^\perp}\times N_{\kk_\alpha}$.


The overlap $\cal S$ has dimensions $n_b N_\kk\times n_b N_\kk$, where $n_b$ is the number of doubly occupied bands. However, from the properties of the Bloch functions and by imposing they satisfy the so-called ``periodic gauge'' $\phi_{\kk+ \mathbf G} = \phi_{\kk}$, it follows that the integrals in Eq.~\eqref{sint} are different from zero only if $\kk' -\kk = \qq_\alpha$.  Therefore the determinant of $\cal S$ reduces to the product of $N_{\kk_\alpha}$ determinants of overlaps $S$ built out of $v_{\kk,m}$, the periodic part of the occupied Bloch functions:
\be
S_{mn}(\kk , \kk + \qq_\alpha) = \langle v_{\kk,m} | v_{\kk + \qq_\alpha,n} \rangle. 
\label{eq:sovlps}
\ee
This leads to the formula by which we compute the polarisation of the system 
 \begin{equation}
     \mathbf P_\alpha = -\frac{ef}{2 \pi v} \frac{\mathbf a_\alpha}{N_{\kk_\alpha^\perp}} \sum_{\kk_\alpha^\perp} \mbox{Im ln} \prod_{i=1}^{N_{\kk_\alpha}-1}\ \mbox{det } S(\kk_i , \kk_i + \mathbf q_\alpha). \label{berryP2} 
 \end{equation}
Using matrix properties,~\cite{mcookbook} the logarithm of the matrix determinant can be rewritten as the trace of matrix logarithm, and so Eq.~\eqref{berryP2} can be transformed as 
\be 
\mathbf P_\alpha = -\frac{ef}{2 \pi v} \frac{\mathbf a_\alpha}{N_{\kk_\alpha^\perp}} \sum_{\kk_\alpha^\perp} \mbox{Im} \sum_{i=1}^{N_{\kk_\alpha}-1}\ \mbox{tr ln } S(\kk_i , \kk_i + \qq_\alpha) \label{xtrace}, 
\ee
more suitable to derive the EOMs. By taking the thermodynamic limit ($N_\kk \rightarrow \infty$ and $\qq_\alpha \rightarrow 0$ ) of the latter expression one arrives at the King-Smith and Vanderbilt formula for polarisation.~\cite{KSV1} 
Since in a numerical implementation we deal with a finite number of $N_\kk$ and finite $\qq_\alpha$, we stick here to Eq.~\eqref{xtrace} with $\qq_\alpha = \Delta \kk_\alpha$ to derive the EOMs.  

\subsubsection{Equations-of-motion}
Following Ref.~\cite{souza_prb} we start from the Lagrangian of the system in presence of an external electric field $\efield$:
\be
{\cal L}=\frac{i\hbar}{N_\kk}\sum_{n=1}^M \sum_{\kk}\,
\langle v_{\kk n}|\dot{v}_{\kk n} \rangle-E^0 - v \efield\cdot\PP,
	\label{eq:lagrangian_discrete} 
\ee
where $E^0$ is the energy functional corresponding to the zero-field Hamiltonian:
\be
 E^0= \frac{1}{N_\kk} \sum_{n=1}^M \sum_{\kk}\, \langle v_{\kk n}|\hat H_\kk^0 | v_{\kk n} \rangle, 
\ee
with $\hat H_\kk^0 = e^{-i\kk\rr'}H^0 e^{i\kk\rr} $, and the last term $v \efield\cdot\PP$ is the coupling between the external field and the polarization. Notice that $H^0$ does not connect wave-functions with different $\kk$ vectors.
To simplify the notation we do not explicit the time dependence of the $|v_{\kk n}\rangle$, but they should be considered time-dependent in the rest of the chapter.  

We derive the dynamical equations and the corresponding the Hamiltonian from the Euler-Lagrange equations
\bea
\frac{d}{dt}
\frac{\delta \cal L}{ \langle \delta \dot{v}_{\kk,n} |}-\frac{\delta \cal L}{\langle \delta v_{\kk,n} |} &=&0,\\
i\hbar \frac{d}{dt} |v_{\kk n}\rangle -
\hat H_\kk^0 |v_{\kk n}\rangle -N_\kk v \efield\cdot\frac{\delta \PP}{\langle \delta v_{\kk,n} |} &=&0.
\label{eq:euler-lagrange_b}
\eea
To obtain the functional derivative of the polarisation expression in Eq.~\eqref{xtrace} we use that~\cite{souza_prb,gonze} 
\be
\delta\mbox{tr ln} S = \text{tr}\left[S^{-1}\delta S\right] + {\cal O}(\delta S^2), 
\ee
and that exchanging arguments ($\kk\leftrightarrow \kk'$) in $S$ [Eq.~\eqref{eq:sovlps}] brings a minus sign in Eq.~\eqref{xtrace}.
This leads to (see Ref.~\cite{souza_prb} for details):
\bea
\frac{\delta \PP_\alpha}{\langle \delta v_{\kk,n}|} &=& -\frac{ief}{2 \pi} \frac{\lv_\alpha}{2 N_{\kk_\alpha^\perp} v} \left( | \tilde v_{\kk^{+}_\alpha,n} \rangle - | \tilde v_{ \kk^{-}_\alpha,n} \rangle\right) \label{eq:fdpol}\\
| \tilde v_{ \kk^{\pm}_\alpha,n} \rangle &=& \sum_{m} \left(S(\kk,\kk^{\pm}_\alpha\right)^{-1})_{mn} |v_{\kk^{\pm}_\alpha,m}\rangle, \label{eq:vtilde} 
\eea
where $\kk^{\pm}_\alpha = \kk \pm \Delta \kk_\alpha$,
and from which we can define the field coupling operator
\begin{align}
\hat w_\kk(\efield) =\frac{ief}{4 \pi}\sum_m \sum_{\alpha=1}^3 \left(\lv_\alpha\cdot\efield\right)N_{\kk_\alpha} \sum_{\sigma=\pm} \sigma |\tilde v_{\kk^\sigma_\alpha,m} \rangle\langle v_{\kk,m}|. 
\label{eq:fldcpl}
\end{align}
Notice that the field coupling operator in Eq.~\eqref{eq:fldcpl} is nonhermitian. In order to have well defined Hermitian operators in the EOMs we replace $\hat w_\kk(\efield)$ with $\hat w_\kk(\efield) + \hat w^\dagger_\kk(\efield)$. This is possible because at any time $\hat{\rm w}_{\kk}^{\dagger}|v_{\kk n}\rangle=0$~\cite{souza_prb}.  
Finally, by using Eqs.~\eqref{eq:fdpol}-\eqref{eq:fldcpl} in Eq.~\eqref{eq:euler-lagrange_b} and the Hermitian field coupling operator we obtain the EOMs:
\be
i\hbar  \frac{d}{dt}| v_{\kk,m} \rangle = \left(\hat H_\kk^0 + \hat w_\kk(\efield) +\hat w^\dagger_\kk(\efield) \right)| v_{\kk,m} \rangle. \label{eom}
\ee

Note that Eq.~\eqref{eq:fldcpl} contains a term proportional to 
\be
\frac{1}{2\Delta\kk_\alpha}\left( | \tilde v_{\kk^{+}_\alpha,n} \rangle - | \tilde v_{ \kk^{-}_\alpha,n} \rangle\right)
\label{eq:ncvdrv}
\ee
that has the form of the two-points central finite difference approximation of $\partial_{\kk_\alpha}|v_{\kk_\alpha}\rangle$, but for the fact that $|\tilde v_{\kk^{\pm}}\rangle$ are used instead of $|v_{\kk^{\pm}}\rangle$. As explained in Ref.~\cite{souza_prb}, the $|\tilde v_{\kk^{\pm}}\rangle$ are built from the $|v_{\kk^{\pm}}\rangle$ [Eq.~\eqref{eq:vtilde}] in such a way that they transform as $|v_\kk\rangle$ under a unitary transformation $U_{\kk,nn'}$. 

In fact, there is a gauge freedom in the definition of $|v_\kk\rangle$,  that is $|v_\kk\rangle \rightarrow  U_{\kk} |v_\kk\rangle$, and since the Hamiltonian is diagonalized independently at each $\kk$, the gauge is fixed independently and randomly at each $\kk$. Then, standard (numerical) differentiation will be affected by the different gauge choices at two neighbouring $\kk$-points. Instead the (numerical) derivative in Eq.~\eqref{eq:ncvdrv} is gauge-invariant, or more specifically is performed in a locally flat coordinate system with respect to $U_{\kk,nn'}$. In fact, in the thermodynamical/continuum limit, Eq.~\eqref{eq:ncvdrv} corresponds to the covariant derivative. The problem of differentiating $|v_\kk\rangle$  with respect to $\kk$ has been addressed also in Refs.~\cite{PhysRevB.76.035213,springborg,andrew1998computation,korbel2015optical} that use alternative approaches to ensure the gauge-invariance.
In the here-discussed approach the definition of a numerical covariant differentiation originates directly from the definition of the polarisation as a Berry phase.

\subsection{Treatment of electron correlation}\label{ss:correff}
Correlation effects play a crucial role in both linear\cite{Onida} and non linear\cite{PhysRevB.82.235201,PhysRevB.80.155205} response of solids. 
Since we assumed that $|\Psi_0\rangle$ in Eq.~\eqref{limit} can be written as a single Slater determinant, effects beyond the IPA can be introduced in $\hat H^0$ through an effective time-independent one-particle operator that can be either spatially local as in time-dependent density functional theory, or spatially non-local as in time-dependent Hartree-Fock. 

However, both time-dependent density functional theory and time-dependent Hartree-Fock are not suitable approaches to optical properties of semiconductors: the former, within standard approximations for the exchange-correlation approximations, underestimates the optical gap and misses the excitonic resonances; the latter largely overestimates the band-gap and excitonic effects.   

In the framework of Green's function theory a very successful way to deal with electron-electron interaction in semiconductors is the combination of the $G_0W_0$ approximation for the quasi-particle band structure~\cite{PhysRevB.25.2867} with the Bethe-Salpeter equation in static ladder approximation for the response function.~\cite{strinati}  

We recently extended this approach to the real-time domain~\cite{attaccalite} by mean of non-equilibrium Green's function theory and derived a single particle Hamiltonian that includes correlation from Green's function theory. These many-body corrections and their effect on the non-linear properties will be discussed in Chapters~\ref{chaptercorr} and~\ref{chaptertddft}.\\
In this section we will show how the so-colled local field effects and the quasi-particle corrections enter in our EOMs.\\
Here as starting point for our real-time dynamics, we choose the Kohn-Sham  Hamiltonian at fixed density as a system of independent particles,~\cite{PhysRev.140.A1133} 
\be
\hat H^{0,\text{IPA}} \equiv \hat h^{\text{KS}} = -\frac{\hbar^2}{2m}\sum_{i} \nabla_i^2 + \hat V_{eI} + \hat V_{H}[\rho^0]+ \hat V_{\text{xc}}[\rho^0],      
\label{eq:HIPA}
\ee
where $V_{eI}$ is the electron-ion interaction, $V_{H}$ the Hartree potential and $V_{\text{xc}}$ the exchange-correlation potential.
The advantage of such a choice is that the Kohn-Sham system is the independent-particle system that reproduces the electronic density of the unperturbed many-body interacting system $\rho^0$, thus by virtue of the Hohenberg-Kohn theorem~\cite{PhysRev.136.B864} the ground-state properties of the system. Furthermore, no material dependent parameters need to be input, but for the atomic structure and composition. 

As first step beyond the IPA, we introduce the corrections to the independent-particle energy levels by the electron-electron interaction through a (state-dependent) scissor operator 
\be
\Delta \hat H = \sum_{n,\kk} \Delta_{n,\kk} |v^0_{n,\kk}\rangle\langle v^0_{n,\kk}|.
\ee
 The latter can be calculated \ai e.g., via the $G_0W_0$ approach $\Delta_{n,\kk} = (E^{G_0W_0}_{n,\kk} - \varepsilon^{\text{KS}}_{n,\kk} ) $, or can be determined empirically from the experimental band gap  $\Delta_{n,\kk} = \Delta = E^{\text{exp}}_{\text{GAP}} - \Delta\varepsilon^{\text{KS}}_{\text{GAP}}$. We refer to this approximation as the independent quasi-particle approximation (QPA): 
\be
\hat H^{0,\text{QPA}} \equiv \hat h^{\text{KS}} + \Delta \hat H. 
\label{eq-tdqpa}
\ee
Notice that in our approach the inclusion of a non-local operator in the Hamiltonian does not present more difficulties than a local one, while  this is not a trivial task in the response theory in frequency domain\cite{PhysRevB.82.235201}. 
As a second step we consider the effects originating from the response of the effective potential to density fluctuations. By considering the change of the Hartree plus the exchange-correlation potential in Eq.~\ref{eq:HIPA} we will obtain the TD-DFT response. Here we include just ``classic electrostatic'' effects via the Hartree part. We refer to this level of approximation as the time-dependent Hartree (TDH)
\be
\hat H^{0,\text{TDH}} \equiv \hat H^{0,\text{QPA}} + \hat V_{H} [\rho-\rho^0]. 
\label{eq-tdh}
\ee
In the linear response limit the TDH is usually referred as Random-Phase approximation and is responsible for the so-called crystal local field effects.\cite{PhysRev.126.413} 

Beyond the TDH approximation one has the TD-Hartree-Fock that includes the response of the exchange term to fluctuations of the density matrix $\gamma$. As discussed above this level of approximation is insufficient for optical properties of semiconductors, normally worsening over TDH results. 
Correlation effects beyond TD-Hartree will be discussed in chapter~\ref{chaptercorr}.
We want to emphasise again that within this approach many-body effects are easily implemented by adding terms to the unperturbed independent-particle Hamiltonian $\hat H^{0,\text{IPA}}$ in the EOMs [Eq.~\eqref{eom}]. 
Limitations may arise because of the computational cost of calculating those addition terms. In the specific the large number of $\kk$-points needed to converge the SHG and THG spectra makes more correlated approaches impracticable. However, much less $\kk$-points are needed for converging for example the screened-exchange self-energy itself (see Chapter~\ref{chaptercorr}) and currently we are investigating how to exploit this property and devise ``double grid'' strategies similar to the one proposed in Ref.~\cite{kammerlander}. 
In this chapter effects beyond IPA are limited to the QPA and TDH.

Finally, when the wave-function cannot be approximated anymore with a single Slater determinant (as in strong-correlated systems) the evaluation of the polarisation operator [Eq.~\ref{limit} ] becomes quite cumbersome.\cite{stella} Also we are not aware of any successful attempt to combine Berry's phase polarisation with Green's function theory or density matrix kinetic equations beyond the screened Hartree-Fock approximation (i.e. including scattering terms), even if some appealing approaches have been proposed in the literature\cite{restagw,PhysRevB.84.205137,doi:10.7566/JPSJ.83.033708,nourafkan2013electric}.

\section{Computational scheme and numerical parameters}\label{sc:compdet}

\begin{wrapfigure}{r}{0.5\textwidth}
        \begin{center}
\includegraphics[width=0.4\textwidth]{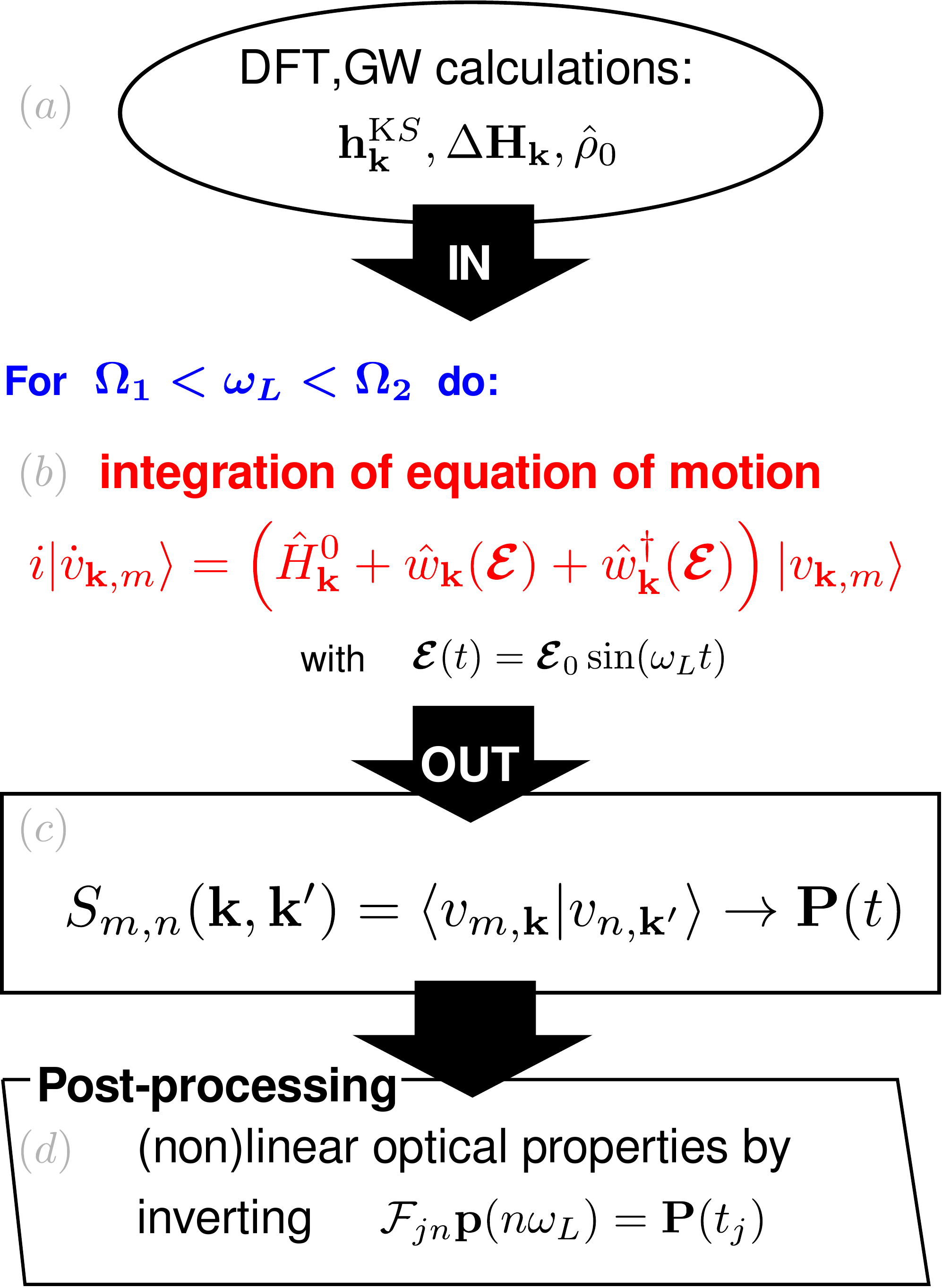}
\caption{\footnotesize{Real-time \ai scheme to compute SHG and THG spectra in the $\[\Omega_1,\Omega_2\]$ energy range for extended systems with PBC: $(a)$ Results from KS-DFT and $G_0W_0$ are input to determine the zero-field Hamiltonian. $(b)$ The EOMs [Eq.~\eqref{eom}] are then integrated,  the polarisation is computed as in Eq.~\eqref{berryP2}. In the post-processing step $(d)$ the nonlinear susceptibilities are obtained by inversion of the Fourier matrix [Eq.~\eqref{eq:fouinv}], see Sec.~\ref{sc:compdet} for details.}}
\label{fg:cmpscm}
\end{center}
\end{wrapfigure}

Figure~\ref{fg:cmpscm} illustrates the computational scheme we use to calculate the SHG and THG spectra. It consists in:
\begin{enumerate}[(a)]
\item we obtain the density, the KS eigenvalues, and eigenfunctions from a planewaves DFT code and then we calculate the quasi-particle corrections within the G$_0$W$_0$ approximation. All these quantities define the zero-field Hamiltonian; 
\item we integrate the equation of motions [Eq.~\eqref{eom}] in presence of  a monochromatic electric field $\efield(t) = \efield_0 \sin(\omega_L t)$ to obtain the $\PP(t)$ from Eq.~\eqref{berryP2}
\item We post-process the $\PP(t)$ to extract the nonlinear susceptibilities. 
\end{enumerate}
The latter two steps are repeated varying the laser frequency $\omega_L$ within the energy range for which we calculate the spectra. 

The scheme in Fig.~\ref{fg:cmpscm} has been implemented in the development version of the {\sc Yambo} code.\cite{yambo} 
Kohn-Sham calculations have been performed using the {\sc Abinit} code,\cite{abinit} and the relevant numerical parameters are summarized in Ref.~\cite{nloptics2013}. All the operators  appearing in the EOMs[Eqs.~\eqref{eom},\eqref{eq-tdh},\eqref{eq-tdqpa}] have been expanded in the Kohn-Sham basis set and the number of bands employed in the expansion is again reported in Ref.~\cite{nloptics2013}. 

Rigorously to have a fully \ai scheme, the scissor operator has to be calculated using e.g., $G^0W^0$. In the examples presented in this chapter we use an empirical values for the scissor operator (reported in Ref.~\cite{nloptics2013}) since the scope is to validate the computational scheme, and to facilitate the comparison with other works in the literature.

The EOMs [Eq.~\eqref{eom}] have been integrated using the following algorithm~\cite{koonin90}
\be
\label{eq:time_evolution}
\ket{v_{\kk n}(t+\Delta t)}= \frac{I-i(\Delta t/2)\hat H^0_{\kk}(t)}{I+i(\Delta t/2) \hat H^0_{\kk}(t)} \ket{v_{\kk n}(t)},
\ee
valid for both Hermitian and non-Hermitian Hamiltonians, and strictly unitary for any value of the time-step $\Delta t$ in the Hermitian case. In all real-time simulations we used a time-step of 0.01 fs.\\

\begin{figure}[ht]
\centering
\includegraphics[width=0.6\textwidth]{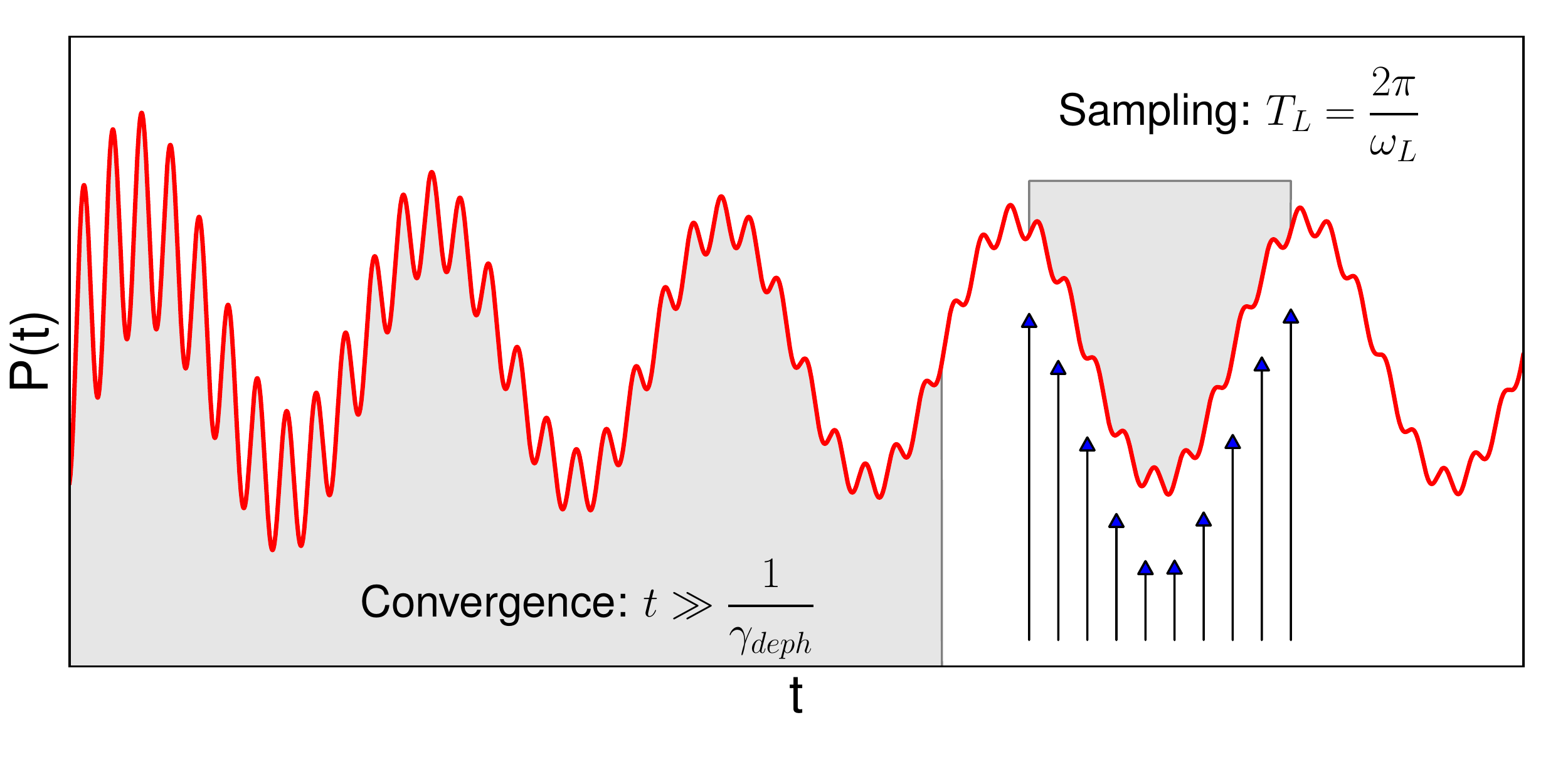}
\caption{\footnotesize{Pictorial representation of the signal analysis in the post-processing step. The signal  $P(t)$ (red line) can be divided into two regions: an initial convergence region (up to $t\gg 1/\gamma_{deph}$) in which the eigenfrequencies of the systems are ``filtered out'' by dephasing and a second region where Eq.~\eqref{eq:frrexp} holds. In this second region the signal $P(t)$ is sampled within a period $T_L=2\pi/\omega_L$ to extract the $P^\alpha_i$ coefficients of Eq.~\ref{eq:fouinv}. Note that $P(t)$ is not a realistic one: for illustration purposes we enhanced the second-harmonic signal that otherwise would not be visible on this scale.}} 
\label{fg:ptanalysis}
\end{figure} 

In our simulations we switch on the monochromatic field at $t=t_0$. This sudden switch excites the eigenfrequencies of the system $\omega^{0}_l$ introducing spurious contributions to the non-linear response.
We thus add an imaginary term into the Hamiltonian $H^{0}_\kk$ to simulate a finite dephasing:   
\be
\Gamma= -\frac{i}{\gamma_{\text{deph}}} \sum_l  \{  | v_{\kk,l}\rangle \langle  v_{\kk,l} | - | v^0_{\kk,l}\rangle \langle  v^0_{\kk,l} | \}
\label{dephterm}
\ee
where $|v^0_{\kk,l}\rangle$ are the valence bands of the unperturbed system and $\gamma_{\text{deph}}$ is the dephasing rate. Then we run the simulations for a time much larger than $1/\gamma_{\text{deph}}$ and sample $\PP(t)$ close to the end of the simulation, see Figure~\ref{fg:ptanalysis}.
Since $\gamma_{\text{deph}}$ determines also the spectral broadening, we cannot choose it arbitrary small. For example in the present calculations we have chosen $1/\gamma_{\text{deph}}$ equal to 6 fs that corresponds to a broadening of approximately 0.2 eV (comparable with the experimental one) and thus we run the simulations for 50-55 fs.\\
Once all the eigenfrequencies of the system are filtered out, the remaining polarisation $\PP(t)$ is a periodic function of period $T_L =\frac{2\pi}{\omega_L}$, where $\omega_L$ is the frequency of the external perturbation and can be expanded in a Fourier series
\be\label{eq:frrexp}
\PP(t) = \sum_{n=-\infty}^{+\infty} \pp_n e^{-i\omega_n t},
\ee  
with $\omega_n = n \omega_L$, and complex coefficients:
\begin{equation}\label{eq:frrcff}
 \pp_n = F\{\PP(\omega_n)\} =\int_{0}^{T_L} dt \PP(t) e^{i\omega_n t}.
\end{equation}
To obtain the optical susceptibilities of order $n$ at frequency $\omega_L$ one needs to calculate the $\pp_n$ of Eq.~\eqref{eq:frrexp}, proportional to $\susc n$ by the $n$-th power of the $\efield_0$. 
However, the expression in Eq.~\eqref{eq:frrcff} is not the most computationally convenient since one needs a very short time step---significantly shorter than the one needed to integrate the EOMs---to perform the integration with sufficient accuracy. As an alternative we use directly Eq.~\eqref{eq:frrexp}: we truncate the Fourier series to an order $S$ larger than the one of the response function we are interested in. We sample $2S+1$ values $\PP_i\equiv\PP(t_i)$ within a period $T_L$, as illustrated in Figure~\ref{fg:ptanalysis}. Then Eq.~\eqref{eq:frrexp} reads as a system of linear equations 
\be
{\cal F}_{in} p^\alpha_n = P^\alpha_i,
\label{eq:fouinv}
\ee 
from which the component $p^\alpha_n$ of $\pp_n$ in the $\alpha$ direction is found by inversion of the $(2S+1)\times(2S+1)$ Fourier matrix ${\cal F}_{in} \equiv \exp(-i\omega_n t_i)$. We found that the second harmonic generation converges with S equal to 4 while the third harmonic requires S equal to 6. Finally we noticed that averaging averaging the results on more periods can slightly reduce the numerical error in the signal analysis. \\
\begin{figure}[ht]
\centering
\includegraphics[width=1\textwidth]{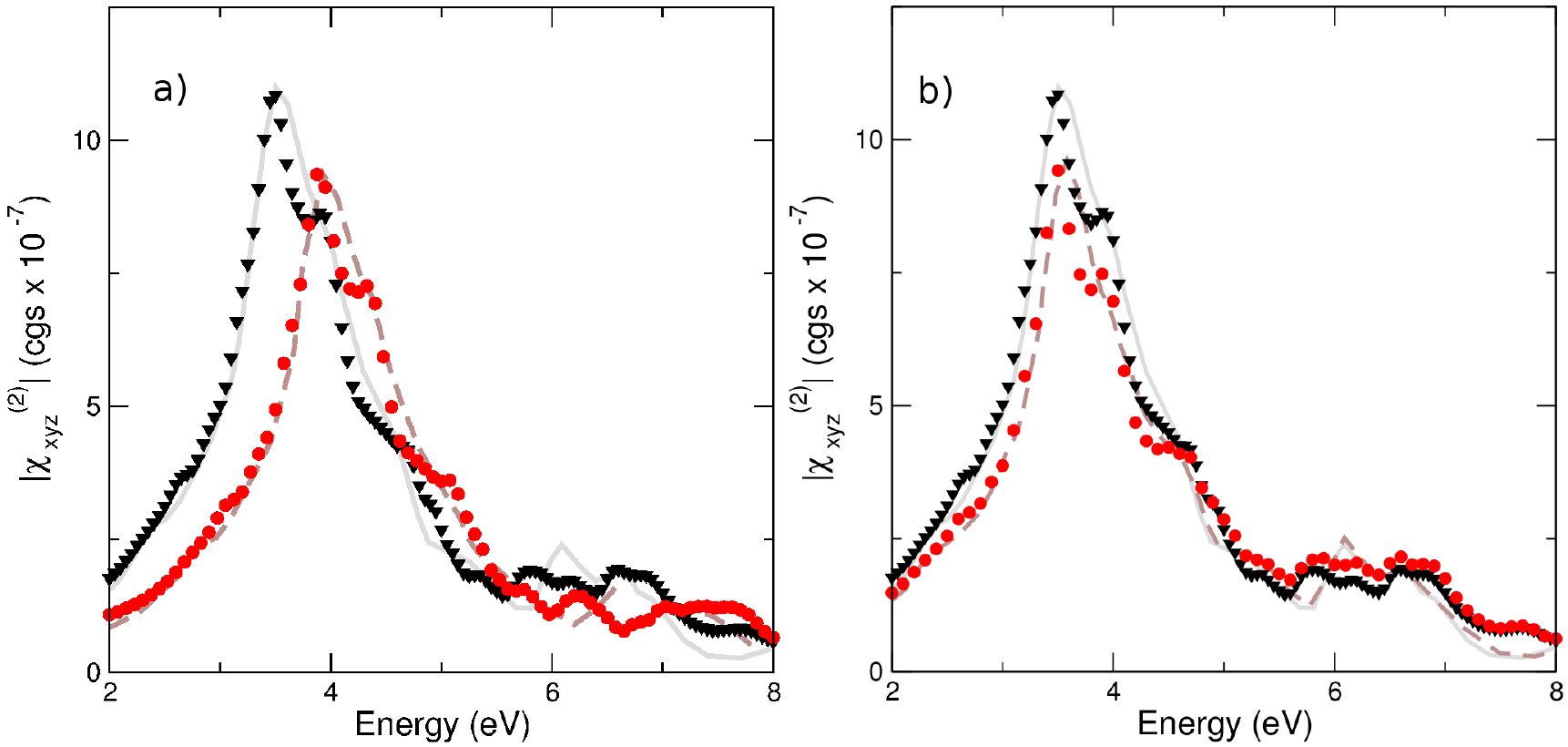}
\caption{\footnotesize{Magnitude of $\chi^{(2)}(-2\omega,\omega,\omega)$ for bulk SiC calculated within the IPA (black triangles) and QPA (red circles) $(a)$ panel and RPA (red circles) $(b)$ panel. Each point corresponds to a real-time simulation at the given laser frequency (see Sec.~\ref{sc:compdet}). Comparison is made with results obtained \ai by direct evaluation of the $\chi^{(2)}$ in Ref.~\cite{PhysRevB.82.235201} in IPA (grey solid line) and QPA (brown dashed line) $(a)$  panel and RPA (brown dashed line) $(b)$ panel.  \label{fg:SiCQPRPA} [Figure from Ref.\cite{nloptics2013}]}}
\end{figure}
Alternatively one can opt for a slow switch on of the electric field as in Takimoto et al.,\cite{takimoto:154114} so that no eigenfrequencies of the system are excited, and avoid to introduce imaginary terms in the Hamiltonian. We found, however, that the latter approach also requires long simulations, and on the other hand, it is less straightforward to extract the $\susc n$.

\section{Results}\label{sc:results}
The main objective of this section is to validate the computational approach described in Secs.~\ref{sc:theory} and \ref{sc:compdet} against results in the literature for SHG obtained by the response theory in frequency domain. In particular we chose to validate against results from Refs.~\cite{PhysRevB.82.235201,PSSB.427.1984} on bulk SiC and AlAs in which the electronic structures is obtained---as in our case---from a pseudo-potential plane-wave implementation of Kohn-Sham DFT with the local density approximation, which makes the comparison easier.
In the following we considered the zinc-blende structure of SiC and AlAs for which the $\susc 2$ tensor has only one independent nonzero component, $\susc 2_{xyz}$ (or its equivalent by permutation).

\begin{figure}[ht]
\centering
\includegraphics[width=1\textwidth]{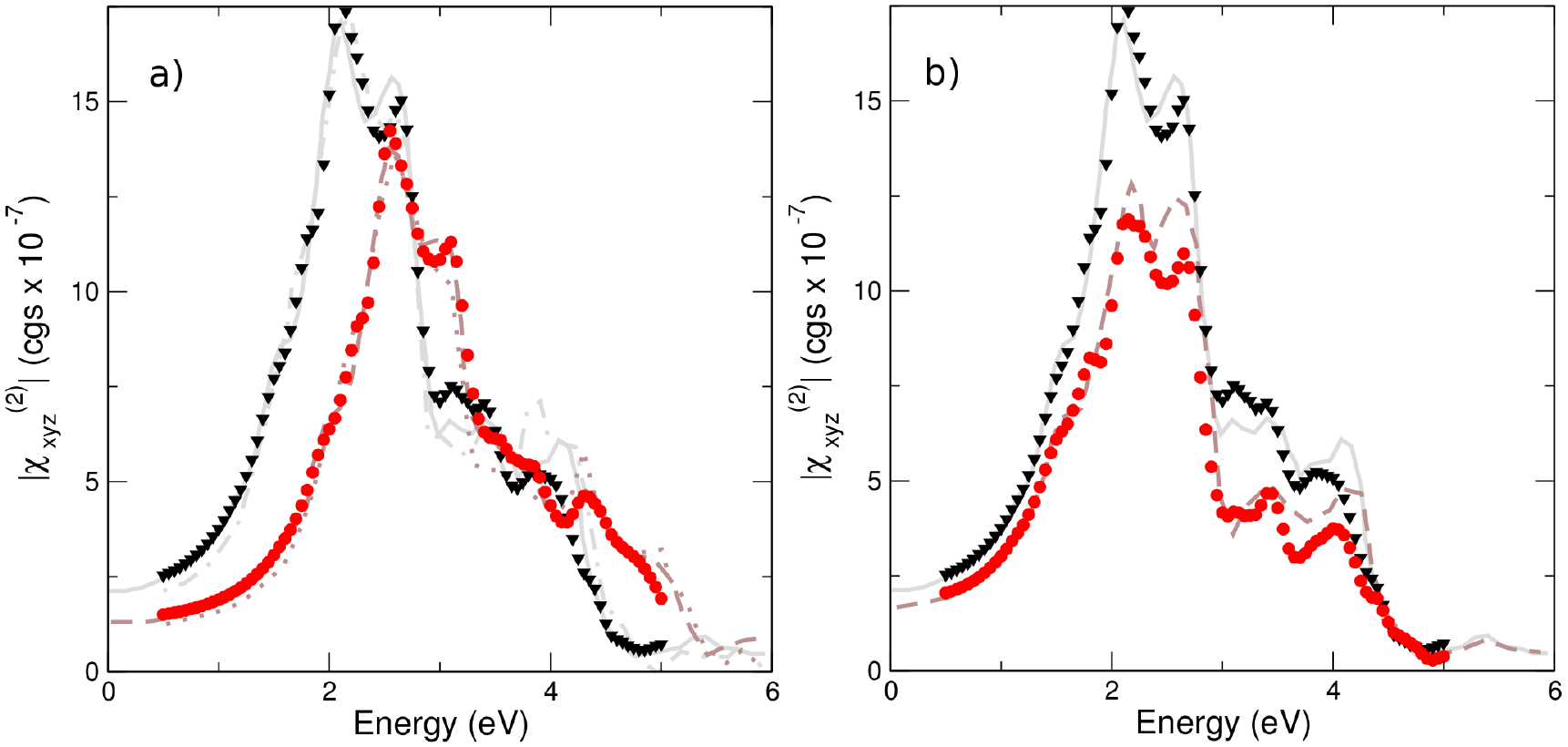}
\caption{\footnotesize{Magnitude of $\chi^{(2)}(-2\omega,\omega,\omega)$ for bulk AlAs calculated within the IPA (black triangles) and QPA (red circles) $(a)$ panel and RPA (red circles) $(b)$ panel. Each point corresponds to a real-time simulation at the given laser frequency (see Sec.~\ref{sc:compdet}). Comparison is made with results obtained \ai by direct evaluation of the $\chi^{(2)}$ in Refs.~\cite{PhysRevB.82.235201,PSSB.427.1984} in IPA (grey solid and dot-dashed line) and QPA (brown dashed and dotted line) $(a)$ panel and RPA (brown dashed and dotted line) $(b)$ panel. \label{fg:AlAsQPRPA} [Figure from Ref.\cite{nloptics2013}] }}
\end{figure}

Figure \ref{fg:SiCQPRPA} show results for the magnitude of SHG in SiC at the IPA, QPA and TDH level of theory. 
At all levels of approximation we obtained an excellent agreement with the results in Ref.~\cite{PhysRevB.82.235201}. The minor discrepancies between the curves are due to the different choice for the $\kk$-grid used for integration in momentum space: we used a $\Gamma$-centred uniform grid (for which we can implement the numerical derivative) whereas Ref.~\cite{PhysRevB.82.235201} used a shifted grid. Figure~\ref{fg:AlAsQPRPA} shows results for the magnitude of SHG in AlAs at the IPA, QPA and TDH level of theory. 
Also in this case results obtained from our real-time simulations agree very well with the reference results and again the small differences between the spectra can be ascribed mostly to the different grid for $\kk$-integration.
As side results we can also observe the effects of different levels of approximation for the Hamiltonian on the SHG spectrum. In order to interpret those spectra note that SHG resonances occur when either $\omega_L$ or $2\omega_L$ equals the difference between two single-particle energies. Then one can distinguish two energy region: below the single-particle minimum direct gap where only resonances at $2\omega_L$ can occur, and above where both $\omega_L$ or $2\omega_L$ resonances  can occur.\\
Regarding the quasi-particle corrections to the IPA energy levels by a scissor operator, below the minimum Kohn-Sham direct band gap the IPA spectrum is shifted by half of the value of the scissor shift (0.4 eV for SiC and 0.45 eV for AlAs) and the spectral intensity reduced by a factor 1.18 (SiC) and  1.25 (AlAs). Above the minimum Kohn-Sham direct band gap instead the QPA spectrum cannot be simply obtained by shifting and renormalizing the IPA one because of the occurrence of resonances at $\omega_{L}$, that are shifted and renormalised differently.\\  
Regarding the crystal local field, their global effect is to reduce the intensity with respect to the IPA. For SiC, the intensity is reduced by about 15\% below the gap, while above the band gap TDH and IPA have similar intensities. For AlAs we observe a reduction of about 30\% in intensity for the whole range of considered frequencies, but for frequencies larger than 4 eV (that is where the $\omega_L$ resonances with the main optical transition occur) for which again the TDH and IPA have similar intensities.

\section{All gauges are equal but some gauges are more equal than others}
In the previous sections we presented a real-time approach based on Berry's phase to calculate the non-linear response in solids. Here we briefely discuss possible alternatives to this formulation together with their advantages and disdvantages. In order to describe polarisation in bulk materials, Berry's phase is a necessary ingredient. However this is true only for the intrinsic polarisation, because the polarisation induced by an external field can be expressed as an integral of the total current:
\be
\Delta \PP (t) = \int \JJ (t) dt.
\label{intj}
\ee
In periodic systems the current operator $\hat \jj$ is always well defined  and therefore we do not need any special trick to carry out its calculation. For this reason, even before the King-Smith-Vanderbilt formula, some researchers\cite{PhysRevB.62.7998} presented real-time implementation of time-dependent density functional theory in periodic systems where the polarisation was obtained by current integration.  Since scalar and vector potentials associated to electromagnetic fields are gauge dependent\cite{Jackson02} it is always possible to choose a particular gauge where all the coupling with the external field is described through the vector potential $\bA$. Therefore we do not need the dipole operator, and we can write the Hamiltonian, in the so-called velocity gauge, as:
\be
\HH_v =  \frac{1}{2m} \left [ \pp + \bA (\rr,t) \right ]^2 + V(\rr).
\label{velh}
\ee
In this formulation one can complitely avoid the use of the Berry's phase. The vector potential that enters in the velocity gauge Hamiltonian is realted to the electric field by the equation:
\be
\EE(\rr,t) = -\frac{1}{c} \partial_t \bA(\rr,t)
\ee
Although at first sight this choice seems to be more convenient and simpler than the direct use of the polarisation, it presents different inconvenients that motivated us to develop the new approach presented in this chapter.\\
The length gauge, we used to define our Hamiltonian [Eq.~\ref{eq:startH}], is obtained from the multipolar gauge within the electric-dipole approximation (EDA).~\cite{Kobe1982}
The multipolar gauge is given by the condition $\Av^{M}\cdot\rr=0$ (the $M$ superscript indicates the multipolar gauge). The scalar and vector potentials are defined directly from the electric and magnetic fields respectivelly:~\cite{Kobe1982,Jackson02}  
\begin{eqnarray}
\Av^{M}(\rr,t) &=& -\rr \times \int_0^1 \,du\,u\BB(u\rr,t), \label{eq:mpgau1}\\
\varphi^{M}(\rr,t) &=& -\rr \cdot \int_0^1 \,du\,\Efield(u\rr,t).\label{eq:mpgau2}
\end{eqnarray}
Then for zero-field, both $\Av^{M}$ and $\varphi^{M}$ are zero. This is in general not true for other gauge choices as the zero-field situation is generally described by $\Av = \nabla \Lambda$ and $\varphi = -\partial_t \Lambda$ where $\Lambda(\rr;t)$ is an arbitrary function of space and time that defines the gauge transformation.  The importance of having zero scalar/vector potentials in the zero-field situation has been emphasised by several authors. Since the vector potential can be different from zero alse in the case of zero-field in the Hamiltonian of Eq.~\eqref{velh}, it means that the eigenvalues and eigenfunctions of the velocity gauge Hamiltonian are gauge dependent and therefore cannot represent physical states not even at zero field.\\
Another advantage of using length gauge can be understood when we pass from  length to the velocity gauge through the gauge transformation:
\bea
\TT (\rr,t) &=& e^{i e \bA(t) \cdot \rr/\hbar}\\
\Psi_v(\rr,t) &=& \TT^{+} \Psi_l(\rr,t) \\
\HH_v &=& \TT^{+} \HH_l \TT\\
\HH_l &=&  \frac{1}{2m} \pp ^2 + V(\rr) +  \hat \rr \cdot \EE(t),\label{hlength}
\eea
where  $\HH_l$ is the original Hamiltonian in lenght gauge [Eq.~\ref{hlength} ], $\HH_v$ is the Hamiltonian in velocity gauge [Eq.~\ref{velh} ] and $\TT (\rr,t)$ is the gauge transformation operator.
In the full many-body Hamiltonian all operators are local and therefore there is not problem to perform this transformation. But in approximated Hamiltonians, non local operators appear that do not commute with the Gauge transformation and therefore they acquire an unusual dependence from the vectors potential:
\be
\hat O_v = e^{-i e \bA(t) \cdot \rr/\hbar} \hat O_l e^{i e \bA(t) \cdot \rr/\hbar}
\ee
One runs in this problem in presence for examples of non-local pseudo-potentials (see section III of Ref.~\cite{PhysRevB.62.7998})  or more seriously in presence of the exchange or self-energies derived from Green's function theory. Therefore real-time dynamics in presence of non-local operators is much more complicated in velocity guage than in the length one. 
Last but not least, the dephasing operator that appears in the Hamiltonian~[Eq.~\ref{dephterm}] do not commute with the gauge transformation, this implies that in velocity gauge the smearing cannot be simply added at the end of the calculations\cite{sangalligauge,PhysRevA.36.2763} but its full dynamics has to be taken in to account.
Notice that also the relation \eqref{intj} between current and polarization is modified by the presence of dephasing terms in the Hamiltonian. For example in presence of a simple quasi-particle life-time $\epsilon_i = e_i + i \gamma/2$ the relation [Eq.~\ref{intj}] becomes \cite{tokman}:
\be
 \JJ (t) = \dot \PP (t) + \gamma \PP (t).
\ee
All these difficulties explain the common agreement in the scientific literature that velocity gauge is not suitable or more difficult to use to study non-linear response\cite{boyd_gauge,PhysRevB.52.14636}.

\section{Conclusions}\label{conclusion}                                        
In this chapter we presented an \ai real-time approach to calculate nonlinear optical properties of extended systems in the length gauge. The key strengths of the proposed approach are first, the correct treatment of the coupling between electrons and the external field and second the possibility to include easily correlation effects beyond the IPA.

Regarding the treatment of the electron-field coupling, following the work of Souza et al.\cite{souza_prb}, we started from the Berry's phase formulation for the dynamical polarisation---a definition consistent with the periodic boundart conditions (PBC)---to derive a covariant numerical expression for the dipole operator in the EOMs.

Note that we worked in the length-gauge even if the velocity gauge may appear a more natural choice. In fact, as opposed to the position operator the velocity operator is consistent with the PBC. However, in the velocity gauge even if the position operator disappears from the Hamiltonian, it reappears in the phase factor for the wave-function~\cite{PhysRevA.36.2763}, so that the problem of re-defining the position operator remains. 
Furthermore, the velocity gauge is plagued by unphysical numerical divergences for the response at low frequencies~\cite{PhysRevB.52.14636}. 
Concerning effects beyond the independent-particle approximation, they are included by simply adding the corresponding single-particle operator to the Hamiltonian. This is an easy task when compared with deriving the corresponding expressions for the nonlinear optical susceptibility.~\cite{PhysRevB.80.155205,PhysRevB.80.165318} As an example, in the present chapter we have included quasi-particle corrections to the band-gap by adding to the Hamiltonian a scissor operator and crystal local-field effects by adding the time-evolution of the Hartree potential. In principle, one can add as well excitonic effects by adding the time-evolution of the screened exchange self-energy (see chapter~\ref{chaptercorr}); or perform a real-time TD-DFT calculations by adding the time-evolution of the exchange-correlation potential (see chapter~\ref{chaptertddft}). Being the focus of this chapter the validation of the proposed approach for calculating nonlinear properties, the inclusion of these correlation effects is discussed in the rest of this work.
We have proved the validity of our approach by comparing our results, obtained from real-time simulations, with results in the literature obtained from direct evaluation of the second order susceptibility in frequency-domain.

\clearemptydoublepage

\chapter{Correlation from non-equilibrium Green's functions} 
\label{chaptercorr}
\section{A simple introduction to non-equilibrium Green's functions}
In this section we present a simple introduction to the non-equilibrium Green's functions, following the lines of Ref.~\cite{kremp}. \\
In quantum mechanics a many-body system is characterised by its Hamiltonian, i. e. by:
\be
H=\sum^N_{i=1} \frac{\pp^2_i}{2m} + \sum_{i}^{N} V_{ei} (\rr_i) + \sum_{i<j}^N V_{ee}(\rr_i - \rr_j ), 
\label{fullH}
\ee
where $V_{ei}$ is the potential generated by the ions and $V_{ee}$ is the electron-electron interaction.
This last term in the previous equation makes the many-body problem impossible to solve in the case of three  or more interacting particles. However we can always try to find a solution of the Hamiltonian [Eq.~\ref{fullH}] as a linear combination of the
single-particle solutions of a non-interacting problem, i. e. the solution of the  above Hamiltonian
without the electron-electron interaction. The non-interacting solution constitutes a complete basis set in the form:
\be
| b_1, b_2,...,b_N \rangle = \frac{1}{\sqrt{N!}} a^+(b_1) .... a^{+}(b_N) | 0 \rangle
\ee
where $|0\rangle$ is the state without particles and the creation operators $a^+(b_i)$ create a particle
in a given state $b_i$ of our single particle Hamiltonian. The properties of the creation/destruction
operators guarantee the correct Fermi statistics. Once we have determined the wave-function, or at least,
an approximate wave-function, all the observables can be expressed in terms of  $a$ and $a^+$ operators.\\
Proceeding in this way is a formidable task, because in a solid the number of particles is of the order
of the Avogadro's number $N \simeq 10^{23}$, i.e. practically an infinite number of particles.\\
But we do not need all this information to characterise a physical system. In fact the mean value of any
single particle operator as dipole, momentum, etc.. can be expressed in terms of the single particle density matrix, without the need of the full wave-functions:
\bea
\gamma (\xx_1,\xx_1') &=& N \int d \xx_2 d \xx_3... d \xx_N \Psi (\xx_1,\xx_2,...,\xx_N) \Psi^* (\xx_1,\xx_2,...,\xx_N)\\
\langle  A \rangle &=& tr \left ( \gamma  A \right ),
\eea
where $A$ is a single particle operator, and $\gamma (\xx_1,\xx_1')$ the one-body density matrix.
Obviously the mean-value of a s-particle operator may be evaluated by means of the s-particle density matrix. \\ If we know the EOMs for the density matrix it will be possible to
follow the full many-body dynamics without passing by the full wave-function.\\
Based on this idea John Von Neumann in 1927 derived an equation for the temporal evolution of the density matrix operator\cite{neumann}:
\be
i \frac{\partial \gamma}{\partial t} = [\HH, \gamma].
\label{densmat}
\ee
This equation can be obtained from the  Schr\"odinger equation, and provides an equivalent description of quantum mechanics. The major problem of Eq.~\ref{densmat} is that it is not a closed equation. If we write down explicitly [Eq.~\ref{densmat}] for the single particle density matrix we immediately realise that the r.h.s. depends from the two-body density matrix, whose EOMs will depend from the three-body density matrix and so on. This set of equations, called the BBGKY hierarchy (Bogoliubov-Born-Green-Kirkwood-Yvon hierarchy), describes the full  dynamics of a system with a large number of interacting particles.\cite{bonitz} \\
The solution of the BBGKY hierarchy has the same complexity of the initial  Schr\"odinger equation. 
For this reason different scientists searched for a closed form of Eq.~\ref{densmat} that involves only a limited order of density matrices. The simplest decoupling of the hierarchy is achieved by the application of the Hartree-Fock (HF) approximation:
\be
\gamma (\rr_1,\rr_2,\rr_1',\rr_2') =  \gamma (\rr_1,\rr_1') \gamma (\rr_2,\rr_2') -  \gamma (\rr_1,\rr_2') \gamma (\rr_2,\rr_1').
\ee
In the HF the two-particle density matrix is expressed in terms of the single particle one, and therefore the EOMs for single-particle density matrix are closed. Clearly HF is not a satisfying approximation to describe electronic properties of real-systems. In the literature many different ways to close the BBGKY equations have been proposed that are able to treat correlations at different levels (for a discussion see Ref.~\cite{bonitz,RevModPhys.74.895} and references there in).\\
The density matrix formalism is a very powerful tool to study the many-body problems in a large spectra of situations, however it presents two important limitations that motivated the development of Green's function theory. First of all, there are  different situations where we are interested in time-dependent (or frequency dependent) correlation functions that are not directly accessible from the density matrix. Second the decoupling of the BBGKY hierarchy equations is not an easy task, especially in systems with many particles as for instance bulk materials. Now we will see how it is possible to generalise the density matrix approach to solve these two issues. \\
If we write the density matrices in term of field operators:
\be
\gamma (\rr_1 ... \rr_s,\rr_1'... \rr_s) = \langle \psi^\dagger(\rr_1',t) ... \psi^\dagger(\rr_s',t) \psi(\rr_1,t) ... \psi(\rr_s,t)  \rangle,
\ee
it follows that Green's functions are a generalisation of s-particle density matrices with field operators at different times:
\be
G^<(\rr_1,t_1 ... \rr_s,t_s,\rr_1',t_s' ... \rr_s, t_s') = \left(-\frac{1}{i} \right)^s \langle \psi^\dagger(\rr_1',t_1') ... \psi^\dagger(\rr_s',t_s') \psi(\rr_1,t_1) ... \psi(\rr_s,t_s)  \rangle.
\ee
Different arrangements of the field operators correspond to different time-correlation functions, i.e. different Green's functions: greater, casual, retard or advanced.
As in the case of density matrices the properties of a full many-particle systems can be described in terms of Green's functions. But differently from density matrices we have a direct access to dynamical properties, as for instance response functions, excitation energies and so on (see Chapter 3 of Ref.~\cite{kremp}).\\
A central task of the theory is the determination of these functions. The dynamics of the Green's functions can be derived from the EOMs of the field operators. For example for the $G^<$ we obtain:
\be
\left( i \frac{\partial}{\partial t_1} + \frac{\nabla_i^2}{2m} - U(1) \right) G^<(1,1') - i \int d2 V(1,2) G^<(12,1'2^+) = 0
\ee
where $1,2$ are indexes for both time and position, the $2^+$ is the limit of $2' \rightarrow 2$ with $t_2'> t_2$, and $U(1)$ is the external potential.
As in the case of reduced density matrix, the EOMs for the single particle Green's functions are not closed and depend from two particle Green's functions and so on. The problem may, in principle, be solved using equations similar to the BBGKY hierarchy, the so called Martin-Schwinger hierarchy equations. However Green's function presents a major advantage that is the possibility to construct a single particle operator, the self-energy $\Sigma$, that is not local in time and space and allows us to close the EOMs in the form:
\be
\int_C d \bar 1 \left [ G_0^{-1} (1 \bar 1) - U(1\bar1) - \Sigma (1 \bar 1) \right ] G(\bar 1 1)  = \delta(1- 1'),
\label{kbeq}
\ee
where $G_0$ is the independent particle Green's function and the integral is performed on the Keldysh contour.\footnote{In this introduction we have no space to derive and explain Eq.~\ref{kbeq}, for more details see the textbooks \cite{kremp,schafer,kadanoff,haug2008quantum}.}\\
This famous equation was derived for the first time by Kadanoff and Baym and by Keldysh. It is a generalisation of the Dyson equation from the field theory to quantum statistics. Equation~\ref{kbeq} describes time evolution of real-time Green's functions under equilibrium and non-equilibrium conditions. \\
Of course, all problems of the hierarchy equations are now transferred to the self-energy construction.
In the literature different approaches have been proposed to construct self-energies. In this chapter we will use the so-called GW approximation. The GW self-energy is the first order correction, in term of screened Coulomb interaction, obtained from many-body perturbation theory\cite{aryasetiawan1998gw}. This approximation has been derived by Hedin\cite{hedin1999correlation} for the electron gas and then applied to semiconductors by Hanke, Sham and Strinati\cite{PhysRevB.25.2867}.\footnote{As it is often the case in  physics, the GW self-energy originates from previous studies on the electron gas, based on perturbation theory in terms of screened Coulomb interaction performed by DuBois, Kadanoff, Baym and Bonch-Bruevich and Tyablikov.} The GW corrections have shown a  clear  improvement for band gaps\cite{aryasetiawan1998gw,Aulbur19991}, level ordering\cite{faber2011first} and gradients of the electronic levels respect to the atomic displacements\cite{faber2015exploring}, when compared to available experimental data.\\
In the next sections, starting from a non-equilibrium formulation of the GW self-energy\cite{PhysRevB.69.205204}, we will derive an effective single-particle Hamiltonian for our real-time dynamics. 

\section{General Introduction}

Real-time methods have proven their utility in calculating optical properties of finite systems mainly within time-dependent density functional
theory (TDDFT).\cite{PhysRevB.62.7998,PSSB:PSSB200642067,sun:234107} On the other hand extended systems have been mostly studied by using many-body  perturbation
theory (MBPT) within the linear response regime~\cite{strinati}. 
The different treatment of correlation and nonlinear effects marks the range of applicability of the two approaches. The real-time TDDFT makes
possible to investigate nonlinear effects like second harmonic generation\cite{takimoto:154114} or hyperpolarizabilities of 
molecular systems\cite{PSSB:PSSB200642067}.
However, as we will see in Chapter~\ref{chaptertddft} TD-DFT is not a correct theory to describe excitations at zero momentum $\qq=0$ in extended systems.
For this reason, even with the exact exchange correlation functional TD-DFT is not suppose to reproduce the exact response functions in solids.
On the contrary MBPT allows to include correlation effects using controllable and systematic approximations for the self-energy $\Sigma$,
that is a one-particle operator non-local in space and time.
$\Sigma$ can be evaluated within different approximations, among which one of the most successful is the so-called GW
approximation\cite{Aulbur19991}.
Since its first application to semiconductors\cite{PhysRevLett.45.290} the GW self-energy has been shown to
correctly reproduce quasi-particle energies and lifetimes for a wide range of materials\cite{Aulbur19991,faber2013many}.
Furthermore, by using the static limit of the GW self-energy as scattering potential of the
Bethe-Salpeter equation (BSE)\cite{strinati}, it is possible to calculate response functions including electron-hole interaction
effects.

In recent years, the MBPT approach has been merged with density-functional theory (DFT) by using the Kohn-Sham Hamiltonian as zeroth-order term in the
perturbative expansion of the interacting Green's functions. This approach is parameter free and completely \emph{ab-initio}~\cite{Onida}.  
However MBPT is a very cumbersome technique that, based on a perturbative concept, increases its level of complexity with
the order of the expansion. As an example, this makes the extension of this approach beyond the linear response regime quite complex, though there have been recently some applications in nonlinear optics.~\cite{Chang2002,Leitsmann2005,PhysRevB.82.235201}
A generalisation of MBPT to non-equilibrium situations has been proposed by Kadanoff and Baym and Keldysh\cite{kadanoffbaym}. 
In their seminal works the authors derived a set of equations for the real-time Green's functions, the Kadanoff-Baym equations (KBE's), that provide the basic tools of the non-equilibrium Green's Function theory and allow essential advances in non equilibrium statistical mechanics\cite{kadanoffbaym}.   
Both the standard MBPT and non-equilibrium Green's Function theory are based on
the Green's function concept. This function describes the time propagation of a single particle excitation under the action of an external perturbation.  
In the equilibrium MBPT, due to the time translation invariance,
the relevant variable used to calculate the Green's functions is the frequency $\omega$. Instead, out of equilibrium, in all non steady-state situations, the time variables acquire a special role and much more attention is devoted to the their propagation properties. 
The time propagation avoids the explosive dependence, beyond the linear response, of the MBPT on high order Green's functions. Moreover the KBE are
non-perturbative in the external field therefore weak and strong fields can be treated on the same footing. 

One of the first attempts to apply the KBE's for investigating optical properties of semiconductors was presented in the seminal paper of Schmitt-Rink and co-workers.~\cite{PhysRevB.37.941} 
Later the KBE's were applied to study quantum wells,~\cite{PhysRevB.58.2064} laser excited
semiconductors,~\cite{PhysRevB.38.9759} and luminescence~\cite{PhysRevLett.86.2451}. However, only recently it was possible to simulate the Kadanoff-Baym
dynamics in real-time.~\cite{Kohler1999123,PhysRevLett.103.176404,PhysRevLett.84.1768,PhysRevLett.98.153004} 
In this chapter we combine a simplified version of the KBE's with DFT in such a way to obtain a parameter-free theory that is able to reproduce and predict
ultra-fast and nonlinear phenomena in crystalline solids and nano-structures 
(Sec.~\ref{tdbse_section}). This approach, that we
will address as real-time BSE (RT-BSE), reduces to the standard BSE for weak perturbations (Sec.~\ref{linear_response}) but, at the same time when coupled
with Modern Theory of Polarisation (see Chapter~\ref{chapterberry}), naturally describes optical excitations beyond the linear regime. 
After discussing some relevant aspects of the practical implementation of our approach (Sec.~\ref{teospectro}), we 
exemplify how it works in practice by calculating the optical absorption spectra of {\it h}-BN, and we'll apply it to the 
second harmonic generation in monolayer {\it h}-BN and MoS$_2$.

\section{The Real-Time Bethe-Salpeter equation}
\label{tdbse_section}                                        
We derive here a novel approach to solve the time evolution of an electronic
system with Hamiltonian coupled with an external field,
\be
\label{hamiltonian}
\hat{H} = \hat{h} + \hat{H}_{mb} + \hat{U},     
\ee
where $U$ represents the electron-light interactions (see
Sec.~\ref{ss:solution} for its specific form). As usually done in MBPT, $\hat{H}$
is partitioned in an (effective) one-particle Hamiltonian
$\hat h$ and a part containing the many-particle effects $\hat{H}_{mb}$. 

In our derivation, we take as starting point the KBE's that we briefly introduce in
Sec.~\ref{ss:KBEs} (see e.g. Refs.~\cite{kremp} for a systematic
treatment). Then, in Sec.~\ref{ss:KS-CHSX} we proceed in analogy with
the equilibrium MBPT: first, we define $\hat h$ as the Hamiltonian of the Kohn-Sham
system, second we introduce the same approximations for the
self-energy operator. As a result we obtain the analogous of the
successful $GW$+BSE approach for the non-equilibrium case. Indeed in
Sec.~\ref{linear_response} we show that our approach, the
real-time BSE (RT-BSE), reduce to the $GW$+BSE in the linear regime.\footnote{Notice that a real-time version
        of the Bethe-Salpeter equation was already proposed in Ref.~\cite{PhysRevB.67.085307} as
an efficient method to solve the BSE, but it was limited to the linear response.}

\subsection{The Kadanoff-Baym equations}
\label{ss:KBEs}
Within the KBE's, the time evolution of an electronic system 
coupled with an external field is described by the equation of motion for the non-equilibrium Green's functions $G\(\rr,t;\rr' t'\)$ ~\cite{kadanoffbaym,kremp,schafer}.  
To keep the formulation as simple as possible and, being interested
only in long wavelength perturbations,
we expand the generic $G$ in the eigenstates $\{\varphi_{n,\kk}\}$ of the $\hat{h}$ Hamiltonian for a fixed momentum point $\kk$:
\be
\label{GFdefinition}
\[\GG_{\kk}\(t_1,t_2\)\]_{n_1 n_2}\equiv G_{n_1 n_2, \kk }(t_1,t_2)=\int
\varphi^*_{ n_1\kk}({\mathbf r}_1) G\(\rr_1,t_1;\rr_2,t_2\) \varphi_{n_2\kk}({\mathbf
r}_2){\mathrm d}^3r_1{\mathrm d}^3r_2.
\ee
As the external field does not break the spatial invariance of the system $\kk$ is conserved.
Notice that both the Green's functions and the self-energies will depends from two times $t_1, t_2$ because we are in an out-equilibrium situation. These two times can be also rewritten in terms of $t = (t_1 + t_2)/2$ and $\tau = t_1 - t_2$, where $t$ is the physical time and $\tau$ a fictitious time related to quantum correlations.\\
Within a second-quantization formulation of the many-body problem, the equation of motion for the Green's function described by Eq.~\eqref{GFdefinition} are obtained
from those for the creation and destruction operators.  However the resulting equations of motion for $\GG_{\kk}$ are not closed:
they depend on the equations of the two-particle Green's function that in turns depends on the three-particle Green's function and so on.
In order to truncate this hierarchy of equations, one introduces the self-energy operator $\SiS_{\kk}(t_1,t_2)$, a non-local and frequency dependent
one-particle operator that holds information of all higher order Green's functions. 
A further complication arises with respect to the equilibrium case because of the lack of time-translation invariance in non-equilibrium phenomena that implies that 
$\SiS_{\kk}(t_1,t_2)$ and $\GG_{\kk}(t_1,t_2)$ depend explicitly on both $t_1,t_2$. Then, one can define an
advanced $\SiS_{\kk}^a$ ($\GG_{\kk}^{\mathrm a}$), a retarded $\SiS_{\kk}^r$ ($\GG_{\kk}^{\rr}$), a greater and a lesser $\SiS_{\kk}^>,\SiS_{\kk}^<$ ($\GG_{\kk}^>,\GG_{\kk}^<$) self-energy
operators (Green's functions) depending on the ordering of $t_1,t_2$ on the time axis. 
Finally, 
the following EOM for the $\GG_{\kk}^<$ is obtained (see e.g. Ch.~2 of
Ref.~\cite{kremp} for more details):
\begin{multline}
 i\hbar  \frac{\partial}{\partial t_1} G^<_{n_1n_2\kk}(t_1,t_2)=\mbox{}  \delta(t_1-t_2)\delta_{n_1n_2}    \\
+  h_{n_1n_1\kk}(t_1) G^<_{n_1n_2\kk}(t_1,t_2) + \sum_{n_3}U_{n_1n_3\kk}(t_1) G^<_{n_3n_2\kk}(t_1,t_2) \\
+  \sum_{n_3} \int \mathrm{d}t_3 \big( \Sigma^{\rr}_{n_1n_3\kk}(t_1,t_3)G^<_{n_3n_2\kk }(t_3,t_2) 
+ \Sigma^<_{n_1n_3\kk}(t_1,t_3) G^{\mathrm a}_{n_3n_2\kk}(t_3,t_2)\big) .
\label{eqmotGone}
\end{multline}
This equation, together with the adjoint one for $ i\hbar  \frac{\partial}{\partial t_2} G^<$, describes the
evolution of the lesser Green's function $\GG_{\kk}^<$ that gives
access to the electron distribution ($\GG_{\kk}^<(t,t)$) and to the
average of any one-particle operator such as for example the electron
density [Eq.~\eqref{eqden}], the linear polarisation and the current.  However, in general $\Sigma^{\rr},\Sigma^<$ and  $\GG_{\kk}^{\mathrm a}$ depend on $\GG_{\kk}^>$, so that in addition to Eq.~\eqref{eqmotGone} the corresponding equation for the $\GG_{\kk}^>$ has to be solved. 

Then, in principle, to determine the non-equilibrium Green's function in presence of an external perturbation one needs  to solve the system of coupled equations for $\GG_{\kk}^>,\GG_{\kk}^<$, known as KBE's. Indeed, this system has been implemented within several approximations for the self-energy in model
systems\cite{Kohler1999123,PhysRevLett.103.176404}, in the homogeneous electron gas\cite{PhysRevLett.84.1768}, and in atoms\cite{PhysRevLett.98.153004}.
The possibility of a direct propagation in time of the KBE's provided, in these systems, valuable insights on the real-time dynamics of the electronic excitations, as their lifetime and transient effects.\cite{Kohler1999123,PhysRevLett.103.176404,PhysRevLett.84.1768,PhysRevLett.98.153004} 
Nevertheless, the enormous computational load connected to the large number of degrees of freedom {\it de facto} prevented the application of this method to crystalline solids, large molecules and nano-structures. In the next subsection we show a simplified approach---grounded on the DFT---that while capturing most of the physical effects we are interest in, makes calculation of ``real-world'' systems feasible. 

\subsection{The Kohn-Sham Hamiltonian and an approximation for the self-energy}
\label{ss:KS-CHSX}

In analogy to MBPT for the equilibrium case,
we choose as $\hat{h}$ in Eq.~\eqref{hamiltonian} the Kohn-Sham Hamiltonian,~\cite{PhysRev.140.A1133} 
\be
\hat{h} = -\frac{\hbar^2}{2m} \sum_i \nabla_i^2 + 
\hat{V}_{eI} + \hat{V}^H[\tilde \rho]   + \hat{V}^{xc}[\tilde \rho], \label{eq:kshamH}
\ee
where $\hat{V}_{eI}$ is the electron-ion interaction, 
$\hat{V}^H$ is the Hartree potential and
$\hat{V}^{xc}$ the exchange-correlation potential.
Within DFT, the Kohn-Sham Hamiltonian corresponds to the independent particle system
that reproduces the ground-state electronic density $\tilde \rho$ of
the full interacting system ($\hat h + \hat H_{\text{mb}}$), that is
\be
\tilde \rho = \sum_{n\kk} f_{n\kk}|\varphi(\rr)|^2,
\label{eq:KSden}
\ee
where $f_{n\kk}$ is the Kohn-Sham Fermi distribution. 

The EOM for the $\GG_{\kk}^<$  [Eq.~\eqref{eqmotGone}] can be greatly simplified by choosing 
a static approximation for the self-energy,
\begin{subequations}
\begin{align}
{\bf\Sigma}^{\rr} (t_1,t_2) & = \left[ {\bf\Sigma}^{\mathrm {cohsex}} (t_1) - {\bf V}_{xc} \right ] \delta(t_1 - t_2) \label{eq:appcsx}\\
{\bf\Sigma}^<(t_1,t_2) & =0 \label{eq:appret}
\end{align}
\end{subequations}
where the usual choice is ${\bf\Sigma}^{\mathrm {cohsex}}$, the so-called Coulomb-hole plus screened-exchange self-energy (COHSEX)\cite{PhysRevB.38.7530}. In Eq.~\ref{eq:appcsx} we subtracted the correlation effects already accounted by Kohn-Sham Hamiltonian $\hat{h}$.
The COHSEX is composed of two parts:
\bea
\Sigma^{\text{sex}}(\mathbf r,\mathbf r',t)=i W(\mathbf r,\mathbf r'; G^<)G^<(\mathbf r,\mathbf r',t), \\
\Sigma^{\text{coh}}(\mathbf r,\mathbf r',t)= -W(\mathbf r,\mathbf r'; G^<) \frac{1}{2}\delta(\mathbf r-\mathbf r'),
\label{coh_anx_sex}
\eea
where $W(\mathbf r,\mathbf{r'}; G^<)$ is the Coulomb interaction in
the random-phase approximation (RPA) and $G^<(\rr,\rr',t)$ is the time-diagonal lesser Green's function $G^<(\rr,\rr',t)=G^<(\rr,\rr',t,t)$. These two terms are obtained as a
static limit of the $GW$ self-energy (see Ch.4 of Ref.~\cite{kremp} and
Refs.~\cite{PhysRevB.38.7530, PhysRevB.69.205204}). 

If we insert the approximated self-energy Eqs.~\eqref{eq:appcsx}--\eqref{eq:appret} in Eq.~\eqref{eqmotGone}, the EOM for the $\GG_{\kk}^<$  does not depend anymore on ${\bf G}^>$.
Moreover, since the COHSEX self-energy is local in time, and depends only on $G^<(\rr,\rr',t)$, Eq.\eqref{eqmotGone} can be combined with the adjoint one for $\frac{\partial}{\partial t_2} G^<_{n_1n_2\kk}(t_1,t_2)$ in such a way to have a closed equation in $t=(t_1+t_2)/2$:

\begin{multline}
\label{eqmotGcohsex}
 i\hbar  \frac{\partial}{\partial t} G_{n_1,n_2,\kk}^<(t)=
 \left [ \hh_{\kk} + \UU_{\kk}(t) +  \VV_{\kk}^H[\rho] -
   \VV_{\kk}^H[\tilde \rho] \right .\\
   \left . + (\SiS_{\kk}^{\mathrm{cohsex}} (t)
   -\VV_{\kk}^{xc}[\tilde \rho]), \GG_{\kk}^<(t)\right ]_{n_1,n_2}.
\end{multline}
where the term $\VV_{\kk}^H[\rho] - \VV_{\kk}^H[\tilde \rho]$ is the induced variation of the Hartree term that generates the exchange in the BSE, and $\rho$ is the density obtained from the $G^<$ as 
\be
\label{eqden}
\rho(\mathbf r,t) = \frac{i}{\hbar} \sum_{n_1n_2\kk} \varphi_{n_1 \kk} (\mathbf r)  \varphi^*_{n_2 \kk}  (\mathbf r) G^<_{n_2n_1\kk}(t).
\ee
After reducing to a single time, \eqref{eqmotGcohsex} is equivalent to a density matrix approach with a potential local in time. 
%
However, despite the full real-time COHSEX dynamics [Eq.~\ref{eqmotGcohsex}] is an appealing
option considerably simplifying the dynamics with respect to the KBE's,
it neglects the dynamical dependence of the self--energy operator. This, in practice, induces a consistent renormalization 
of the quasi-particle charge\cite{PhysRevLett.45.290} in addition to an opposite enhancement of the optical properties.~\cite{PhysRevLett.91.176402}
In the COHSEX approximation both effects are neglected.
At the level of response properties for
most of the extended systems dynamical effects are either negligible or very small\footnote{
Recently it has been shown that dynamical effects in the BSE can be important for finite
systems, see Refs.~\cite{bsedynamic,PhysRevB.77.115118}} and it has been shown that they  
partially cancel with the  quasi-particle renormalization factors.~\cite{PhysRevLett.91.176402} 

Therefore we modify Eq.~\ref{eqmotGcohsex} in order to include only the effect
of the dynamical self-energy on the renormalization of the quasi-particle energies, that is the most
important effect.
Also in this case, our idea is to proceed in strict analogy with equilibrium MBPT and to derive a real-time equation that reproduces the fruitful combination of the $G_0W_0$
approximation---for the one-particle Green's function---and of the BSE with
a static self-energy---for the two-particle Green's function. 
Indeed the $G_0W_0$+BSE is the state-of-the-art approach to study optical
properties within MBPT.~\cite{Onida} 
To this purpose Eq.~\eqref{coh_anx_sex} is modified as:
\begin{multline}
\label{tdbse}
 i \hbar  \frac{\partial}{\partial t} G^<_{n_1n_2\kk}(t)= 
\left [ \hh_{\kk} + \Delta \hh_{\kk} + \UU_{\kk} +\VV_{\kk}^H[\rho] - \VV_{\kk}^H[\tilde \rho] \right . \\
 \left . + \SiS_{\kk}^{\text {cohsex}}[G^<] - \SiS_{\kk}^{\text{cohsex}}[\tilde G^<], \GG_{\kk}^<\(t\) \right ]_{n_1n_2}. 
\end{multline}
$\Delta \hh$ is a scissor operator\cite{Onida} that
applies the $G_0 W_0$ correction to the Kohn-Sham eigenvalues, $e_{n_1\kk}^{KS}$,
\be
\label{eq:sciss}
\[\Delta \hh_{\kk}\]_{n_1,n_2} =  \left ( e_{n_1\kk}^{G_0W_0}  -e_{n_1\kk}^{KS}  \right ) \delta_{n_1,n_2},
\ee
and $\tilde G^<_{n n'}$ is the solution of Eq.~(\ref{tdbse}) for the
unperturbed system ($U=0$)
\be
\tilde G^<_{n n' \kk} = i \hbar f_{n\kk} \delta_{n n'}, \label{gtilde}
\ee
where we assume that the Kohn-Sham Fermi distribution is not changed by the
scissor operator. Note further that in Eq.~\ref{tdbse}, $V^{xc}[\tilde \rho]$ cancels out because it is independent of $G^<(t)$.

Equation~\eqref{tdbse} is the key result of this work. It is equivalent to assume that the
quasi-particle corrections modify only the single particle eigenvalues
leaving unchanged the Kohn-Sham wave functions.  Within equilibrium MBPT this approximation
is very successful for a wide range of materials characterised by weak correlations (see e.g. Refs.~\cite{Onida,Aulbur19991}).\\
Equation~\eqref{tdbse} can rewritten in term of density matrix as:
\bea
i \hbar  \frac{\partial}{\partial t} \gamma_{n_1n_2\kk}(t) &=& \left [ \HH_{mb}(t),\gamma(t) \right ]_{n_1n_2\kk} \\
\HH_{mb} (t) &=& \hh_{\kk} + \Delta \hh_{\kk} + \UU_{\kk} +\VV_{\kk}^H[\rho-\rho_0]+ \SiS_{\kk}^{\text {cohsex}}[\gamma-\gamma_0]. 
\label{mbhamiltonian}
\eea
where $\rho_0$ and $\gamma_0$ are the density and the single particle density matrix at equilibrium. Eq.~\ref{mbhamiltonian}  is the Hamiltonian we will use
in combination with dynamical Berry phase to study the non-linear response in low dimensional systems (see Sec.~\ref{ss:solution}).



\subsection{The linear response limit}
\label{linear_response}

When an external perturbation $U(t)$ is switched on in Eq.~\eqref{tdbse}, it induces a variation of the Green's function, 
$\Delta \GG_{\kk}^<(t) = \GG_{\kk}^<(t) -  \tilde{\GG}_{\kk}^<$.
In turns, this variation induces a change in the 
self-energy and in the Hartree potential.
In the case of a strong applied laser field these changes depend on
all possible orders in the external field. However for weak fields the
linear term is dominant. 
In this regime it is possible to show analytically that Eq.~\eqref{tdbse}  reduces to the $G_0W_0$+BSE approach\cite{strinati,Aulbur19991}.
Proceeding similarly to Ref.~\cite{bsedynamic} we consider the retarded density-density correlation function:
\be
\label{chi-rr}
\chi^{\mathrm{r}}(\rr,t;\rr',t') =
-i\[\langle \rho(\rr,t)\rho(\rr',t')\rangle 
- \langle \rho(\rr,t)\rangle \langle \rho(\rr',t')\rangle\]\theta\(t-t'\).
\ee
$\chi^{\mathrm{r}}$ describes the linear response of the system to a weak
perturbation, represented in Eq.~\eqref{hamiltonian}  by $U$,
\be
\label{eq:phychi}
\chi^{\mathrm{r}}(\rr,t;\rr',t') = \left. \frac{\langle
  \delta\rho(\rr t) \rangle}{\delta U(\rr^\prime t^{\prime})}\right\vert_{U=0}.
\ee
We start by expanding  $\chi (\mathrm{r})$ in terms of the Kohn-Sham orbitals:
\be
\label{basisChi}
\chi^{\mathrm r}(\rr,t; \mathbf{r'},t'; \mathbf q) =
  \sum_{\substack{i,j,\kk \\ l,m,\kk^\prime} } \chi^{\mathrm r}_{\substack{i,j,\kk \\ l,m,\kk^\prime} }(t,t^\prime; \mathbf q)
\times \varphi_{i,\kk} (\rr)\varphi^*_{j, \mathbf {k+q}} (\rr) \varphi^*_{l, \mathbf {k^\prime}} (\rr')
\varphi_{m, \mathbf {k^\prime+q}} (\rr'),
\ee
where $\qq$ is the momentum, and we define the matrix elements of $\chi^{\mathrm r}$ as,
\be
\label{eq:chimat}
 \chi^{\mathrm r}_{\substack{ij,\kk \\ lm,\kk^\prime} }(t,t^\prime; \mathbf q) =
\iint \mathbf{d}^3r \mathbf{d}^3r^\prime \varphi^*_{i,\kk}(\rr)
\varphi^*_{m, \kk^\prime+\qq} (\rr^\prime)
\varphi_{j,\kk+\qq} (\rr) \varphi_{l, \kk^\prime} (\rr^\prime).
\ee
Since we are interested only in the optical response, in what follows
we restrict ourselves to the case $\qq =0$ and drop the $\qq$
dependence of $\chi^{\mathrm r}$ (for the extension to finite momentum
transfer see Ref.~\cite{PhysRevLett.84.1768}). 
Inserting the expansion for $\chi$ [Eq.~\eqref{basisChi}], $\rho$ [Eq.~\eqref{eqden}] and $U$ ($U_{m,n \kk}\equiv \langle m \kk | U | n \kk \rangle$) in Eq.~\eqref{eq:phychi} we obtain the following relation linking the matrix elements of $\chi^{\mathrm r}$ to the matrix elements of $G^<$:
\be
\label{DDcorFun_n}
\chi^{\mathrm r}_{\substack{ij,\kk \\ lm, \pp }}(t,t^\prime) =
\left. \frac{\delta\langle i G^<_{ji,\kk}(t)\rangle}{\delta U_{lm,\pp}(t^{\prime}
  )}\right\vert_{U=0}.
\ee
Then, we can obtain the equation of motion  for the matrix elements of $\chi^{\mathrm r}$
by taking the functional derivative of Eq.~\eqref{tdbse} with respect to $U_{l,m,\kk}\(t\)$,
\begin{multline}
\label{dtGtt}
-i\hbar \frac{\partial}{\partial t} \chi^{\mathrm r}_{\substack{ij, \kk \\ lm,\pp }}(t,t^\prime)= 
\frac{\delta}{ \delta U_{lm, \pp}(t^{\prime})} [\hh_{\kk} + \Delta \hh_{\kk} + \UU_{\kk}(t) +  \VV_{\kk}^H[\rho(t)] -\VV_{\kk}^H[\tilde \rho]  \\
+  \SiS_{\kk}[G^<(t)] - \SiS_{\kk}[\tilde G^<], \GG_{\kk}^<(t) ]_{\substack{ji}}. 
\end{multline}
Making use of the definitions in Eqs.~\eqref{eq:sciss} and \eqref{gtilde}, together with Eq.~\eqref{DDcorFun_n}, it can be verified that the functional derivative of the one-electron Hamiltonian and 
of the external field give the contribution 
\begin{multline}
\label{diag_contr}
\left. \frac{\delta}{ \delta U_{l,m, \pp}(t^{\prime})} [\hh_{\kk}+\Delta \hh_{\kk} + \UU_{\kk},\GG_{\kk}^<(t)]_{\substack{ji}} \right\vert_{U=0}=\\
(e^{G_0W_0}_{j\kk} - e^{G_0W_0}_{i\kk})  \chi^{\mathrm r}_{\substack{ji, \kk \\ lm,\pp }}(t-t^\prime) + i(f_{i\kk}-f_{j\kk})\delta_{jl}\delta_{im}\delta_{\kk\pp} \delta(t-t'). 
\end{multline}
Note that $\chi^{\mathrm r}$  is invariant with respect to time translations ( $\chi^{\mathrm r}$  depends only on $t-t'$) since the functional derivative in Eq.~\eqref{diag_contr}, as in the rest of the section, is evaluated at equilibrium ($U=0$), and the unperturbed Hamiltonian does not depend on time. The term in Eq.~\eqref{dtGtt} containing the Hartree potential, which is not directly depending on the external perturbation. This term is expanded with respect to $U_{l,m,\kk}(t)$ by using the functional derivative chain rule and the definition of  $\chi^{\mathrm r}$  given by Eq.~\ref{DDcorFun_n} as
\begin{multline}\label{V_exp}
\delta V^H_{ij,\kk}\[\rho\(t\)\]= 
 \sum_{\substack{n,n',\pp\\l,m,\kk^\prime}} \iint\,dt^\prime\,dt^{\prime\prime} \frac{\delta V^H_{ij,\kk}\[\rho\(t\)\]}{\delta G^<_{n'n,\pp}\(t^\prime\)} \times \chi^{\mathrm r}_{\substack{n,n',\pp\\lm,\kk^\prime}}(t^\prime,t^{\prime\prime})\delta U_{lm,\kk^\prime}\(t^{\prime\prime}\),
\end{multline}
A similar equation can be obtained for $\Sigma^{\text{cohsex}}_{ij,\kk}[G^<(t)]$.
Equation~\eqref{V_exp} for Hartree potential and its analogous for the self-energy can be made explicit by using 
\begin{align}
        V^H_{mn,\kk}(t)=&-2i \sum_{ij,\kk'} G^<_{ji,\kk'}\(t\) v_{\substack{mn, \kk \\ ij,\kk' }}, \label{eq:hrtrG}\\
\Sigma^{\text{cohsex}}_{mn,\kk}(t)=& i \sum_{ij,\qq} G^<_{ji,(\kk-\qq)}\(t\) W_{\substack{m \kk,i (\kk-\qq) \\ n\kk,j (\kk-\qq) }},\label{eq:chsxG}
\end{align}
where the matrix elements of  $v$ and $W$ are labelled accordingly to Eq.~\eqref{eq:chimat}.
In Eq.~\eqref{eq:hrtrG} $v$ is the bare Coulomb potential, responsible for the local field effects in the
BSE.   
Then by inserting Eq.~\eqref{eq:hrtrG} in Eq.~\eqref{V_exp} the functional derivative for the Hartree term is
\begin{multline}
\label{H_contr}
\left. \frac{\delta}{ \delta U_{lm, \pp}(t^{\prime})} \left[\VV_{\kk}^H[\rho(t)] -\VV_{\kk}^H[\tilde \rho],\GG_{\kk}^<(t)\right]_{\substack{ji}} \right\vert_{U=0}=\\
\(2i^2\)\(f_{i\kk} - f_{j\kk}\) \sum_{st, \kk'} v_{\substack{ji, \kk \\ st,\kk' }} \chi^{\mathrm r}_{\substack{st, \kk' \\ lm,\pp }}(t-t^\prime).
\end{multline}

\begin{figure*}[t]
\centering
\begin{subfigure}[b]{1\textwidth}
\includegraphics[width=0.9\textwidth]{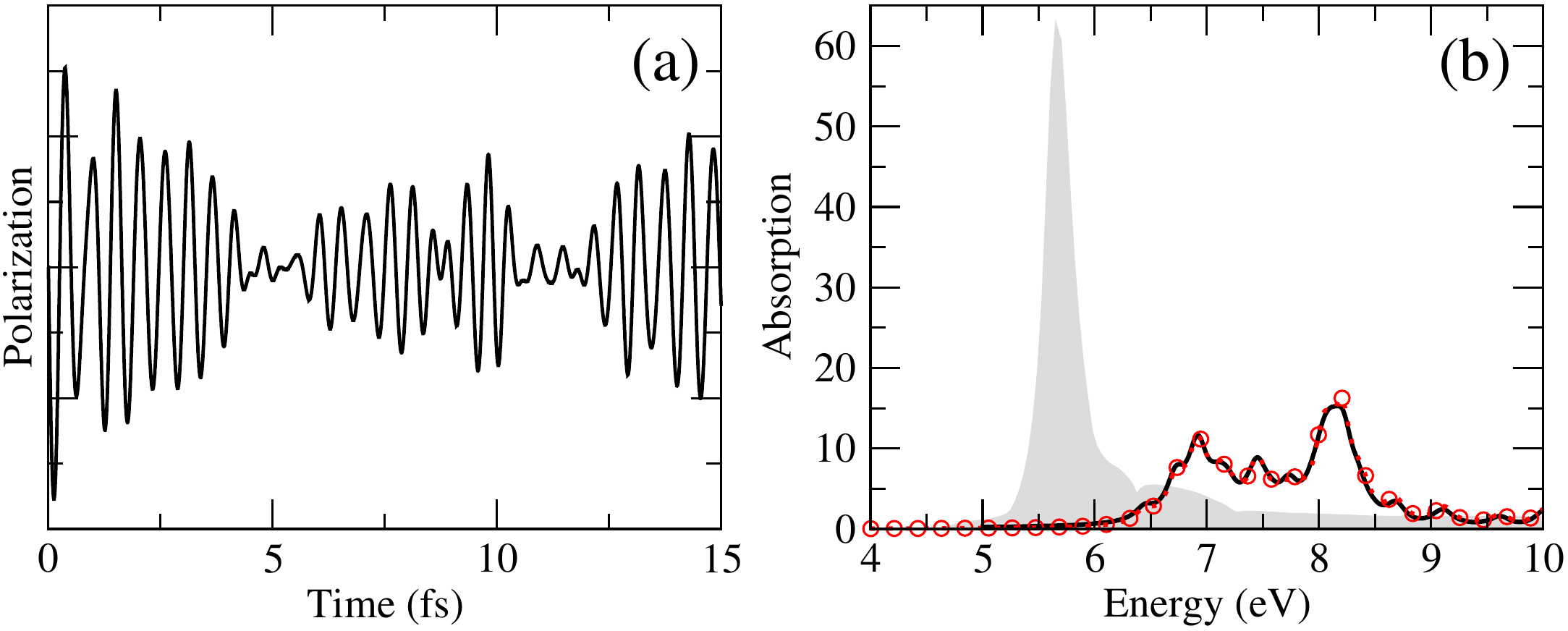}
\end{subfigure}
\begin{subfigure}[b]{1\textwidth}
\includegraphics[width=0.9\textwidth]{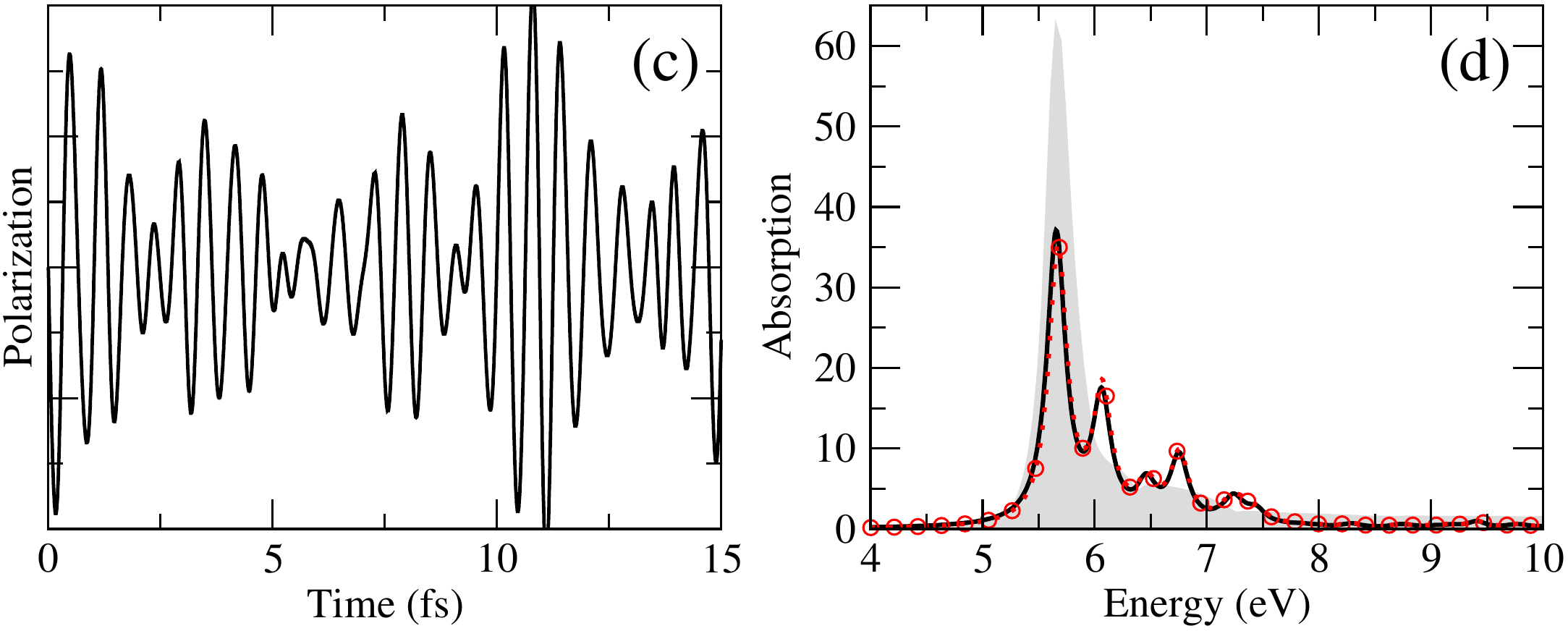}
\end{subfigure}
\caption{\footnotesize{
{\it h}-BN: Comparison between the real-time approach and the
standard RPA and BSE approaches based on the equilibrium MBPT. {\bf(a)},{\bf(c)}: polarisation $\mathbf P(t)$ generated by an electric field
$\mathbf E(t)=\mathbf E_o \delta(t)$ within the TDH [{\bf(a)}] and
RT-BSE [{\bf(c)}] approximations.{\bf(b)},{\bf(d)}: the corresponding
absorption spectra (red circles) are compared with the RPA [{\bf(b)}]
and with the BSE [{\bf(d)}] results (black line). The experimental
absorption spectrum (grey shadow) is also shown as reference. [Figure from Ref.\cite{attaccalite}]}}
\label{hbn}
\end{figure*}

Similarly, an analogous equation is obtained for the self-energy (see also Appendix \ref{fastcohsex}), 
\begin{multline}
\label{xc_contr}
\left. \frac{\delta}{ \delta U_{lm, \pp}(t^{\prime})} \left[\SiS_{\kk}[\GG^<(t)] -\SiS_{\kk}[\GG^<(t)],\GG_{\kk}^<(t) \right]_{\substack{ji}}\right\vert_{U=0}=\\
\(-i^2\)\(f_{i\kk} - f_{j\kk}\) \sum_{st,\qq}  W_{\substack{j \kk,s (\kk-\qq) \\ i\kk,t (\kk-\qq) }} \chi^{\mathrm r}_{\substack{st,(\kk-\qq) \\ lm,\pp }}(t-t^\prime),
\end{multline}
where we neglected the part containing the functional derivative of the screened interaction with respect to the external perturbation. This is a basic assumption of the standard BSE that is introduced in order to neglect high order vertex corrections.\cite{strinati}

Finally, we insert Eqs.~\eqref{diag_contr},~\eqref{H_contr} and~\eqref{xc_contr} in Eq.~\eqref{dtGtt}, and by Fourier transforming with respect to $(t-t^\prime)$
we obtain
\begin{multline}
\label{bse}
 \left[  \hbar \omega- \left(\epsilon_{j \kk}^{\mathrm{G_0W_0}} -\epsilon_{i \kk}^{\mathrm{G_0W_0}}\right) \right]
\chi^{\mathrm r}_{\substack{ij,\kk\\lm \mathbf p}}(\omega) = 
i \(f_{i \kk}-f_{j \kk}\) \left[ \delta_{jl} \delta_{im} \delta_{\kk,\pp} + \right. \\ \left.
+i\sum_{st,\qq}\{ W_{\substack{j \kk,s (\kk-\qq) \\ i\kk,t (\kk-\qq) }} -2   v_{\substack{ji, \kk \\ st,\kk-\qq }}  \}  
\chi^{\mathrm r}_{\substack{st, \kk-\qq \\ lm,\pp }}\(\omega\) \].
\end{multline}
formally equivalent to the standard BSE. 


\section{Optical properties from a real-time BSE}
\label{ss:solution}
In this section we describe how to combine the many-body Hamiltonian [Eq.~\ref{mbhamiltonian}] with dynamical Berry's phase. Then we validate our approach against standard GW+BSE results and then apply it to the second harmonic generation in low dimensional materials. 
\label{teospectro}      
\subsection{Practical implementation and validation}
As we have seen in Chapter \ref{chapterberry}, in order to study non-linear phenomena in solids and periodic nanostructures we cannot employ the standard dipole operator used in the linear response formalism. In fact dipole matrix elements are correct only for bands that do not cross, as for instance valence and conduction bands.\cite{blount} But all physics beyond the linear regime includes intra-band transitions and transition between crossing bands.\\
Modern Theory of Polarisation\cite{souza_prb,nloptics2013} provides the correct way to couple electrons and external electric fields beyond the linear regime. In presence of any single-particle Hamiltonian, as for instance the one in Eq.~\ref{mbhamiltonian}, it is possible to formulate an effective time-dependent Schr\"odinger equation as:
\bea
i\hbar  \frac{d}{dt}| v_{m\kk} \rangle &=& \left( H^{\text{mb}}_{\kk} +i \efield \cdot \tilde \partial_\kk\right) |v_{m\kk} \rangle \label{tdbse_shf}.
\eea
Where $| v_{m\kk} \rangle$ are the periodic part of the Bloch states that determines the system polarisation [Eq.~\ref{berryP2}], and the last term in the r.h.s describes the coupling with the external field (see Chapter~\ref{chapterberry}).
Notice that other choices consistent with the periodic boundary
conditions are possible, as for example an external field
with the cell periodicity\cite{PhysRevLett.87.036401}, or an electric
field  with a finite momentum $\mathbf q = \kk-\kk'$~\cite{PhysRevLett.84.1768}. In these two cases we do not need the Berry's phase formulation even in the length gauge.\\
In order to illustrate and validate the real-time BSE approach and
our numerical implementation, 
we present an example on {\it h}-BN. This is a wide gap insulator whose optical properties are
strongly re-normalised by excitonic effects and for which all the
parameters necessary in DFT, $G_0W_0$ and response calculations,
are known from previous studies\cite{PhysRevLett.96.126104,PhysRevLett.100.189701,attaccalite}. 
In this example we used Eq.~\eqref{tdbse_shf}, with and without
the self-energy terms. We refer to the
former approximation as RT-BSE, and to the latter as
TDH. Within equilibrium MBPT these two approximations
would correspond to the BSE and RPA, and in fact they reduce to BSE and RPA
within the linear response limit (Sec.~\ref{linear_response}).   

In the example (Fig.~\ref{hbn}), we simulated {\it h}-BN interacting with a
weak delta-like laser field. A delta-like laser field probes all frequencies of the system and
the Fourier transform of the macroscopic polarizability provides directly the susceptibility, and thus the dielectric
constant:
\bea
\mathbf P(\omega) &=& \epsilon_0 (\hat \epsilon(\omega) -\hat I)  \mathbf E( \omega)\\ 
\hat \chi(\omega) &=& \frac{\mathbf P(\omega)}{\epsilon_0 \mathbf E(\omega)}.
\eea
Since we use a weak field, we
expect negligible nonlinear effects. Then accordingly with Sec.~\ref{linear_response},  the results from
RT-BSE and TDH can be directly compared with the BSE and RPA within
the standard MBPT approach. All computational details are reported in Ref.~\cite{attaccalite}. Indeed, in Figs.~\ref{hbn}{\bf (b)},~\ref{hbn}{\bf (d)} the imaginary part of the dielectric constant
(optical absorption), obtained by Fourier transform of the polarisations
in Figs.~\ref{hbn}{\bf (a)},~\ref{hbn}{\bf (c)}, is indistinguishable from that
obtained within equilibrium MBPT, validating our numerical
implementation.
Beyond the linear regime we use the scheme described in Sec.~\ref{sc:compdet} to extract the non-linear coefficients that describe high order response functions.
\subsection{Second Harmonic Generation in h-BN and MoS$_2$ monolayers}
In this section we apply the RT-BSE to study second harmonic generation(SHG) in two dimensional crystals.\cite{attaccalite2015strong,gruning2014erratum,PhysRevB.89.081102}
In recent years hexagonal boron nitride (h-BN) and the rich family of transition metal di-chalcogenides, has attract the attention of the scientific community for the possible applications to optoelectronics and photonics.\cite{doi:10.1021/nn403159y,PhysRevB.87.201401}
Low dimensionality is responsible for the unique electronic and optical properties of 2D crystals, but at the same time is their limiting factor: because of the extremely short optical absorption length the light-matter interaction in absolute terms is inherently weak, though relatively very strong.\cite{doi:10.1021/nn403159y} Therefore ongoing research focuses on finding out physical mechanisms to enhance the SHG in those materials.

\begin{wrapfigure}{l}{0.4\textwidth}
    \vspace{-0.7cm}
\begin{center}
\includegraphics[width=0.4\textwidth]{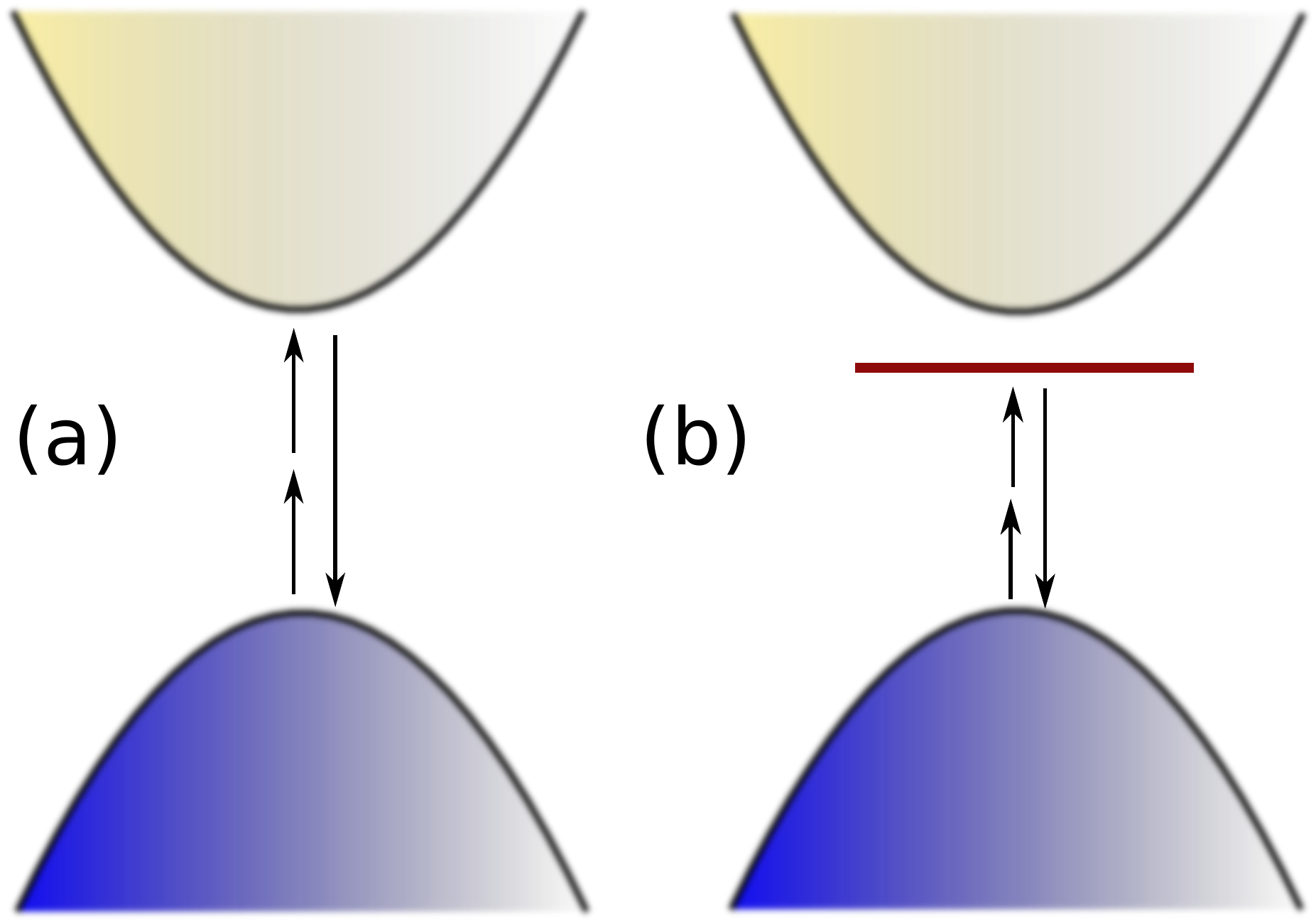}
\caption{\footnotesize{Schematic representation of  (a) the SHG within the IPA, or (b) accounting for electron-hole interaction. In (a) SHG is given simply by transitions between the valence (blue) and conduction (yellow) manifolds; in (b) electron-hole may lead to the formation of a bound exciton, an atomic-like level (dark red) into the fundamental band gap that strongly modifies the SHG. \label{schemeshg}}}  
\end{center}
\end{wrapfigure}   

On a more fundamental level, SHG is extremely interesting since it allows to probe symmetries and excitations not visible in the linear regime\cite{doi:10.1021/nl401561r,kumar2013second}.   In both cases it is 
important to have the support and guide from the theory through accurate and reliable numerical simulations.

Using the methodology developed in the previous chapters we investigate the SHG in h-BN and MoS$_2$ monolayers, two materials with promising applications in optoelectronic and photonic devices. These two materials share an hexagonal 2D structure but their electronic and optical properties are quite different. The optical absorption of h-BN is dominated by two very bright and strongly bound excitons.\cite{PhysRevLett.96.126104,PhysRevLett.100.189701} Instead the optical absorption of MoS$_2$ is characterised by a weakly bound exciton, split by spin-orbit coupling,\cite{PhysRevLett.105.136805} and by a brighter exciton in the continuum at about 3~eV.\cite{molina2013effect} 

The interpretation of the linear optical properties of these materials has been possible through \emph{ab initio} calculations (e.g. Refs.~\cite{molina2013effect,PhysRevLett.100.189701}). These computational studies captured the peculiar nature of excitons in nanostructures\cite{scholes2006excitons} by the inclusion of the relevant correlation effects.

In contrast, large part of calculations of nonlinear optical properties employ the independent particle approximation\cite{guo2005second,margulis2013optical} (IPA) which are inadequate for low dimensional systems (see Fig.~\ref{schemeshg}).\cite{scholes2006excitons} 
In order to investigate the contribution of correlation effects on the SHG spectra of h-BN and MoS$_2$ monolayers, we use the RT-BSE approach developed in this chapter. For both materials we disclose the signature of bound excitons and show that excitonic effects not only significantly modify the shape of the spectrum with respect to the IPA, but strongly enhance its intensity. In the conclusions we comment how this finding may open the possibility of engineering the SHG signal in these materials. 



\subsubsection{Results}
In order to simulate isolated hexagonal-BN and MSo$_2$ monolayers we used a supercell approach with a large distance between the sheets. Then we propagate the time-dependent Schr\"odinger equation with the Hamiltonian derived in Eq.~\ref{mbhamiltonian}. Non-linear coefficients are obtained from the real-time polarisation [Eq~\ref{berryP2} ] as described in chapter~\ref{chapterberry} and all numerical details are presented in Ref.~\cite{PhysRevB.89.081102}. 

\subsubsection{h-BN monolayer}
Hexagonal Boron-Nitride is a transparent insulating material with a large band gap of about $6~eV$. Its optical properties are dominated by strong bound excitons and they are nearly independent from the layers arrangement.\cite{PhysRevLett.96.126104,PhysRevLett.100.189701} The single layer h-BN inherits all these properties from its bulk counterpart.
\begin{figure}[H]
    \vspace{-0.32cm}
    \centering
\includegraphics[width=0.5\textwidth]{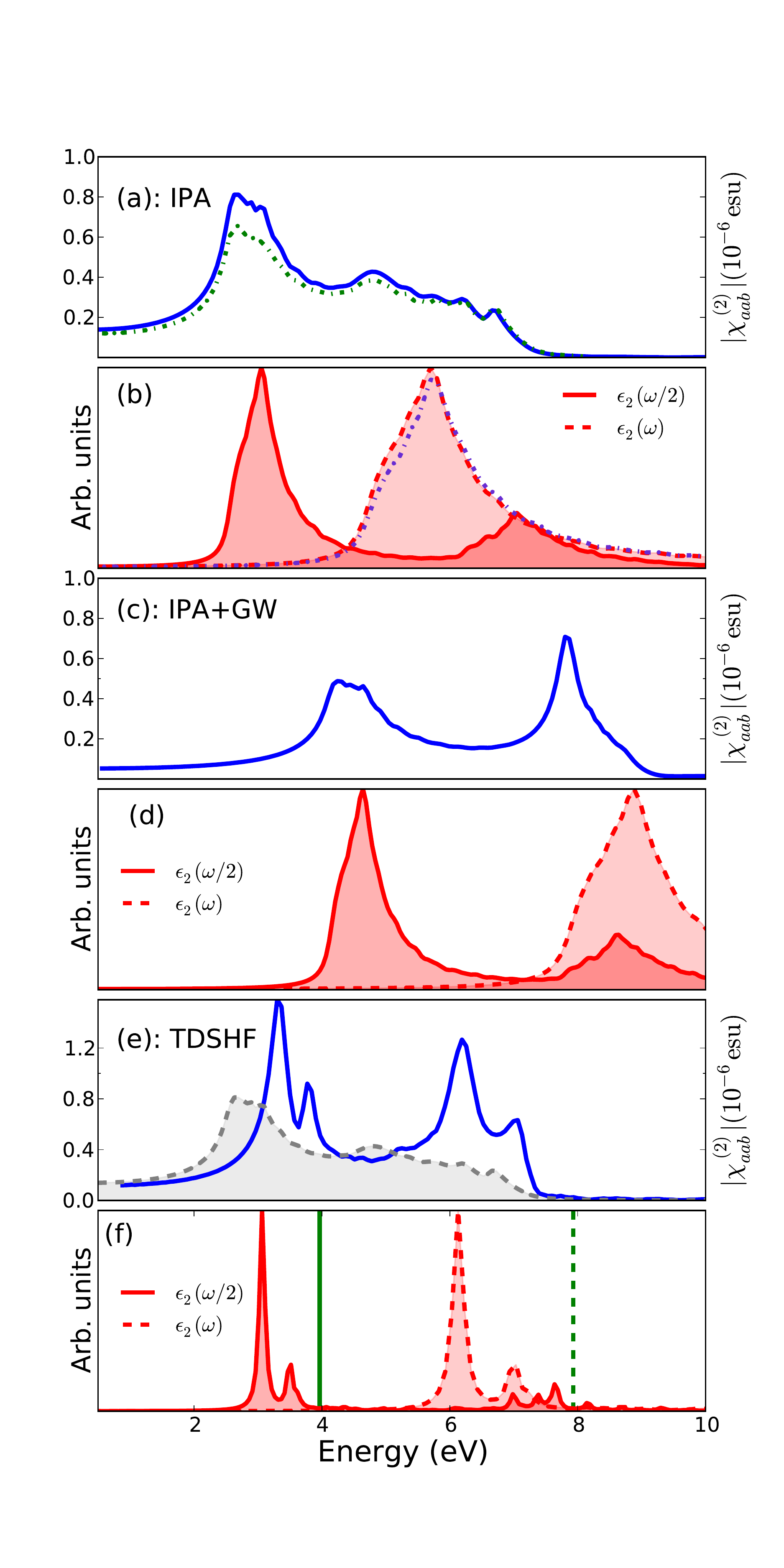}
    \vspace{-0.85cm}
\caption{\footnotesize{SHG spectra for the h-BN monolayer at different levels of theory [Eq.~\eqref{mbhamiltonian}]: (a) IPA (blue continuous line) and TDH (green dashed line); (c) IPA + GW correction (blue continuous line); (e) RT-BSE (blue continuous line) and IPA (grey dashed line). The imaginary part of the dielectric constant at both $\omega/2$ (red continuous line) and $\omega$ (red dashed line) is reported in (b), (d) and (f) for IPA, TDH and RT-BSE respectively. (f) also reports the LRC spectrum (blue dotted-dashed line) whose intensity has been reduced by a factor 0.5 for presentation reasons. The vertical lines represent the $GW$ fundamental gap (green dashed line) and half of the $GW$ fundamental gap (green continuous line). \label{absX2bn} [Figure from Ref.\cite{PhysRevB.89.081102}]}}
\end{figure}

\begin{figure}[h]
\centering
\includegraphics[width=0.6\textwidth]{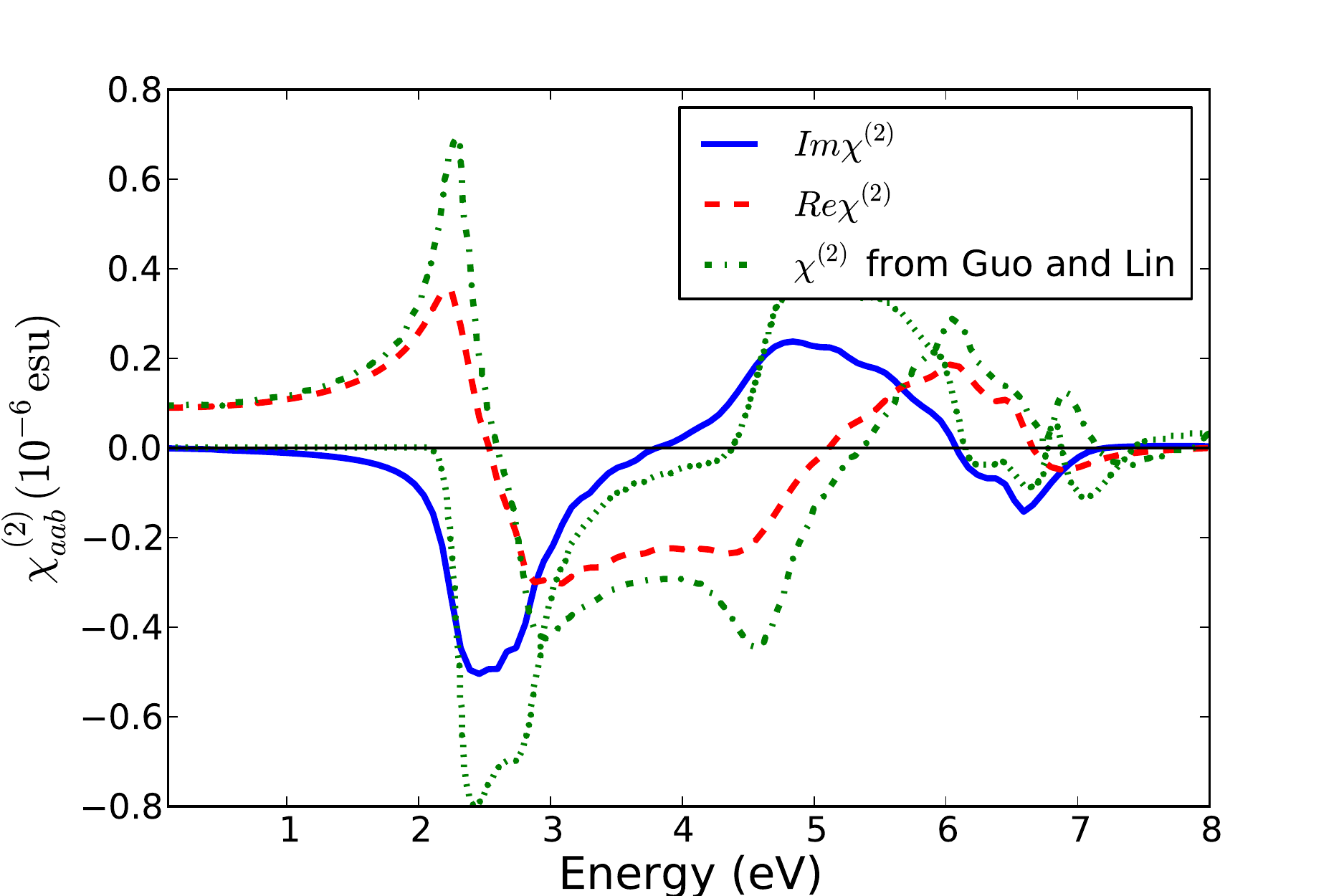}
\caption{\footnotesize{SHG for the h-BN monolayer within the IPA: the real (red dashed line) and the imaginary (blue continuous line) part of the calculated $\chi^{(2)}_{aab}$ are compared with the real (green dashed-dotted line) and imaginary part (green dotted line) of the $\chitwo$ obtained by Guo and Lin.\cite{guo2005second}[Figure from Ref.\cite{PhysRevB.89.081102}]}\label{X2bn}}
\end{figure}
In Fig.~\ref{absX2bn} we report the calculated absolute value of $\chi^{(2)}_{aab} (\omega)$ at different levels of approximation. $\chi^{(2)}_{aab} (\omega)$, where $a$ and $b$ are the in-plane Cartesian directions, is the only independent in-plane component of $\chi^{(2)} (\omega)$: all other components can be obtained from the $\chi^{(2)}_{aab} (\omega)$ with simple symmetry considerations, for instance $\chi^{(2)}_{bbb} (\omega)=-\chi^{(2)}_{aab} (\omega)$. 

At IPA level [Fig.~\ref{absX2bn}(a)], the SHG presents a peak at $2.3~eV$ and a broad structure between $4 - 7 eV$. By comparison with the imaginary part of the dielectric constant $\epsilon_2$ both at $\omega/2$ and  $\omega$ [Fig.~\ref{absX2bn}(b)] calculated at the same level of theory, we can attribute the peak at $2.3~eV$ to two-photon resonances with $\pi \to \pi^*$ transitions, and the broad structure mostly to one-photon resonances with $\pi \to \pi^*$ transitions, with contributions around $7 eV$ of two-photon resonances with $\sigma \to \sigma^*$ transitions.

This level of theory is the one usually employed in theoretical calculations of SHG: in fact results for the h-BN monolayer were previously obtained by Guo and Lin:~\cite{guo2005second} Fig.~\ref{X2bn} shows a very good agreement between our results and those obtained in Ref.~\cite{guo2005second}. In the following we show how effects beyond the IPA---that is the additional terms in Eq.~\eqref{mbhamiltonian}---modify the SHG spectrum.

\begin{figure}[h]
\centering
\includegraphics[width=.7\textwidth]{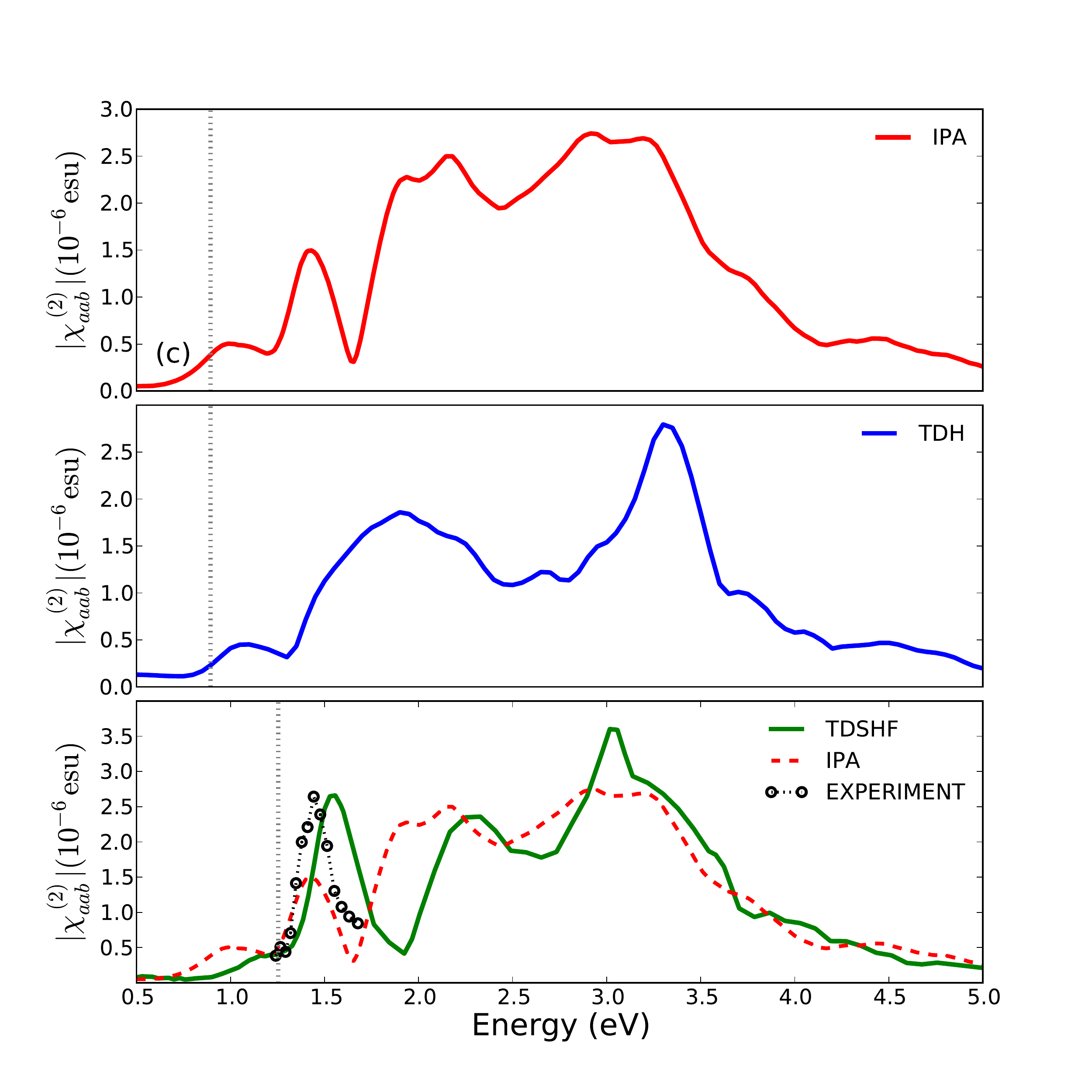}
\caption{\footnotesize{SHG in the MoS$_2$ monolayer at different levels of the theory [Eq.~\eqref{mbhamiltonian}]: (a) IPA (red squares), (b) TDH (blue circles) and (c) RT-BSE (green squares). The latter is compared with IPA (red dashed line) and experimental results of Malard et al.\cite{PhysRevB.87.201401} (black circles). Since the experimental SHG is measured relatively to the substrate, we renormalized the experimental spectrum to match the intensity of the $~ 1.5 eV$ peak in the calculated spectrum. The grey dotted vertical lines indicate the energy of half of the Kohn-Sham band gap in (a) and (b), and of  half of the $GW$ band gap in (c).[Figure from Ref.\cite{PhysRevB.89.081102}] }\label{fg:SHMoS2}}
\end{figure}

We start by adding crystal local field effects, included at the TDH level [Fig.~\ref{absX2bn}(a)]. Because of the weak in-plane inhomogeneity of the h-BN, local field effects are quite small---though they are not negligible as for the absorption spectrum  [Fig.~\ref{absX2bn}(b)]---and results in the reduction of about 20\% of the peak at $2.3~eV$. Next we consider the renormalization of the band structure by quasiparticle corrections within the $GW$ approximation (IPA+$GW$) [Fig.~\ref{absX2bn}(c)]. For h-BN this renormalization can be safely approximated by a rigid shift of the conduction bands. Differently from the absorption spectrum [Fig.~\ref{absX2bn}(d)], the SHG is not simply shifted by $GW$ corrections, but its shape changes remarkably as a consequence of the more involved poles structure of the second order susceptibility.~\cite{PhysRevB.82.235201,hughes1996calculation}
In fact, the IPA+$GW$ shows two peaks: the first at about $4~eV$ is the shifted two-photon $\pi \to \pi^*$  resonances peak which is attenuated by 40\% with respect to IPA  [Fig.~\ref{absX2bn} (a)]; the second very pronounced peak at about $8~eV$ comes from the interference of  $\pi \to \pi^*$  one-photon resonances and  $\sigma \to \sigma^*$ two-photon resonances.  

Finally, in Fig.~\ref{absX2bn}(e) we consider the full Hamiltonian in Eq.~\eqref{mbhamiltonian}. In particular we add the self-energy terms that introduces an attractive interaction between the excited electrons and holes\cite{strinati}. The SHG spectrum presents four sharp and strong peaks and its onset is red-shifted by about $1~eV$ with respect to the the IPA+$GW$ [Fig.~\ref{absX2bn} (c)]. By comparing with the imaginary part of the dielectric constant $\epsilon_2$ both at $\omega/2$ and  $\omega$ [Fig.~\ref{absX2bn} (f)] calculated at the same level of theory, the two couples of peaks can be identified respectively as the two- and one-photon resonances with the excitons at $6$ and $7~eV$.  Fig.~\ref{absX2bn}(e) also shows again the SHG within the IPA to emphasise the striking difference between the two spectra: the RT-BSE spectrum presents features that are missing in IPA and more importantly is twice as strong than IPA at the exciton resonances. 


\begin{center}
\begin{table}[h]
\small
\begin{tabular}{c|cc|c|c|c}
\hline
$|\chi^{(2)}_{aab} (0)| $ (pm/V) & \multicolumn{2}{c|}{IPA} & TDH & IPA+G$_0$W$_0$ & RT-BSE  \\
\hline
 h-BN &  57.1(3)  & [40.7] & 48.5(5) &  21.8(1)  & 46.9(2)  \\
\hline
\end{tabular}
\caption{$\omega\rightarrow 0$ limit of $\chi^{(2)}_{aab} (-2\omega,\omega,\omega)$ of the h-BN monolayer at different levels of the theory [Eq.~\eqref{mbhamiltonian}]. As a comparison, for the IPA we report in square brackets also the value obtained in Ref.~\cite{guo2005second}.\label{tab1}}  
\end{table}
\end{center} 

In  Table~\ref{tab1} we report  the value of the second optical susceptibility at $\omega=0$, $\chi^{(2)}(\omega \to 0)$, extrapolated from the SHG  behaviour at small frequencies.    
Again, at the IPA level our result agrees with the one of Guo and Lin\cite{guo2005second} within the error bar. Adding the effects beyond IPA, modifies the $\chi^{(2)}(\omega \to 0)$ value, and in particular within RT-BSE we found a value smaller by about 10\% than at the IPA level.

Excitonic effects in SHG spectrum have been treated as well in a TD-DFT framework\cite{PhysRevB.82.235201} by using the so-called long-range-corrected (LRC) approximation,\cite{LRC} a semi-empirical simple model for the screened electron-hole attraction, that includes only the long-range part of the interaction. In Fig.~\ref{absX2bn}(f) we see the $\epsilon_2$ calculated within the LRC approximation: as earlier recognised, this approximation fails for strong excitons. In fact by tuning the empirical parameter for the screening we could get the position of the first exciton, though its intensity is strongly overestimated (see caption of Fig.~\ref{absX2bn}), but in no way we could get the second excitonic peak. Those pitfalls would be reflected in the SHG spectrum, though we did not test it in our approach. Then clearly the h-BN monolayer and similar low dimensional materials with strong excitonic effects cannot be treated within the approach proposed in Ref.~\cite{PhysRevB.82.235201}.

\subsubsection{MoS$_2$ monolayer}
MoS$_2$ differs from h-BN in several aspects. First, while the h-BN has an indirect minimum band gap as its bulk counterpart, in MoS$_2$ an indirect-to-direct band gap transition occurs passing from the bulk to the monolayer due to the vanishing interlayer interaction.  Second, spin-orbit coupling plays an important role in this material, splitting the top valence bands, as visible from the absorption spectrum, presenting a double peak at the onset.\cite{PhysRevLett.105.136805} Third, Mo and S atoms in the MoS$_2$ monolayer are on different planes resulting in a larger inhomogeneity than for the h-BN.
Figure~\ref{fg:SHMoS2} presents the SHG spectra $|\chi^{(2)}_{aab}|$  at the different level of approximations of Eq.~\eqref{mbhamiltonian}. At the lowest IPA level [Fig.~\ref{fg:SHMoS2} (a)], the SHG presents three main features: a small peak at 1~eV, which originates from two-photon resonances with transitions close to the minimum gap at the $K$ point; a larger peak around $1.5$~eV, which originates from two-photon resonances with transitions along the high symmetry axis between $\Gamma$ and $K$ where the highest valence and lowest conduction bands are flat and there is a high density of states; a broad structure between $2-3.5$~eV which originates from one-photon resonances with transitions at $K$ and along $\Gamma-K$ and two-photon resonances with transitions to higher conduction bands. Note that we do not include spin-orbit coupling in Eq.~\eqref{mbhamiltonian}. 
The latter is expected to split the lowest peak into two weaker sub-peaks~\cite{molina2013effect}, but to leave unaffected the second peak, the most important when it comes to applications and central to our analysis.\cite{PhysRevB.87.201401}  
Because of the quite strong inhomogeneity of the MoS$_2$ monolayer, the addition of crystal local field effects within the TDH strongly modifies the SHG  [Fig.~\ref{fg:SHMoS2} (b)]. 
In particular the main peak at $1.5$~eV merges with the plateau at $2$~eV while a peak appears around $3.3$~eV.
Finally, within the RT-BSE, we add the quasi-particle effects and electron-hole interaction  [Fig.~\ref{fg:SHMoS2} (c)]. The small shoulder around $1$~eV, below half of the $GW$ gap (1.25 eV, grey dotted vertical line in the figure), originates from two-photon resonances with the bound excitons around $2$~eV which are well visible in the experimental absorption spectra.\cite{PhysRevLett.105.136805}  The main peak at about $1.5$~eV, present in the IPA spectrum but washed out by local field effects within the TDH, is restored by the electron-hole interaction and its intensity is two times larger than in the IPA case. This peak corresponds to a two-photon resonance with the bright exciton at 3~eV observed in the absorption spectrum,\cite{molina2013effect} and its position and shape agrees well with the experimental measurements of Malard et al.\cite{PhysRevB.87.201401} for the SHG between $1.2-1.7$~eV, also reported in Fig.~\ref{fg:SHMoS2}(c). The calculated spectrum also shows a strong one-photon resonance with this exciton at 3~eV.
With respect to h-BN, where the IPA is clearly inadequate when compared with the RT-BSE [Fig.~\ref{absX2bn}(e)], here IPA and RT-BSE  presents similar features, at approximately the same position [Fig.~\ref{fg:SHMoS2} (c)]. In fact, this different behaviour could be expected from the linear optical response of these materials, as discussed in the Introduction. Nevertheless, also in this case the electron-hole interaction proved to be key for SHG as it doubles the intensity of the main peak in the visible range, the one that is relevant for applications in nonlinear optics.      
\section{Conclusions}
We presented a novel approach for  \emph{ab-initio} calculation of linear and non-linear
optical properties in bulk materials and nano-structures that uses a
real-time extension of the BSE.
The proposed approach combines the flexibility of a real-time approach
with the strength of MBPT in capturing electron-correlation.  It
allows to perform computationally feasible simulations beyond the
linear regime.
Being the approach based on the non-equilibrium Green's Function theory, it is possible to
include effects such as lifetimes, electron-electron
scattering\cite{bsedynamic} and electron-phonon
coupling\cite{giustino2016electron} in a systematic way.
We validated the RT-BSE in the case of {\it h}-BN calculating
the optical absorption and comparing it with the results from equilibrium
GW+BSE, then we applied the same methodology to study the non-linear response of two-dimensional
crystals.
We found that electron-hole interaction greatly enhances the SHG signal in 2D crystals with respect to the independent-particle picture. Specifically, for the h-BN monolayer one- and two-photons resonances with bound excitons produce strong signatures in the SHG spectrum with intensities two times larger than expected from the IPA. In MoS$_2$, though the shape of the spectrum is not strikingly modified by excitonic effects as for h-BN, the electron-hole interaction enhances, again by about 200\%, the SHG signal in the visible range with respect to the IPA. 
This finding may provide a spin-off for the quest of materials with high SHG. In fact---given that the SHG signal depends largely on the electron-hole interaction that in turn depends on the electronic screening---the SHG intensity can be tuned by changing the electronic screening. Then, it may be possible, as proposed in Ref.~\cite{gao2012artificially}, to engineer meta-materials with a high SHG by combining layers of different 2D crystals\cite{gao2012artificially} so to change the electronic screening, and further enhance the electron-hole interaction effects.
As side finding, our results emphasise that it is critical for theoretical and computational approaches to accurately include electron-hole interaction, together with quasiparticle and local field effects, in order to predict non-linear optical response in low dimensional materials. In this regard, our recently proposed approach~\cite{nloptics2013,attaccalite} is quite promising as it imports into the very flexible real-time framework---apt to treat nonlinear optics---the combination of BSE+$GW$ successfully applied to the linear optical response of low-dimensional materials.\\ 
The interested reader can find other applications of the RT-BSE in Refs.~\cite{attaccalite2015strong,attaccalite1d}.

\clearemptydoublepage

\chapter{Density functional polarisation theory}
\label{chaptertddft}
\section{Time-dependent Density Functional Theory in periodic systems}
Density functional theory (DFT)\cite{PhysRev.140.A1133,PhysRev.136.B864} is a standard approach for calculating ground-state properties of extended and finite systems\cite{doi:10.1021/jp960669l,RevModPhys.87.897}. Time-dependent density functional theory (TDDFT) is an extension of the ground-state formalism that allows to investigate the properties and dynamics of many-body systems in the presence of time-dependent potential. TDDFT is based on the Runge-Gross (RG) theorem\cite{PhysRevLett.52.997}  that establishes a one-to-one correspondence between time-dependent densities and time-dependent one-body potentials. For example when we consider an isolated molecule and an electric filed as  perturbation by means of TDDFT we have access its optical response. As for the DFT case, the RG theorem just guarantees the existence of the mapping, but do not provide a way to construct it. Different approximations for the time-dependent (or frequency dependent) exchange-correlation functional have been proposed in the literature. Exchange correlation functional have been obtained from the ones of the DFT, including long range corrections or derived from more accurate methods\cite{Onida,faber2014excited}. The response equations of TDDFT have been encoded in standard quantum chemical packages\cite{valiev2010nwchem}, and real-time TDDFT is currently used to simulate the short time dynamics of excited electrons\cite{PSSB:PSSB200642067}.\\
Despite the success of TDDFT in molecular systems, the situation is more complicated in extended system. 
In fact, although other methods, as for instance Green's function theory\cite{strinati}, provide a similar accuracy in extended\cite{Aulbur19991} and finite systems\cite{blase2011charge,faber2012electron}, this is not the case of TDDFT.\\
Due to the lower computational cost of TDDFT respect to Green's function formalism it would be desirable to have the same accuracy in extended and finite systems.\\
The first idea one can have it is that the available exchange correlation functionals, for some reason, are not enough precise to describe excitation in solids. Unfortunately the problem is more serious.\cite{maitra2003current}
In order to understand from where the problem originates, let's go back to the original statement of the RG theorem.
In the first part of the proof, RG established a one-to-one correspondence between potentials and currents.  Then in the second part of their work, they used the continuity equation to relate currents to densities in order to prove the mapping between densities and potentials.  The problem in the RG proof stays in the use of the continuity equation. As we saw in the introduction to chapter~\ref{chapterberry} there is not a univoque mapping between current and density in periodic systems. 
In fact the mapping between currents and densities requires that certain surface integral involving the density and the potential vanishes. For finite systems this condition  can be given rigorously at the surface in which the density vanishes. For a periodic system, one might try to choose a surface around which the density and potential are periodic but for a uniform field the linearity of the potential prevents this, and TDDFT does not apply. \\
The problem of TDDFT in periodic system was also illustrated  by means of a simple example in the paper of Maitra et. al.\cite{maitra2003current}. They considered a free electron gas on a ring subjected to a constant uniform electric field. In this case it is possible to write down the exact solution of the problem. One finds that the electric field modifies only the phase of the single particle orbitals, leaving the density unchanged. This result shows that different electric fields  give rise to exactly the same density and therefore there is not an unique mapping between density and external field.\\
In order to solve the problems with TDDFT in periodic systems and extension was presented some years ago, the Time-Dependent Current Density Functional Theory(TDCDFT)\cite{PhysRevA.38.1149}. This formulation uses the direct mapping between the external potential and the current density, without reels on the continuity equation.\\
In this section we will use a simplified versions of TDCDFT, i.e. the Density-Functional Polarization Theory (DFTP).  In DFTP one uses the relation between polarization and current to construct a theory that relies on density and polarization instead of current density. The possibility to use the polarisation as additional variable besides the density it is a valid approximation when it is possible to disregard the transverse term of the current. Since we are interested in the optical response in the limit of long-wave length limit, this is a valid approximation.\\
In the rest of this chapter we will introduce DFTP, and explain how to build functionals in term of polarization and density. Then we will use these functionals in our real-time approach to study non-linear response of bulk semiconductors.
\section{General Introduction}
In periodic systems in presence of a time-dependent homogeneous electric field only the one-to-one correspondence between the time-dependent currents and potentials (scalar and vector) can be established and time-dependent current density functional theory (TD-CDFT) is then the correct theoretical framework.~\cite{maitra2003current,PhysRevA.38.1149} In particular it is the optical limit, i.e. the case in which the transferred momentum $\qq\ra 0$,  which cannot be described starting from the density only. 
One could still work with functionals that depends on the density--only, but there is a price to pay. All the equations have to be worked out with a finite but very small momentum  and the $\qq\ra 0$ limit can be performed only at the end of the calculation. 
Furthermore in order to describe excitonic effect the exchange--correlation functionals have to be ultra--nonlocal and to diverge as $\qq \ra 0$.~\cite{PhysRevLett.88.066404} Such an approach is used within the linear response framework but it is not feasible within a real time framework since for practical reasons calculations have to be performed directly at $\qq=0$. Thus one needs to go beyond the density--only treatment. As a clear indication of this, the macroscopic polarisation and the response functions cannot be calculated within a density--only scheme at $\qq=0$.~\cite{PhysRevB.9.1998}
Problems are not limited to the time--dependent case. Even in the static limit, e.g. for dielectrics in a static homogeneous electric field, Gonze and coworkers proved that {\em ``the potential is not a unique functional of the density, but depends also on the macroscopic polarisation''}.~\cite{Gonze1995} In this case then the theory has to be generalised to consider functionals of both the density and the polarisation in what is called density--polarisation functional theory (DPFT). The latter can be obtained from TD-CDFT in the static limit.

Here we propose a real--time approach based on DPFT for calculating the optical response properties of dielectrics, thus considering functionals of both the time--dependent density and the macroscopic bulk polarisation.
Real--time approaches allows in principle to calculate the optical response at all order so to access nonlinear properties\cite{takimoto:154114}, including nonperturbative extreme nonlinear phenomena\cite{lee2014first} and to simulate real--time spectroscopy experiments.\cite{otobe2015femtosecond} It is highly desirable then to have computational inexpensive first principles real--time approaches, such as TD-DFT, that include excitonic effects.            
In particular here we consider an effective electric field which is a functional of the macroscopic polarisation. We employ simple local functionals of the polarisation~\cite{maitra2003current,PhysRevLett.115.137402} either fitted to reproduce the linear optical spectra\cite{LRC} or derived from the jellium with gap model kernel.~\cite{jgm}

In the following, we review DPFT and we extend it to the case of time-dependent electric fields. We discuss briefly the approximations for the effective-electric field and we present how the relevant response functions are calculated from the macroscopic polarisation.  Then, we show that for the optical absorption, the second- and third-harmonic generation of semiconductors  the simple local functionals of the polarisation account for excitonic effects similarly to the ultranonlocal kernel within the density-only response framework. In the conclusion we discuss the proposed approach as an alternative to existing schemes based on TD-DFT and TD-CDFT.

\section{Density polarisation functional theory}

\subsection{Static case}
An infinite periodic crystal (IPC) in a macroscopic electric field $\Efield^{\text{ext}}$ does not have a ground-state. Therefore the Hohenberg-Kohn theorem cannot be applied and DFT cannot be used. In particular the density does not suffice to describe the system as the one-to-one mapping between density and external potential does not hold: one can devise an external macroscopic electric field that applied to a system of electrons in an IPC does not change its density $n$.
The works of Gonze Ghosez and Godby,~\cite{Gonze1995} Resta~\cite{Resta1996}, Vanderbilt~\cite{Vanderbilt1997} and of Martin and Ortiz~\cite{Martin1997} established that in addition to the density, the macroscopic (bulk) polarisation $\PP$ is needed to characterise IPC in a macroscopic electric field. With some cautions the proof of the Hohenberg-Kohn theorem can be extended~\cite{Martin1997} to demonstrate the existence of the invertible mapping
$$(n(\rr),\PP)\leftrightarrow (\bar v^{\text{ext}}(\rr),\Efield^{\text{ext}}) $$
where $ \bar v_{\text{ext}}$ is the periodic microscopic part of the external potential.
Then the total energy of an IPC is a functional of both the electron density $n$ and the macroscopic polarisation $\PP$:
\be \label{eq:dpften}
E[n,\PP] = \bar F[n,\PP]+ \int_\Omega n(\rr) \bar v^{\text{ext}}(\rr)\,d\rr  -\Omega\, \Efield^{\text{ext}} \cdot \PP,
\ee 
where $\bar F$, the internal energy, is a universal functional of both $n$ and $\PP$ (see Ref.~\cite{Martin1997} for details),
and is defined in the usual way
as the sum of the expectation the kinetic and electron-electron interaction operators
\be
\bar F[n,\PP] = \langle \Psi | \hat T + \hat V_{ee} |\Psi \rangle.
\ee 
The difference with the internal energy within standard DFT is that the $N$-particle wavefunction $\Psi$ is not an eigenstate of the original Hamiltonian (which does not have a ground state), but of an auxiliary Hamiltonian which commutes with the translation operator (see Ref.~\cite{Martin1997} for details). Notice that DPFT is not the only way to treat IPC in a electric field within a density functional framework: as an alternative Umari and Pasquarello proposed $\Efield$-DFT, a density functional theory depending on the electric field.~\cite{Umari2005}   

The Kohn-Sham equations can be extended as well to treat IPC in a macroscopic electric field.~\cite{Martin1997}
In particular the Kohn-Sham crystal Hamiltonian takes the form:
\be \label{eq:dpftks}
H^{s}_\kk
= -\frac{1}{2}\left(\nabla + i\kk\right )^2 + \bar v^{s}(\rr) -\Omega\Efield^{s}\cdot \nabla_\kk 
\ee
which is a functional of both the density and the polarisation.
In Eq.~\eqref{eq:dpftks}  the Kohn-Sham microscopic (periodic) potential $\bar v^{s}$ is defined as
\be\label{eq:kspot}
\bar v^{s}(\rr) = \bar v^{\text{ext}}(\rr) +  \bar v^{\text{H}}(\rr) + \bar v^{\text{xc}}(\rr) 
\ee
$\bar v^{\text{ext}}(\rr)$, $\bar v^{\text{H}}$ are respectively the microscopic external and Hartree potential.
The total classical potential is defined as $\bar v^{\text{tot}}(\rr)=\bar v^{\text{ext}}(\rr)+\bar v^{\text{H}}(\rr)$.
$\bar v^{xc}$ is the functional derivative of the exchange--correlation energy with respect to the density.
$\bar v^{\text{ext}}(\rr)$ here describes the field generated by the ions, i.e. the electron--ion interaction
in the Coulomb gauge and neglecting retardation effects.
The last term of the RHS of Eq.~\eqref{eq:dpftks}---that originates from the last term in the RHS of Eq.~\eqref{eq:dpften}---constitutes
the key difference with respect to the zero-field KS equations.
$\nabla_\kk$~is the polarisation operator derived by functional-differentiating
$\PP$ [Eq.~\eqref{berryP2}] with respect to the KS eigenstates.
$\Efield^{s}$ is the KS macroscopic field  
\be
\Efield^{s} = \Efield^{\text{ext}} + \Efield^{\text{H}}  + \Efield^{\text{xc}},
\label{eq:ksfld}
\ee
that contains the corresponding macroscopic components of $\bar v_s$. Note that these macroscopic components cannot be included via the potential which would be ill defined when imposing periodic boundary conditions.  
The $\Efield^{\text{xc}}$ is defined as the partial derivative of the xc energy with respect to the polarisation density field.
The sum of the macroscopic external and Hartree fields defines the total classical field:
\be
\Efield^{\text{tot}} = \Efield^{\text{ext}} + \Efield^{\text{H}}. 
\label{eq:totfld}
\ee    

At zero-field, that is when no macroscopic external electric field $\Efield^{\text{ext}}$ is applied,
the macroscopic component of the ionic potential and of the Hartree component exactly cancel as a consequence of the charge neutrality of the system and  
the macroscopic xc component vanishes.
In this situation  standard density-only functional theory can be used.

As $\bar v^{s}$ and $\Efield^{s}$ are functionals of the density and the polarisation, the Kohn-Sham equations
for the KS orbitals $\{\phi_{n\kk}\}$  have to be solved self-consistently with the density (spin unpolarized case)
\be
n(\rr) = 2 \sum^{\text{occ}} |\phi_{n\kk}(\rr)|^2 
\ee
and the polarisation expressed in terms of a Berry phase [Eq.~\eqref{berryP2}].

\subsection{Time-dependent case}
The Runge-Gross theorem \cite{PhysRevLett.52.997} is the basis of TD-DFT. It establishes the one-to-one mapping between the time-dependent scalar potential and the time-dependent density. For the case in which a time-dependent vector potential is present Ghosh and Dhara~\cite{PhysRevA.38.1149} showed that the mapping can be established between the current-density and the vector potential. More recently Maitra and co-workers~\cite{maitra2003current} showed that TD-CDFT is the correct framework for IPC in homogeneous electric fields.

The time-dependent change in the polarisation density field $\pp$ is related to the time-dependent current-density $\jj$ by 
\be
\pp (\rr;t) = \int^t_{-\infty} dt' \jj(\rr;t') 
\ee
In a dielectric we can then use either $\pp(\rr,t)$ or $\jj(\rr, t)$ as main variable to describe an IPC in a time-dependent finite homogeneous electric field. Furthermore we can consider separately the microscopic and the macroscopic components of $\pp(\rr, t)$: $\PP (t)$ and $\bar \pp(\rr, t)$. The latter quantity is fully determined by the density through the continuity equation.
Then we can extend to the time-dependent case the one-to-one mapping
$$(n(\rr,t),\PP(t))\leftrightarrow (\bar v^{\text{ext}}(\rr,t),\Efield^{\text{ext}}(t)). $$
The time--dependent Kohn-Sham crystal Hamiltonian has the same form of the equilibrium KS Hamiltonian [Eq.~\eqref{eq:dpftks}] with the only difference that now potentials and wavefunctions are time--dependent:
\be \label{eq:tdksh}
H^{s}_\kk(t)
= -\frac{1}{2}\left(\nabla + i\kk\right )^2 +  \bar v^{s}(\rr,t) -\Omega\, \Efield^{s}(t)\cdot \nabla_\kk
.
\ee
We rewrite the external field and potential as the contribution at equilibrium, $\Efield^{\text{ext},0}$ and $\bar v^{\text{ext},0}(\rr)$  plus the time-dependent perturbation:
\bea
\Efield^{\text{ext}}(t)&=&\Efield^{\text{ext},0}+\Delta \Efield^{\text{ext}}(t), \\
\bar v^{\text{ext}}(\rr,t)&=&\bar v^{\text{ext},0}(\rr) +\Delta v^{\text{ext}}(\rr,t) .
\eea  
Then,
\bea
\bar v^{s}(\rr,t) &=& v^{s,0}(\rr) +\Delta  \bar v^{s}(\rr,t) \\
\Efield^{s}(t)     &=&\Efield^{s,0}+\Delta \Efield^{s}(t),
\eea
where the $0$ superscript denotes that the functional is evaluated 
in presence of the equilibrium fields, thus at equilibrium density and polarisation.  We then restrict ourselves to consider  the case with no external macroscopic electric field at equilibrium, i.e. ${\Efield^{\text{ext},0}=\mathbf{0}}$, and to a macroscopic-only time dependent perturbation, i.e. $\Delta \bar v^{\text{ext}}(\rr,t)=0$.
Therefore
\bea
\Delta \bar v^{s}(\rr, t ) &=& \Delta\bar v^\text{H}+\Delta\bar v^{\text{xc}} \\
\Delta\Efield^{s}(t)&=&\Efield^{s}(t)
\eea
Finally, the TD-KS equations for the periodic part $u_{n\kk}$ of the Bloch function 
can be expressed as
\be
  i\partial_t u_{n\kk} =\left( H^{s,0}_\kk +  \Delta \bar  v^{s}(\rr,t) -\Omega\, \Efield^{s}(t)\cdot \nabla_\kk \right)u_{n\kk}
,
\label{eq:kseom}
\ee
and have to be solved consistently with the time-dependent density and polarisation.
The latter has the same form of the static polarisation [Eq.~\eqref{xtrace}] with
the difference that $|v_{\kk n}\rangle$ are the time-dependent valence bands.

In the time dependent case and within the EDA,
it can be shown straightforwardly that the Hamiltonian
in Eq.~\eqref{eq:tdksh} can be derived from the KS Hamiltonian of TD-CDFT
with a gauge transformation from the velocity to the length gauge.\cite{maitra2003current}

\section{Expressions for the Kohn-Sham electric field}
\label{sc:ksef}

The KS electric field in Eq.~\eqref{eq:ksfld} is the sum of three components. It seems natural to consider the external component $\Efield^{\text{ext}}$ as an input of the calculation, i.e. $\Efield^{\text{ext}} = \Efield^{\text{inp}}$. The total classical field $\Efield^{\text tot}$ is then calculated from Eq.~\eqref{eq:totfld} by adding the Hartree component that in the EDA is the polarisation $\Efield^{H} = 4 \pi \PP$. This is not the only possible choice nor always the most convenient. When calculating linear and nonlinear optical susceptibilities, which do not depend on the total or external fields, it is numerically more convenient to choose the total classical field as input field. As this work objective is the calculations of optical susceptibilities we adopt indeed $\Efield^{\text{inp}} = \Efield^{\text{tot}}$.  The two choices for the input field, i.e. either the total or external field, have been referred as ``longitudinal geometry'' and ``transverse geometry'' by Yabana and coworkers\cite{PhysRevB.85.045134} and are discussed in more length in Appendix~\ref{appA}. 

While the choice of the input field is a matter of computational convenience, the choice of the expression for the xc macroscopic electric field is critical to the accuracy of the results. Like the microscopic xc potential no exact expression is known and one should resort to an approximation for the functional form of the xc field. Contrary to the microscopic xc potential for which hundreds of approximations exist,\cite{libxc} except for the work of Aulbur and coworkers~\cite{aulbur1996polarization} we are not aware of approximations for the xc macroscopic field. What does exist in the literature are xc kernels within the TD-DFT and TD-CDFT that give a non-zero contribution to the response in the optical limit. In what follows we link the xc kernel with the macroscopic field (similarly to Refs.~\cite{maitra2003current,PhysRevLett.115.137402}). 
In fact in the linear response limit the exchange-correlation electric field is related to the polarisation $\pp$ (see for example Refs.~\cite{maitra2003current},~\cite{PhysRevLett.115.137402})
through the xc kernel $\newtensor{F}^{\text{xc}}_{}$. The latter describes how the xc electric field (both microscopic and macroscopic) changes when the polarisation is perturbed. $\newtensor{F}^{\text{xc}}_{}$ can be defined independently through the Dyson equation connecting 
the polarisation response function of the physical system, $\newtensor{\chi}$,
to the polarisation response function of the KS system, $\newtensor{\chi^s}$. 
By rewriting the relation between $\Efield^{\text{xc}}$ and $\newtensor{F}^{\text{xc}}_{}$ in reciprocal space~\footnote{In general the expression of the xc electric field in terms of $\tensor{F}^{\sss XC}_{}$ is an integral along a path in the infinite dimensional space of the densities/macroscopic polarizations. Such integral does not depend on the path if $\tensor{F}^{\sss XC}_{}$ is defined as a functional derivative of some function. Furthermore here we consider the case where $\tensor{F}^{\sss XC}_{}$ is a local (or semi--local) functional. Then the infinite dimensional integral reduces to a simple three-dimensional integral which, in reciprocal space, can be represented as a sum over the $\GG$ vectors.} one obtains~\cite{maitra2003current} for the macroscopic component ($\GG=0$) 
\begin{multline}
\Efield^{\text{xc}}(t)= \int dt' \Big[ \newtensor{F}^{\text{xc}}_{\sss 00}(t-t') \PP(t')  \\
               -{i} \sum_{\GG' \neq 0} \newtensor{F}^{\text{xc}}_{\sss 0\GG'}(t-t')\frac{ n_{\sss \GG'}(t')}{{\rm G'}^2}\GG' \Big] 
\label{Excmac}
\text{,}
\end{multline}
and for the microscopic components $\efield^{\text{xc}}_{\sss \GG}$ ($\GG\neq 0$)
\begin{multline}
\efield^{\text{xc}}_{\sss \GG}(t)= \int dt' \Big[ \newtensor{F}^{\text{xc}}_{\sss \GG 0}(t-t') \PP(t')  \\
                    -{i} \sum_{\GG' \neq 0} \newtensor{F}^{\text{xc}}_{\sss \GG\GG'}(t-t')\frac{ n_{\sss \GG'}(t')}{{\rm G'}^2}\GG' \Big] 
\label{Excmic}
\text{.}
\end{multline}
The first term on the RHS of Eq.~\eqref{Excmac} is directly proportional to the macroscopic polarisation, the second term involves the density and is the microscopic contribution to the macroscopic field. Note that as we assume the EDA we do not have the contribution from the microscopic transverse current as in Maitra and coworkers.~\cite{maitra2003current}
The variation of the microscopic xc potential $\Delta \bar v^{\text{xc}}$ can be written in terms of the microscopic components $\efield^{\text{xc}}_{\sss \GG}$ as
\be
\Delta \bar v^{\text{xc}}_{\sss \GG}(t) =  i \frac{\GG \cdot  \efield^{\text{xc}}_{\sss \GG}(t) }{{\rm G}^2}.
\label{eq:pot_from_exc}
\ee
  
Berger~\cite{PhysRevLett.115.137402} has recently proposed an approximation for $\newtensor{F}^{\text{xc}}_{00}$ from current-density functional theory. The approximation however requires the knowledge of the Random-Phase-approximation (RPA) static dielectric function: while within a linear response approach this does not require any additional calculation, within a real-time approach the RPA static dielectric function needs to be computed beforehand. Furthermore Berger~\cite{PhysRevLett.115.137402} neglects the microscopic contribution.

An alternative way to derive approximations for $\newtensor{F}^{\text{xc}}_{}$ is to rely on the standard TD-DFT xc kernel $f^{\text{xc}}_{}$. The latter describes how the xc potential changes when the density is perturbed and is defined from the Dyson equation relating the density-density response of the physical and the KS system. The two kernels can be related via the equation
\be 
f^{\text{xc}}_{\sss \GG \GG'}(\qq\rightarrow 0; t - t') = \lim_{\qq\rightarrow 0} \frac{\newtensor{F}^{\text{xc}}_{\sss \GG \GG'}(t-t') \cdot \newtensor{g}}{|\qq + \GG||\qq + \GG'|}.  
\label{eq:k2k}
\ee 
where $\newtensor{g}$ is the metric tensor.

For example the long-range corrected (LRC) approximations $f^{\text{xc}}\approx f^{\sss LRC}$, which take the form
\be
f^{\sss LRC}_{\sss \GG\GG'}(\qq\rightarrow 0; t-t') =
  \lim_{\qq\rightarrow 0} \frac{-\alpha^{\sss LRC}\delta_{\sss \GG,\zero}\delta_{\sss \GG',\zero}}{|\qq|^2}\delta(t-t'),   
\label{jgmqzero}
\ee
can be used. Then $\newtensor{F}^{\text{xc}}_{\sss \zero\zero} \cdot \newtensor{g}$ can be approximated with any of the $\alpha^{\sss LRC}$ (we assume $\alpha > 0$) proposed in the literature. Unfortunately all the approximations proposed so far~\cite{LRC,PhysRevB.72.125203} neglect the dependence of $\alpha$ on the reciprocal lattice versors. Furthermore most of the approximations relies on empirical parameters, with the exception of the family of bootstrap kernels~\cite{PhysRevLett.114.146402,PhysRevLett.107.186401} that relate $\alpha$ to the electronic screening in an expression equivalent to that derived by Berger from TD-CDFT.

In this work, we derive the $\newtensor{F}^{\text{xc}}$ needed in Eq.~\eqref{Excmac} from the Jellium with Gap Model (JGM) kernel proposed by Trevisanutto and coworkers.\cite{jgm}
The latter kernel is a functional of the electronic density $n$ and of the fundamental gap of the material $E_{\text{gap}}$. In the optical limit the JGM kernel takes the form of a long-range corrected approximation
with $\alpha^{\sss LRC}$ defined as the cell average~\cite{jgm} of
\be
\alpha^{\sss JGM}(\rr;t) = 4\pi \tilde B \left[1 - \exp{\left( -\frac{E_{\text{gap}}^2}{4\pi n \tilde B} \right)} \right].  
\label{eq:alpha}
\ee
In the equation above $\tilde B = (B + E_{\text{gap}})/(1 + E_{\text{gap}})$, where $B=B[n]$ is a functional of the density found by fitting the local field factor of the homogeneous electron gas from Quantum Montecarlo data.~\cite{PhysRevB.57.14569}   
 
For cubic systems we thus consider $\newtensor{F}^{\text{xc}}\approx\newtensor{F}^{\sss JGM}$ with
\begin{subequations}
\begin{gather}
\newtensor{F}^{\sss JGM}_{\sss \zero\GG}(t-t') = -\frac{1}{2}\alpha^{\sss JGM}_{\sss \GG}(t) \newtensor I \delta(t-t') \\
\newtensor{F}^{\sss JGM}_{\sss \GG\zero}(t-t') = -\frac{1}{2}\alpha^{*\sss JGM}_{\sss \GG}(t) \newtensor I \delta(t-t').
\end{gather}
\label{eq:krnl}
\end{subequations}
where $\alpha^{\sss JGM}_{\sss \GG}(t)$ is the Fourier transform of Eq.~\eqref{eq:alpha} and $\newtensor I$ is the identity tensor. 
Notice that we symmetrized ${F}^{\sss JGM}_{\sss \GG,\GG'}$ so to obtain a Hermitian kernel. Other strategies of symmetrization have been proposed in the literature, see Ref.~\cite{jgm} and reference therein.
Like standard approximations for the kernel this approximation neglects memory effects (i.e. the macroscopic field at time $t$ depends on the values of the density and polarisation only at time $t$) and it is thus frequency independent.

Finally, inserting this approximation for the kernel [Eq.~\eqref{eq:krnl}] in the expression for the xc fields [Eq.~\eqref{Excmac}--\eqref{Excmic}] and using Eq.~\eqref{eq:pot_from_exc} we obtain 
\bea
\Efield^{\sss JGM}(t)&=&\alpha^{\sss JGM}_{\zero}(t)
{\cal \PP}(t)
-\frac{i}{2}\sum_{\sss \GG \neq 0} \alpha^{\sss JGM}_{\sss \GG} (t)\frac{ n_\GG(t)}
{{\rm G}^2}\GG \nonumber \\
\Delta \bar v^{\sss JGM}_{\sss \GG}(t)&=&\frac{i}{2}\sum_{\sss \GG \neq 0}  \frac{\alpha^{*\sss JGM}_{\sss \GG} (t)}{G^2} \GG \cdot {\cal \PP}(t), 
\label{eq:Excapp}
\eea
where the second term in the RHS of Eq.~\eqref{Excmic} is zero due to our symmetrization strategy [Eq.~\eqref{eq:krnl}]. In our calculations we will use either Eq.~\eqref{eq:Excapp} and or an $\alpha^{\text{opt}} \PP$ approximation for the macroscopic xc electric field in which $\alpha^{\text{opt}}$ is a parameter which gives the best agreement between the computed and experimental optical absorption spectra. The two approximations will be referred as JGM polarisation function (JGM-PF) and optimal polarisation functional (opt-PF).

\section{Computational details}

The eigensolutions $\{|\phi^0_{m\kk}\rangle \}$ of the zero-field Hamiltonian are calculated using the plane-wave pseudopotential density-functional code {\sc abinit}~\cite{abinit} within the local density approximation for the exchange-correlation energy. All the numerical details regarding the atomic structure, the number of bands, cutoff and  pseudopotential are detailed in Ref.\cite{gruningtddf1}. 
Respect the previous chapters, here we used a finite difference five-point midpoint formula, proposed by Nunes and Gonze,~\cite{gonze} to calculate the $\kk$-derivative appearing the EOM [ Eq.~\ref{eq:fldcpl}].
\begin{equation}
        \hat{\rm w}_{\kk}(\boldsymbol{\cal E}) = \frac{ie}{4\pi} \sum_{i=\alpha}^3\,N_\alpha^\parallel (\Efield \cdot {\bf a}_\alpha) \frac{ 4 D(\Delta \kk_\alpha) - D( 2 \Delta \kk_\alpha)}{3}, 
\label{eq:wkhat2}
\end{equation}
where $N_{\kk_\alpha^\parallel}$ is the number of $\kk$-points along the reciprocal lattice vector $\bb_\alpha$ and  
\bea
D(\Delta \kk_\alpha) &=& \frac{1}{2} \left (\hat{P}_{\kk_i + \Delta \kk_\alpha} - \hat{P}_{\kk_i - \Delta \kk_\alpha} \right ), \label{eq:wkhat} \\
\hat{P}_{\kk_i + \Delta \kk_\alpha} &=& \sum_n^{\text{occ}} | \tilde u_{n \kk_i + \Delta \kk_\alpha} \rangle\langle u_{n \kk_i}| \label{eq:proj}
\eea
In the definition for the projector [Eq.\eqref{eq:proj}] $| \tilde u_{n \kk_i + \Delta \kk_\alpha} \rangle$ are gauge-covariant,\cite{souza_prb} i.e. are constructed so that transform under unitary transformation in the same way as $|u_{n \kk_i} \rangle$ (see Section~\ref{ss:correff}).
The truncation error in this expression converges as ${\cal O}(\Delta \kk^4)$ whereas the three-point midpoint formula proposed in Ref.~\cite{souza_prb} and used in our previous works~\cite{nloptics2013,PhysRevB.89.081102} converges as ${\cal O}(\Delta \kk^2)$. Though more cumbersome, we prefer Eq.~\eqref{eq:wkhat2}, since we noticed that when using polarisation dependent functionals the EOMs are very sensitive to numerical error.
The final EOM we propagate is 
\be
i\partial_t | u_{n\kk} (t)\rangle =
  \left[ \HH^{s}_\kk(t)+\Delta H^{QP}_\kk + i R_\kk(t)\right] 
     | u_{n\kk}  \rangle \label{eq:def1} .
\ee
where $\HH^{s}_\kk(t)$ is the time-dependent Kohn-Sham crystal Hamiltonian [ Eq.\ref{eq:tdksh}], $\Delta H^{QP}_\kk$ is the scissor operator and $i R_\kk(t)$ is a phenomenological dephasing operator defined in Sec.~\ref{sc:compdet}.

%
%
%

\section{Results} \label{rt-tddft}

We considered the optical properties of bulk Si, which has a diamond structure, and GaAs, AlAs and CdTe, which have zincblende structure. The two structures are similar, both are face-centred cubic systems with a two atom basis (at the origin, and at 1/4 of the unit cell in each direction). In silicon the two atoms are identical, in the zincblende structures are the different atoms of the II-VI (CdTe) or III-V (GaAs and AlAs) compound. In terms of crystal symmetries this implies that at variance with silicon they miss the inversion symmetry, and therefore have a dipole-allowed SHG. In what follows we study linear and nonlinear optical properties contrasting the standard TD-LDA with the real-time DPFT approach. 

\begin{figure}[t]
\centering
\includegraphics[width=0.7\textwidth]{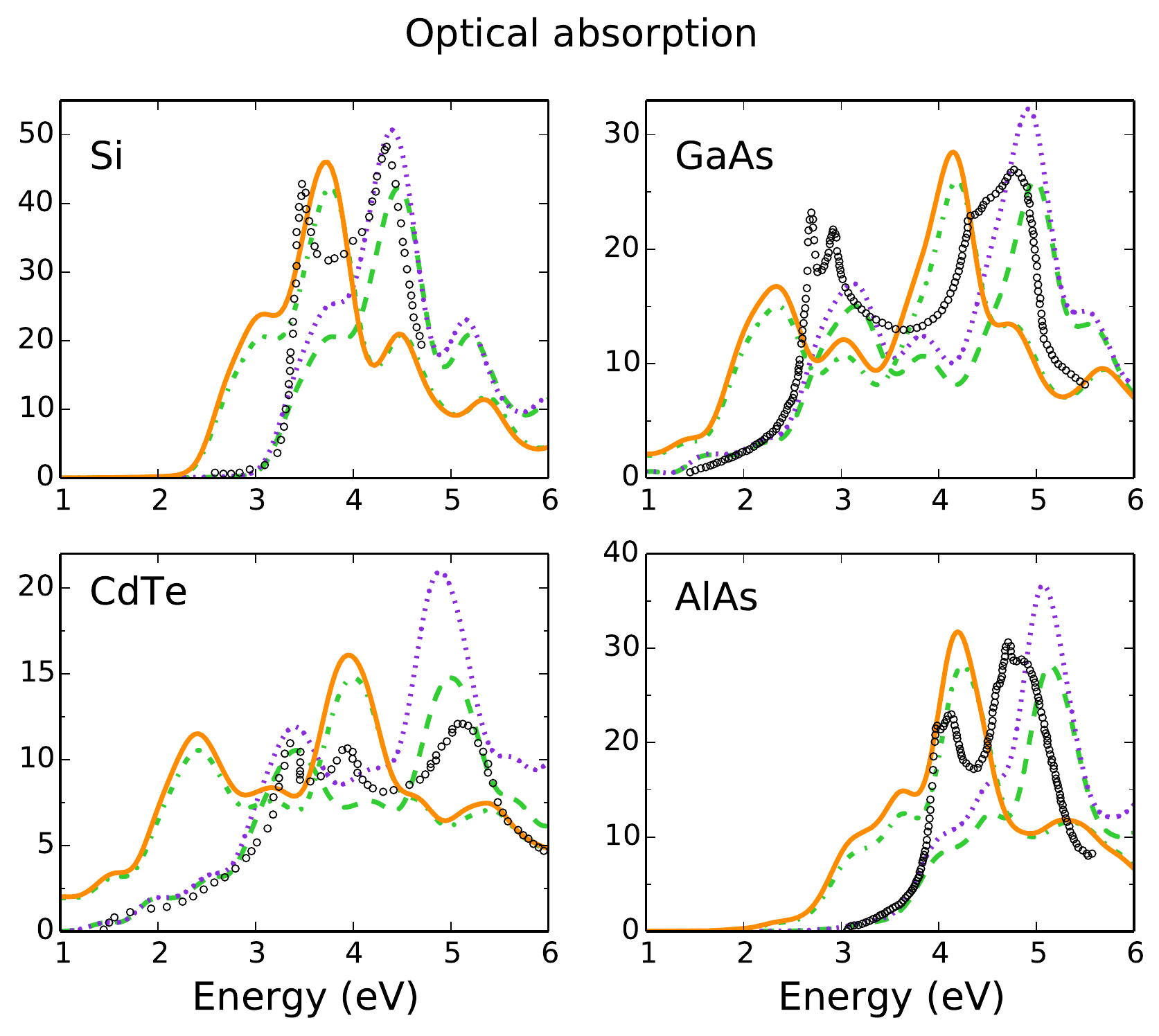}
\caption{\footnotesize{Optical absorption in bulk Si (top left), GaAs (top right), CdTe (bottom left) and AlAs (bottom right): experimental optical absorption spectra (open circles) are compared with real-time simulations at different levels of approximation: TD-LDA (continuous orange line), RPA (green dash-dotted line), both without the scissor correction, and the IPA (violet dotted line) and RPA (green dashed line) with scissor correction.[Figure from Ref.~\cite{gruningtddf1}]}} \label{fg:epsblk}
\end{figure}

\subsection{Optical absorption}

The experimental optical spectra on Si~\cite{PhysRevB.36.4821}, GaAs~\cite{PhysRevB.35.9174}, CdTe~\cite{Adachi} and AlAs~\cite{GARRIGA} (Fig.~\ref{fg:epsblk}, black dashed lines) show qualitative similarities. They all present two main features, a peak at about 3-3.5 eV (referred as $E_1$) and stronger second peak at 4.5-5.0 eV (referred as $E_2$). In GaAs and CdTe, containing heavier third/fourth rows atoms, the $E_1$ peak is split because of the spin-orbit interaction.
Note that we do not include spin-orbit in the Kohn-Sham Hamiltonian and therefore we do not reproduce the splitting at any level of the theory. 

Figure~\ref{fg:epsblk} compares the experimental spectra with results obtained within the RPA and the TD-LDA (without scissor correction). For the considered systems the two approximations produce very similar spectra. As the only difference between the TD-LDA and the RPA is the microscopic xc potential, one can infer that the effect of the latter is minor as already discussed in the literature.~\cite{botti2007time,Onida} 
The most striking difference between the experimental and calculated spectra is the onset that is underestimated by 0.5--1.0~eV. When a scissor operator is added (see Ref~\cite{gruningtddf1}) the agreement is improved though for Si, GaAs and AlAs the $E_2$ peak is slightly blue-shifted and more importantly the $E_1$ peak is either underestimated or appears as a shoulder. Indeed the underestimation of the $E_1$ peak intensity in semiconductor by TD-LDA (and similar TD-DFT approximations) is well known and a signature of missing long-range correlation (see for example Refs.~\cite{PhysRevLett.43.387,PhysRevB.21.4656,botti2007time,Onida}). 
Comparison of the RPA spectra and the independent particle approximation (IPA) spectra shows that crystal local fields effects mostly reduces the intensity of the $E_2$ peak by 15--25\%.  
The experimental optical spectrum of CdTe is well caught within the RPA, but for the overestimation of the $E_2$ peak intensity.  

\begin{figure}[t]
\centering
\includegraphics[width=0.7\textwidth]{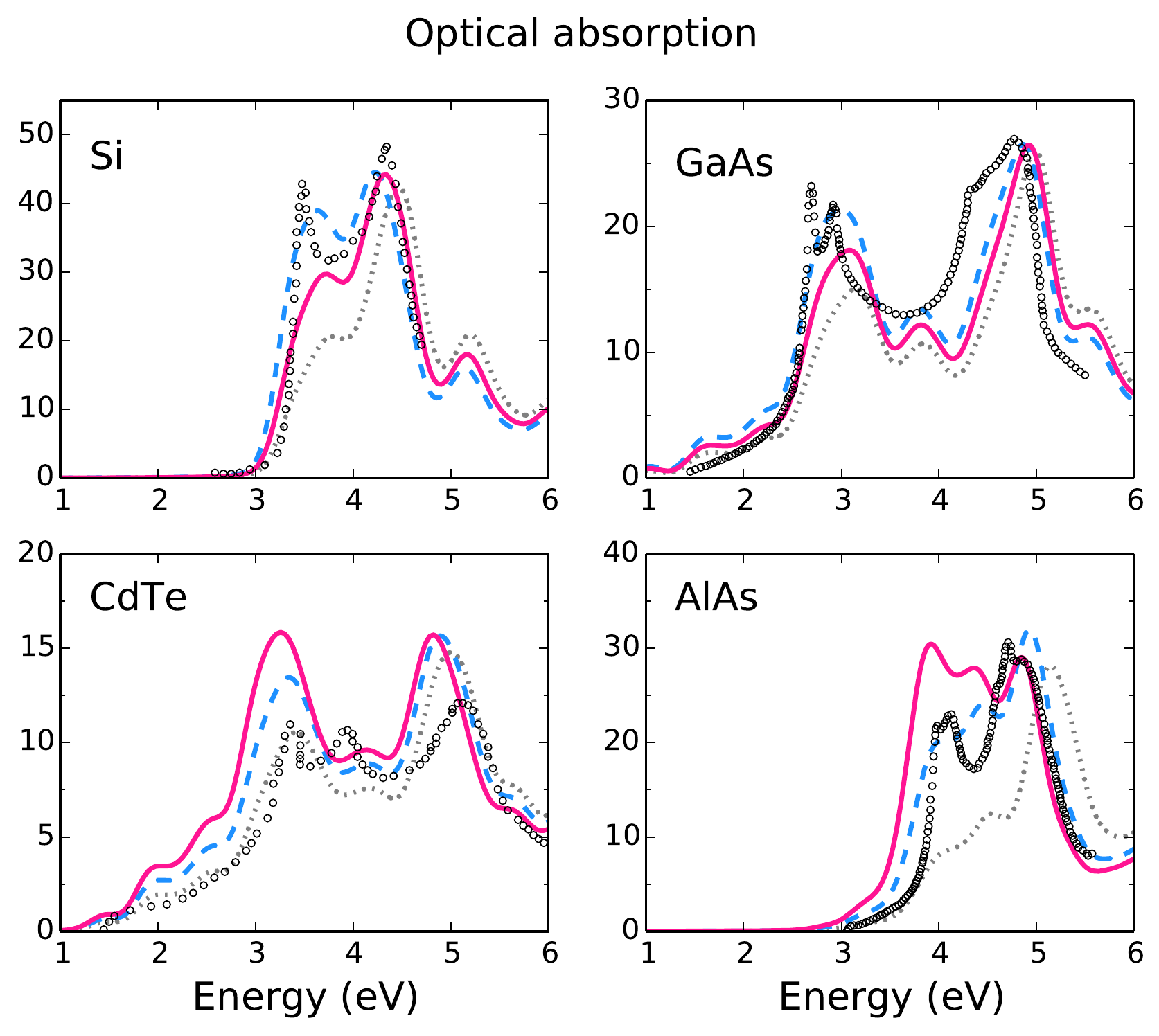}
\caption{\footnotesize{Optical absorption in bulk Si (top left), GaAs (top right), CdTe (bottom left) and AlAs (bottom right): experimental optical absorption spectra (open circles) are compared with real-time simulations at different levels of approximation: opt-PF (blue dashed line), JGM (pink continuous line), RPA(gray dotted line). All approximations include the scissor operator correction.[Figure from Ref.~\cite{gruningtddf1}] }} \label{fg:epsblk1}
\end{figure}

\begin{figure}[ht]
\centering
\includegraphics[width=0.6\textwidth]{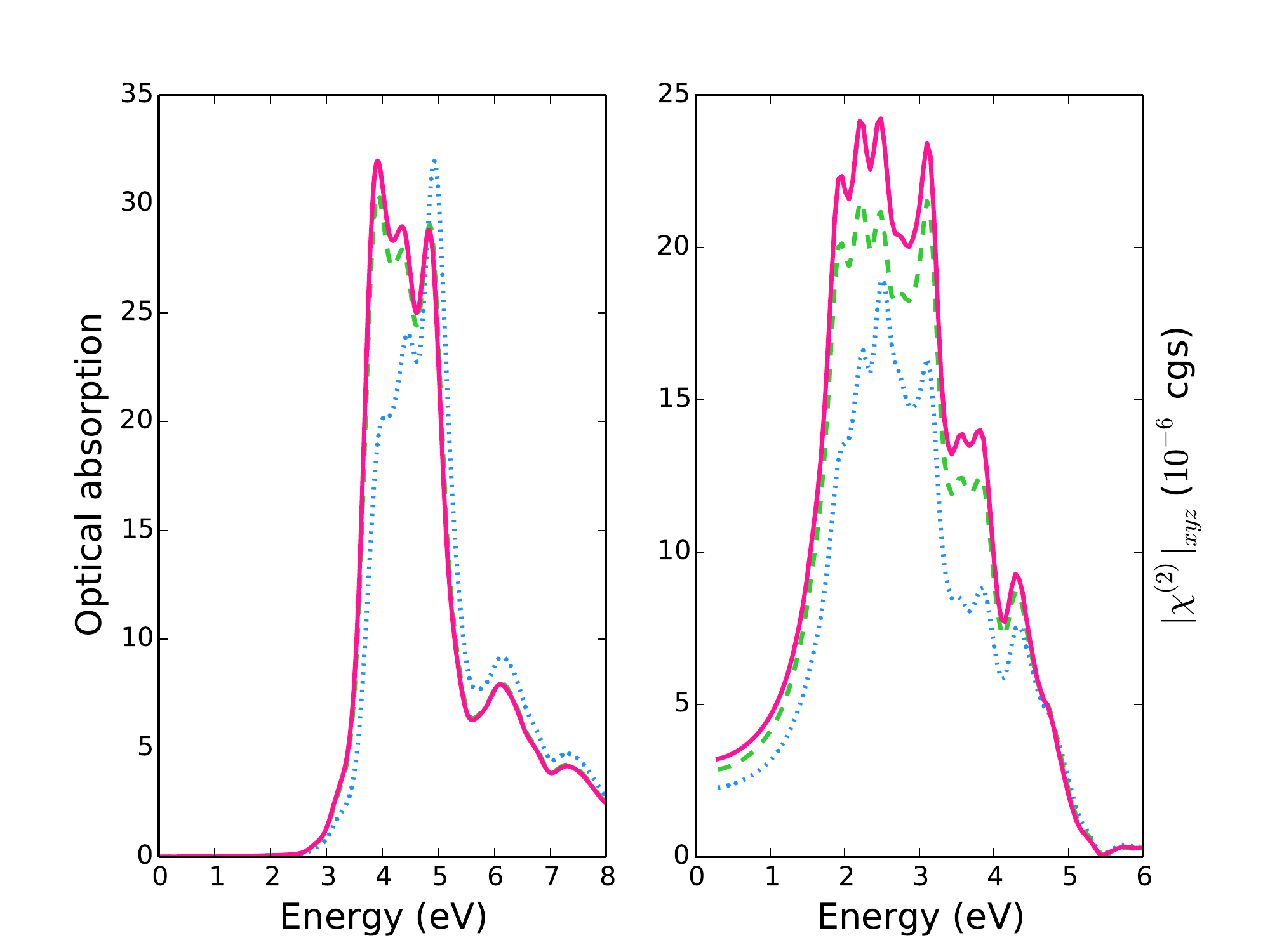}
\caption{\footnotesize{Effect of microscopic components in the JGM-PF on the optical absorption (right panel) and SHG (left panel) of AlAs. The plots compare JGM-PF spectra with (green dashed line) and without microscopic effects (magenta continuous line) and the opt-PF (blue dotted line). [Figure from Ref.\cite{gruningtddf1}] }}
\label{fg:effG}
\end{figure}

Figure~\ref{fg:epsblk1} shows the effects of the macroscopic xc field that is added through the approximated PFs discussed in Sec.~\ref{sc:ksef}. For Si, GaAs and AlAs a clear improvement is observed for the opt-PF: both intensity and position of the peaks are reproduced reasonably well. For CdTe adding the xc macroscopic field lead to an overestimation of the $E_1$ peak intensity which was well caught within the RPA. On the other hand the $E_1/E_2$ intensity ratio is better reproduce by the PFs than within RPA.
For the JGM-PF the agreement is in general less satisfactorily. In particular for Si the $E_1$ peak intensity is still visibly underestimated, while for AlAs it is overestimated. The main difference between the two approximation is the value of $\alpha$: in the opt-PF, $\alpha$ is a parameter optimised to reproduce the optical spectra; in the JGM-PF $\alpha$ is determined from the jellium with a gap model. The model does not reproduce the optimal value. For Si, $\alpha^{\sss JGM} \approx 0.11$ and for AlAs $\alpha^{\sss JGM} \approx 0.52$ respectively smaller and larger than the optimal value reported in Ref.~\cite{gruningtddf1}. It is worth to notice that the xc macroscopic field in the JGM-PF has as well a microscopic contribution. For AlAs this contribution is singled out in the right panel of Fig.~\ref{fg:effG} where it is shown to reduce slightly the absorption. For silicon (not shown) the microscopic contribution to the macroscopic field is negligible.

\subsection{SHG of GaAs, AlAs and CdTe}

\begin{figure}[H]
\centering  \includegraphics[width=0.8\textwidth]{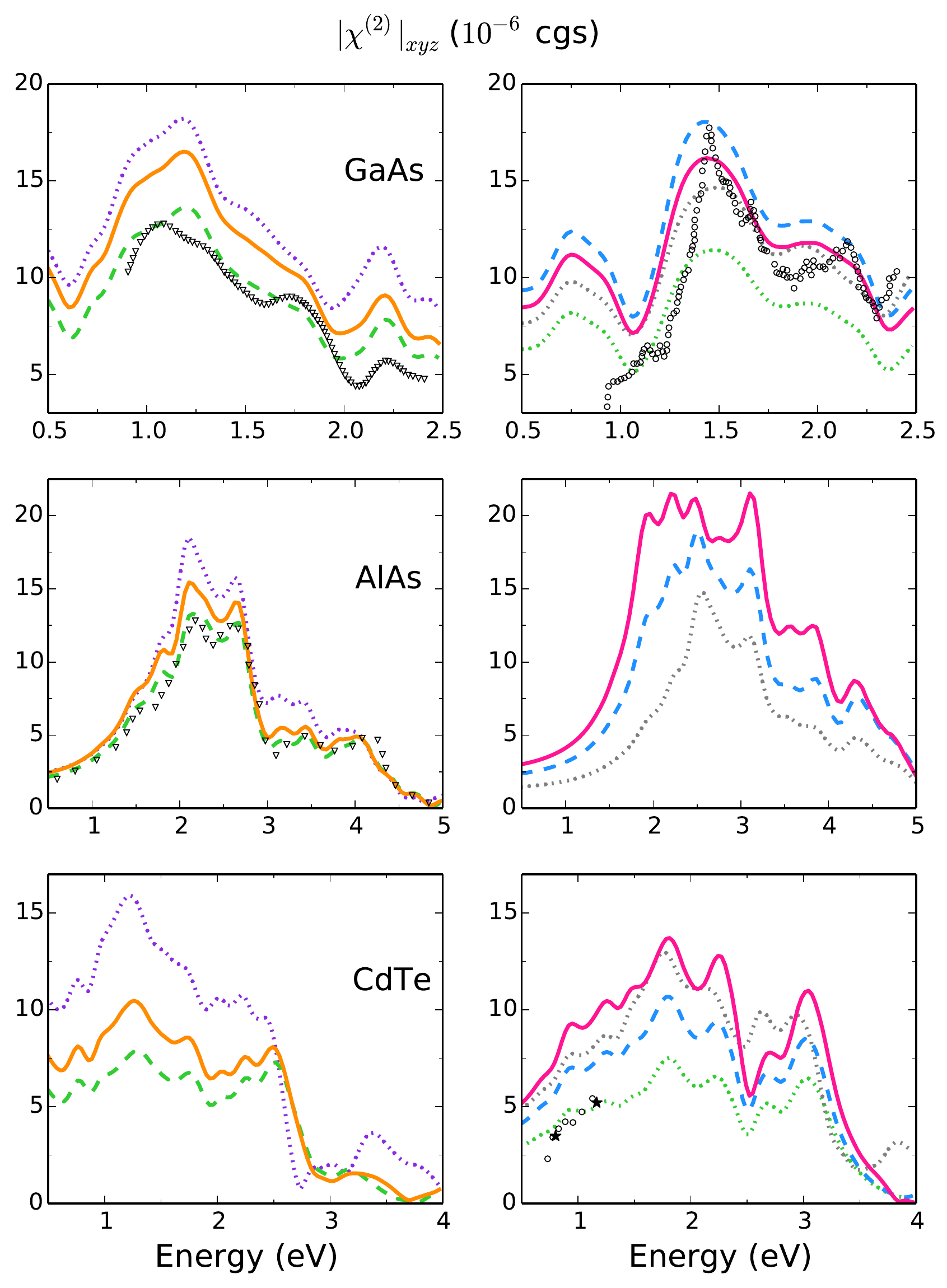}
\caption{\footnotesize{SHG spectra of GaAs (top panels),  AlAs (middle panels) and CdTe (bottom panels) obtained from real-time simulations at different levels of approximation. Left panels: IPA (dotted violet), RPA (dashed green) and TD-LDA (continuous orange)---all without scissor operator correction. For comparison we included the RPA spectrum of GaAs and AlAs calculated by Luppi et al.\cite{PhysRevB.82.235201} (open triangles) \label{fg:shgblk}. Right panels: opt-PF (dashed blue) and JGM-PF (continuous pink) are compared with IPA (dotted gray) and RPA for CdTe and GaAs (dotted green). Available experimental results are shown for GaAs (open circles)~\cite{bergfeld2003second} and CdTe (open circles~\cite{Shoji:97} and stars~\cite{Jang:13}).\label{fg:shg} [Figure from Ref.\cite{gruningtddf1}] }}
\end{figure}

In zincblende structures the only independent non-zero SHG component~\cite{boyd} is $\chi^{(2)}_{xyz}$ (or its equivalent by permutation). The module of the calculated $\chi^{(2)}_{xyz}$ for the systems under study is reported in Fig.~\ref{fg:shg} and compared with experimental values where available.
Note that when the energies are not corrected by a scissor (left panel) for both GaAs and CdTe a large part of the energy range of the SH spectra is in the absorption region where both one-photon and two-photon resonances contribute to the intensity.  For AlAs the part of the SH spectra below 2~eV is instead in the transparency region of the material (only two-photon contributions).
When the scissor-correction to the energy is applied (right panel), the transparency region for GaAs and CdTe is below 1~eV and for CdTe below 3~eV. In the transparency region only two-photon resonances contribute. 
Comparing the TD-LDA with the RPA and the independent particle (IP)  spectra (left panel) shows that crystal local field effects (that tend to reduce the overall SH intensity) are partially compensated by the microscopic exchange-correlation effects (that tend to increase the SH intensity). In general both effects are relatively stronger than for the optical absorption. Applying the scissor correction does not correspond to a simple shift (like in the optical absorption case) but changes the spectra. Firstly the SH intensity is reduced overall (because of sum rules), secondly the intensity is redistributed as the scissor modifies the relative position of one-photon and two-photon resonances (that are shifted by a half of the scissor value). For GaAs and CdTe the addition of macroscopic correlation through the approximated PF leads to an enhancement of about 40\% in GaAs and 80\% in CdTe with respect to the RPA. On the other hand as discussed for those systems local field effects are very large and in fact the spectra form the PF are not significantly different than at the IP level, meaning an almost exact cancellation of the crystal local effects and the macroscopic xc effects as describe by the approximated PFs. Only in the case of AlAs, the macroscopic correlation enhances significantly the SH, adds features and redistributes relative weights with respect to the IPA. 
Regarding the comparison with experiment (right panel), in GaAs the peak at 1.5~eV and the feature at 2.2~eV in the experimental SHG are fairly reproduced by the opt-PF and  JGM-PF approximations. All approximations significantly overestimate the SH for energies below 1~eV. A similar breakdown of the opt-PF approximation (that within the response theory context corresponds with the long-range corrected kernel) has been observed by Luppi and coworkers and traced back to the errors in the theoretical macroscopic dielectric function.~\cite{PhysRevB.82.235201} For CdTe, the approximation that is closer to experimental results (which however are available only around 1~eV) is the RPA while both PF approximations overestimate the experimental SH. This is consistent with the results for optical absorption for which the RPA gives the best agreement among all approximations considered.

We have also compared our results from real-time simulations with those obtained from a response approach by Luppi and co-workers~\cite{PhysRevB.82.235201} and we found a good agreement, slightly better than our previous work\cite{nloptics2013} thanks to the higher order approximation for the covariant derivative [Eq.~\eqref{eq:wkhat2}]. In the left panel of Fig.~\ref{fg:shg} we show for example the comparison for the RPA. There is a very good correspondence between the two spectra for AlAs. For GaAs there are small, but still visible differences which we argue are due to the different pseudopotentials used. In fact we obtain a similar variation in our results when repeating the calculations with different pseudopotentials. It is known that SHG is very sensitive to changes in the electronic structure and that is turn changes when using different pseudopotentials. This is particularly true in the case of GaAs and the sensitivity on the pseudopotential choice was also observed in the referenced calculations.  
Note that in the pseudopotentials we used $d$ orbitals are considered as core electrons, whereas they are included as valence electrons in the calculation of Luppi and coworkers.~\cite{PhysRevB.82.235201} On the other hand pseudopotentials including $d$ electrons that we were testing did not provide a much better agreement.  

\subsection{THG of Si}

\begin{figure}[ht]
\centering
\includegraphics[width=0.9\textwidth]{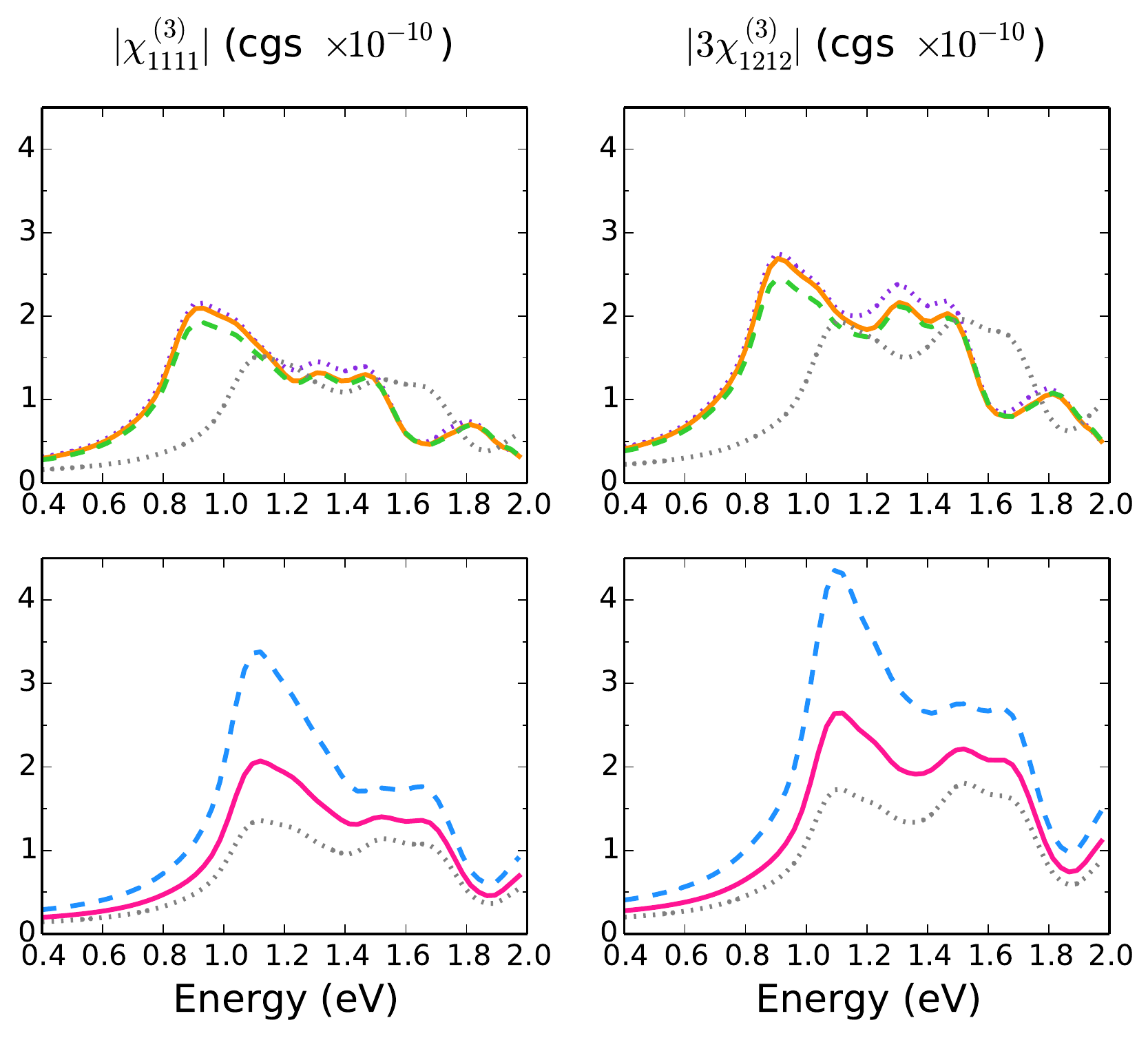}
\caption{\footnotesize{THG of Si: $|\chi^{(3)}_{1111}|$ and $|3\chi^{(3)}_{1212}|$ components (see text). Spectra obtained from real-time simulations at different levels of approximation. Top panels: TD-LDA (continuous orange line), RPA (green dashed line), IPA (dotted violet line) without scissor operator correction are compared with and IPA (gray dotted line) with scissor operator correction. Bottom panels: JGM-PF (continuous pink line), opt-PF (blue dashed line) and RPA (gray dotted line) with scissor operator correction. [Figure from Ref.\cite{gruningtddf1}] }} \label{fg:siX3ab}
\end{figure}

Figure~\ref{fg:siX3ab} shows the calculations for $A=|\chi^{(3)}_{1111}|$ and $B=|3\chi^{(3)}_{1212}|$, the modules of the $1111$ and $1212$ components of the THG of Si~\cite{Moss:89}. These were deduced from calculations with the input field either along the $x$ or along the $xy$ direction.

The TD-LDA spectra (top panels) both present two main features, a peak around 0.9~eV (three-photon resonance with $E_1$) and a shoulder around 1.4  eV (three-photon resonance with $E_2$). Both features are more intense and pronounced in the $|3\chi^{(3)}_{1212}|$. Results within TD-LDA resemble closely those obtained within the RPA and IP approximation. For the $E_1$ three-photon resonance the microscopic exchange-correlation effects cancel with the local-field effects, so that TD-LDA almost coincides with the IP approximation. For higher energies instead, the TD-LDA and RPA spectra are practically identical. Applying a scissor operator does not simply shift the peaks by an amount of about 1/3 of the scissor value. The overall intensity of the spectra is reduced (as expect from sum rules) and as well the relative intensity of the $E_1$/$E_2$ resonances changes. Specifically the ratio is close to or even smaller than $1$ in the scissor corrected spectra, while is $\approx 1.2-1.3$ in the uncorrected spectra. The macroscopic exchange-correlation introduced with the approximations for the PF (bottom panels) enhances the intensity of the spectra and as well the $E_1$/$E_2$ ratio. Consistently with what observed for the linear response, the largest $\alpha$ (opt-PF for silicon) produces the largest correction.

\begin{figure}[hb]
\centering
\includegraphics[width=1\textwidth]{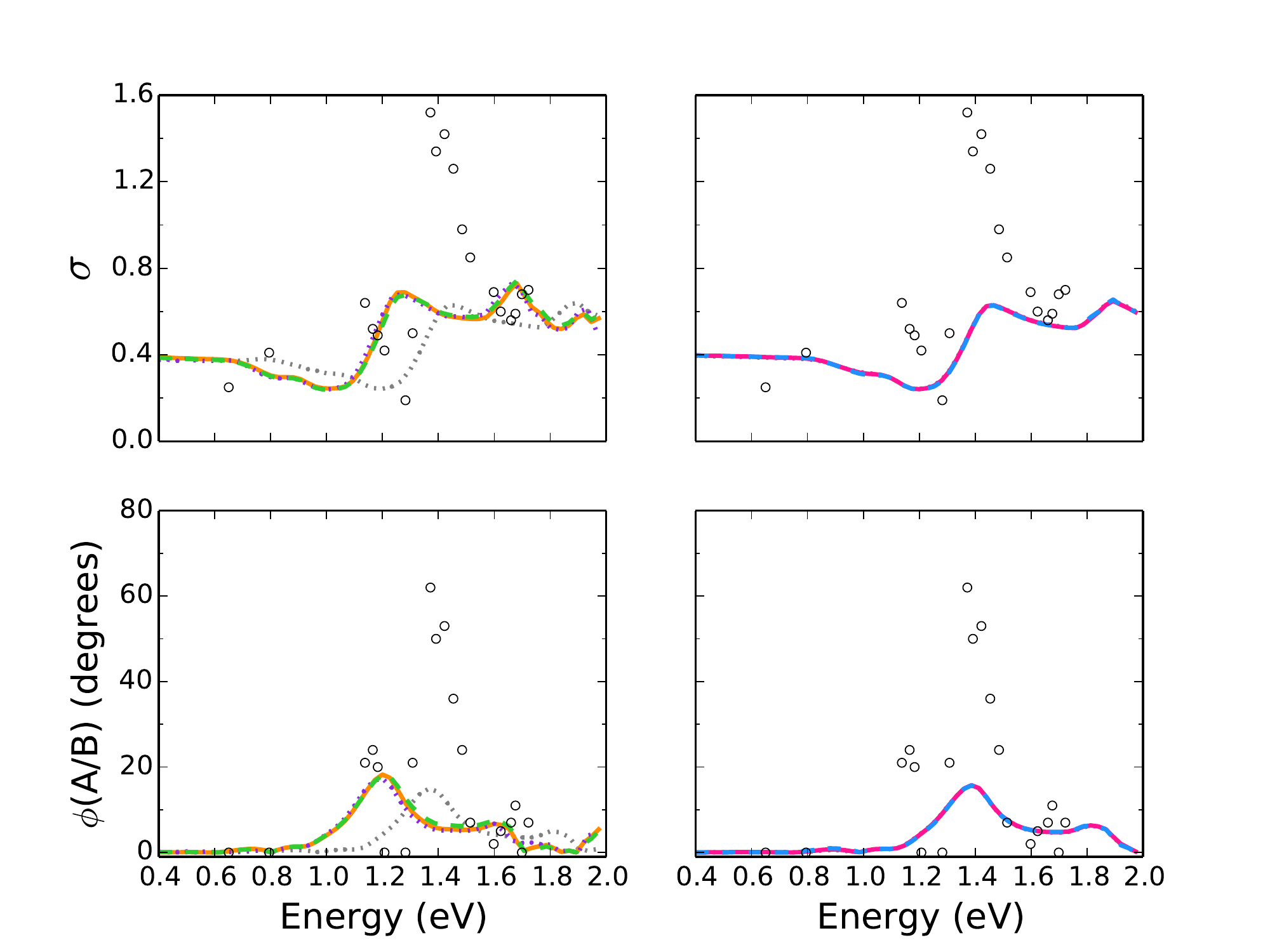}
\caption{\footnotesize{THG of Si: anisotropy parameters $\sigma$ and $\phi$ (see text). Experimental data (open circles)\cite{Moss:89} compared with results obtained from real-time simulations at different levels of approximation. Left panels: TD-LDA (continuous orange line), RPA (green dashed line), IPA (dotted violet line) without scissor operator correction are compared with the IPA (gray dotted line) with scissor operator correction. Right panels: JGM-PF (continuous pink line), opt-PF (blue dashed line) and RPA (gray dotted line) with scissor operator correction.[Figure from Ref.\cite{gruningtddf1}] }} \label{fg:siX3an}
\end{figure}

Experimental measurements are available for the ratio $R_1$ between the THG signal obtained with $45$ and $0$ incident angles and for the ratio $R_2$ between the THG signal obtained with circularly polarised light and linearly polarised light at $0$ incident angle. From those measurements then $\sigma = |1 - B/A|$ and the phase $\phi(B/A)$ can be deduced.~\cite{Moss:89} The experimental results are reported in Fig.~\ref{fg:siX3an}. Both $\sigma$ and $\phi(A/B)$ present two features at about 1.1~eV and 1.4~eV in correspondence of the three-photon $E_1$ and $E_2$ resonances. All the theoretical results are very similar irrespective of the approximation used and the differences observed for the $A=|\chi^{(3)}_{1111}|$ and $B=|3\chi^{(3)}_{1212}|$ in Fig.~\ref{fg:siX3ab}. The results from the scissor corrected approximations (right panels) are just shifted by 1/3 of the scissor operator. When compared with the experiment all the approximation reasonably reproduce the behaviour at energies lower than 1~eV. However for both $\sigma$ and $\phi(A/B)$ (we consider here only the scissor corrected approximations which have resonances at the correct energies) the peak in correspondence of the $E_1$ resonance is missing and the feature in correspondence of the $E_2$ resonance much less pronounced than in experiment. When compared with calculations from Moss and coworkers~\cite{Moss1990} at the independent particle level from the electronic structure calculated either with empirical tight-binding and semi-ab-initio band-structure techniques, the intensity we found for $A$ and $B$ are similar to the latter, but the main spectral features are similar to the former. To notice that the THG based on empirical tight-binding shows in the $\sigma$ and $\phi$ spectra a peak at 1.1~eV.    

\section{Discussion}

\begin{figure}[ht]
\centering
\includegraphics[width=0.6\textwidth]{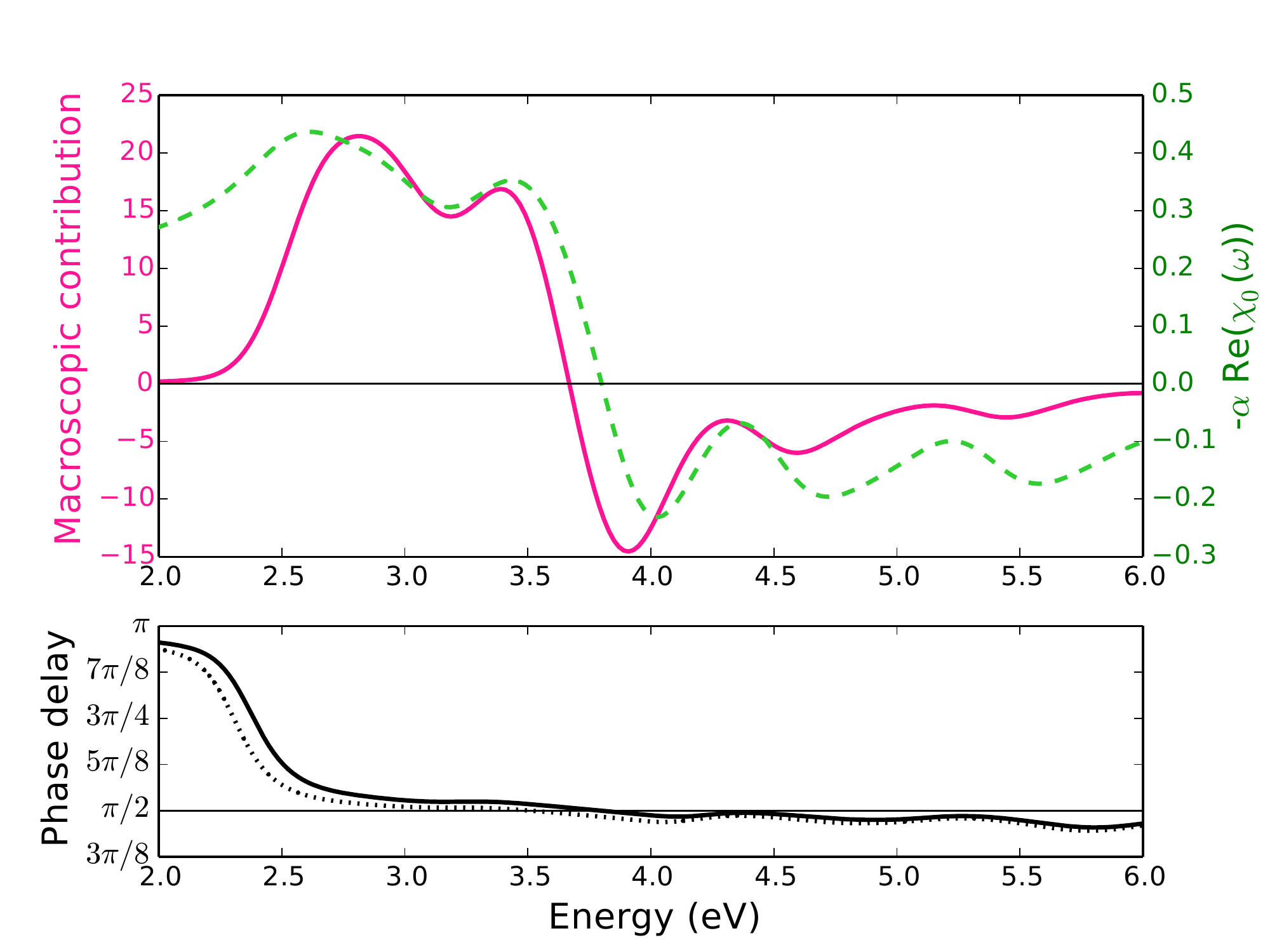}
\caption{\footnotesize{Upper panel: Macroscopic contribution to the optical spectrum of Si calculated as the difference between the the opt-PF  and the RPA optical absorption spectra (pink continuous line) compared with $-\alpha \text{Re}(\chi_0)$ (green dashed line). Bottom panel: phase delay $\phi$ between the polarisation and the applied electric field as a function of the applied field frequency at the RPA (dotted line) and opt-PF (continuous line) level of approximation. The horizontal line highlight the $\phi =\pi/2$ delay. See text.[Figure from Ref.\cite{gruningtddf1}] }} \label{fg:epsanl}
\end{figure}

It is interesting to analyse how an apparently simple approximation such as $\alpha \PP$ correctly ``distinguishes'' where to increase the optical absorption spectrum at RPA level. This information is ``encoded'' in the macroscopic polarisation. In fact, in the linear response limit the effective Kohn-Sham electric field within the proposed PF approximations takes the form $$\Efield^{s} (\omega)= [1 -\alpha \chi(\omega)] \Efield^{\text{tot}} (\omega).$$ That is, the intensity of the applied field is either amplified or reduced depending on the sign of $\text{Re}\chi(\omega)$---as $\text{Im}\chi(\omega) \ge 0$ for any positive $\omega$. In Fig.~\ref{fg:epsanl} (upper panel) we see that indeed the sign of $-\text{Re}(\chi_0)$ follows closely that of the correction induced by $-\alpha P$. 
To gain an insight on how the sign of $\text{Re}(\chi_0)$ is linked to the localisation of the excitation we consider the phasor representation of $\chi_0(\omega) = |\chi_0(\omega)|\exp{(i\phi)}$: the complex argument $\phi$ (see bottom panel of Fig.~\ref{fg:epsanl}) gives the phase delay between $\PP$ and $\Efield$. In particular a delay of $\phi = \pi/2$ corresponds to in-phase oscillation of the macroscopic polarisation current $\JJ$ ($-\partial \PP/\partial t$) with $\Efield$: where the optical absorption is negligible those oscillations are plasmons; in regions with non-negligible optical absorption they can be considered as a signature of delocalized excitations (note that in fact the optical absorption it as a maximum at $\phi=\pi/2$).  Heuristically, for more localized excitations we may expect a phase delay larger than $\pi/2$, and for delocalized excitations a phase delay smaller than $\pi/2$.
Then, the $\cos\phi$, and $\text{Re}(\chi)$ which is proportional to it, are negative for localized excitations and positive for the more delocalized ones. A correction proportional to  $-\text{Re}(\chi)$ then increases the absorption in correspondence of more localised excitation and decreases it for more delocalized excitations. Note as well that in the RPA the phase delay is overestimated. Then the absorption, proportional to $\sin\phi$ is too small for $\phi > \pi/2$ (localized excitation) and too large for $\phi < \pi/2$ (delocalised  excitation).

\section{Summary and conclusions}
We have implemented a real-time density functional approach suitable for infinite periodic crystals in which we work within the so-called length gauge and calculate the polarisation as a dynamical Berry phase.~\cite{souza_prb}        
This approach, in addition to the electron density considers also the macroscopic polarisation as a main variable and extends to the time-dependent case the DPFT introduced in the nineties~\cite{Gonze1995,Resta1996,Vanderbilt1997,Martin1997} to correctly treat IPC in electric fields within a density functional framework. In the corresponding time-dependent KS equations next to the microscopic xc potential also appears a macroscopic xc electric field which is a functional of the macroscopic polarisation (and eventually of the microscopic density).
We have derived approximations for the xc electric field exploiting the connection with long-range corrected approximations for xc kernel within the linear response theory. We have considered two approximations, the optimal polarisation functional, linked to the  long-range corrected xc kernel proposed on Ref.~\cite{LRC} and the Jellium with a gap model polarisation functional linked to the analogous approximation for the xc kernel.\cite{jgm}
We have applied this approach, that we refer to as real-time DPFT, to calculate the optical absorption, second and third harmonic generation in different semiconductors (Si, GaAs, AlAs and CdTe). We have compared results with ``standard'' real-time TD-DFT, namely without macroscopic xc effects, and to experimental results where available. The general trend is an overall improvement over standard TD-DFT as to be expected from the results obtained within the response framework.~\cite{LRC} Of the considered approximations, the opt-PF provides the best agreement with the experiment.
We verified this finding also with other materials, the zinc chalcogenides ZnX (X= S, Se, and Te)\cite{gruningtddf2}, not reported in the present chapter.

The approach here proposed combines the flexibility of a real-time approach, with the efficiency of DPFT in capturing long-range correlation. It allows calculations beyond the linear regime (e.g. second- and third-harmonic generation, four-wave mixing, Fourier spectroscopy or pump-probe experiments) that includes excitonic effects. It is an alternative approach to real-time TD-DFT for extended system proposed by Bertsch, Rubio and Yabana.\cite{PhysRevB.62.7998} At difference with our approach the latter uses the velocity gauge---which has the advantage of using the velocity operator that is well defined in periodic systems---rather than the position operator that requires special attention. On the other hand,  although this approach have shown promising results,\cite{PhysRevB.85.045134,goncharov2013nonlinear} it turns to be quite cumbersome for studying response functions beyond the linear regime due to the presence of divergences that in principle should cancel, but that are difficult to treat numerically.\cite{PhysRevB.52.14636} Furthermore non-local operators---such as pseudo-potentials or the scissor operator---are cumbersome to threat in velocity gauge\cite{tokman} while they are trivial in length-gauge.

Similarly to any density-functional approaches, a delicate point is the approximation of the xc effects. In addition to the xc potential as in standard DFT, in this approach we also need an approximation for the macroscopic xc field. Though for the systems here studied the opt-PF approximation seems to work well, such a good performance cannot be expected in general. For example, based on the experience from linear response calculations, this approximation is not expect to work very well for large gap insulators or systems with a reduced dimensionality (e.g. nanostructures or layers) in which the electronic screening is small.~\cite{PhysRevB.68.205112} Furthermore, in the opt-PF the $\alpha$ is chosen has a material dependent parameter rather than obtained from first-principles. In this respect within the other approximation here studied, JGM-PF, $\alpha$ is determined from first-principles but not always has the optimal value. Further studies then should try to develop universal approximations to the polarisation functional, possibly going beyond the linear response formulation that was here used in the derivation of the polarisation functionals.  

\clearemptydoublepage

\chapter{Conclusions} 
In this thesis we presented a new formalism to study non-linear response of solids and nanostructures. Our approach is based on a real-time solution of an effective Schr\"odinger equation. Our starting point is the parameter free Kohn-Sham Hamiltonian. Then we include correlation effects trought single-particle operators in the Hamiltonian. We consider three important effects that are known to affect  linear and non-linear response in solids: the GW correction to the Kohn-Sham band structure; the local field effects that are due to the density response in inhomogeneous materials; and the electron-hole interaction generated by the screened exchange. We found that inclusion of these three effects is crucial to reproduce and predict non-linear response. In particular in low dimensional materials the electron-hole interaction can double the SHG response. Finally we consider a more efficient approach derived from Density Functional Polarisation Theory, an extension of TD-DFT to periodic systems. We show that simple functionals that depend only from density and polarisation are able to catch large part of these correlation effects. Thanks to the strong efficiency of the TD-DFPT approach, it will be possible to extend this kind of simulations to a large spectra of materials and structures.\\
\section*{Future developments and perspectives} 
The coupling between Modern Theory of Polarisation and correlation effects described by means of NEGF or TD-DFPT was the successful outcome of this thesis. However the story does not end here. There are still some open questions that are waiting for an answer. In particular the inclusion of non empirical dephasing terms in the real-time dynamics, and the application of this approach to complex spectroscopic techniques. A correct dephasing can be obtained by means of non-equilibrium Green's function theory. However a key ingredient of our formalism is the phase difference generated by the external perturbation and this quantity cannot be easily obtained from Green's functions or other perturbtive approaches. This makes the coupling of these two worlds quite cumbersome. Some authors proposed rather complicated solutions to these problems, that didn't find a general application in the scientific community. Another possible route is to pass for the velocity gauge, but in this case  the non-locality of the many-body operators make the dynamics difficult to solve.
The second problem and/or prospective is the application of the present formalism to complex spectroscopic techniques as four-wave-mixing, Fourier-spectroscopy and two-dimensional spectroscopy. These techniques requires multiple laser sources and  a deeper analysis to extract data from real-time simulations. For these reasons we are currently working on new techniques as {\it wave-let analysis} and {\it compress sensing} to reduce simulation time and access to new phenomena in an efficient way.
Finally there all the numerical implementation and code parallelization that we did not discuss in the present manuscript. At present we created a very efficient code for the non-linear response and we hope it will be released under GPL licence at the beginning of the next year.

\section*{Acknowledgements}
\label{ackno}                                        
First of all I want to acknowledge all my collaborators: Myrta Gr\"uning,  Davide Sangalli, Andrea Marini and Elena Cannuccia. Then a special thank is due to Elena for reviewing part of this manuscipt. Finally I am grateful to the \href{http://www.cinam.univ-mrs.fr/cinam/spip.php?rubrique36}{TSN group} for their warm welcome. 
This work used the computing facilities of the Atomistic Simulation Centre--Queen's University Belfast, of the CINaM Aix-Marseille Universit\'e, of the ARCHER UK National Supercomputing Service (http://www.archer.ac.uk)  through EPSRC grant EP/K0139459/1 allocated to the UKCP Consortium, and of the ``Curie" national GENGI-IDRIS supercomputing centre under contract No. x2012096655. I acknowledge also the EUSpec Cost Action MP1306.

\clearemptydoublepage

\setcounter{chapter}{0}
\renewcommand{\chaptername}{Appendix}
\renewcommand{\thechapter}{A}
\renewcommand{\theequation}{A.\arabic{equation}}
\renewcommand{\thesection}{\thechapter.\arabic{section}}
\renewcommand{\thesubsection}{\thesection.\arabic{subsection}}
\appendix

\chapter{An efficient method to update the COHSEX self--energy during the time evolution}
\label{fastcohsex}
In this appendix we show how we store and update the $\Sigma^{\text{cohsex}}$ self-energy in a efficient manner. First of all we neglect the variation of the
screened interaction $W(\mathbf r,\mathbf {r'}; G^<(t))$ with respect to the $G^<(\mathbf r,\mathbf{r'},t)$ by setting to zero the functional
derivative $\partial W/\partial G$ (see Sec.~\ref{linear_response}). Within this approximation the $\Sigma^{\text{coh}}$ does not contribute to the
time evolution, therefore only $\Sigma^{\text{sex}}$ needs to be updated:
\begin{align}
\Sigma^{\text{sex}}(\mathbf r,\mathbf{r'},t) = i W(\mathbf r,\mathbf {r'})\sum_{\substack{n,n'}{\kk}} \varphi^{}_{n, \kk}(\mathbf r) 
\varphi^*_{n', \kk}(\mathbf {r'}) G^<_{n,n',\kk}(t).
\label{sex}
\end{align} 
The KBE
involves the  matrix elements $ \langle m, \kk |\Sigma^{sex} | m', \kk \rangle$:
\begin{align}
\Sigma^{\text{sex}}_{m,m',\kk}(t) = \sum_{\substack{\mathbf G,\mathbf{G'},\mathbf q \\ n,n'}} \rho^{}_{\substack{m,n \\ \mathbf{k,q}}} (\mathbf{G'})
\rho^*_{\substack{m',n' \\ \mathbf{k,q}}}(\mathbf{G}) W_{\mathbf G,\mathbf G'}(\mathbf q) G^<_{\substack{n,n' \\ \mathbf{k-q}}}(t),
\end{align}
where 
\be
\rho^{}_{\substack{m,n \\ \mathbf{k,q}}} (\mathbf{G}) = \int  \varphi^*_{m, \kk}( \mathbf r) \varphi_{n,\mathbf{k-q}}( \mathbf r)  e^{i(\mathbf G+\mathbf q) \mathbf r}.
\ee
In order to rapidly update  $\Sigma^{\text{sex}}$ after a variation of  $G^<(\mathbf r,\mathbf{r'},t)$, we store the matrix elements:
\be
M_{ \substack{m,m',n,n' \\ \mathbf q, \kk}} =  \sum_{\mathbf G,\mathbf {G'}} \rho^{}_{m,n} (\kk,\mathbf q,\mathbf{G'}) \rho^*_{m',n'}( \kk, \mathbf q,
\mathbf G) W_{\mathbf G,\mathbf G'}( \mathbf q ) ,
\ee
in such a way that $\Sigma^{\text{sex}}_{m,m'}$ can be rewritten as
\be
\Sigma^{\text{sex}}_{m,m',\kk}(t) = \sum_{\substack{n,n' \\ \mathbf{q}}} M_{\substack{m,m',n,n' \\ \mathbf q, \kk}} \cdot G^<_{\substack{n,n' \\
\mathbf{k-q}}}(t).
\ee
The $M$ matrix can be very large, but its size can be reduced by noticing that: 
(i) the matrix $M$ is Hermitian respect to the $(m,m')$ indexes; 
(ii) the number of {\bf k} and {\bf q} points is reduced by applying the operation symmetries that are left unaltered by the applied external
field; (iii) for converging optical properties only the bands close to the gap are needed.
As an additional numerical simplification we neglected all terms such that $M_{ \substack{m,m',n,n' \\ \mathbf q,\kk}} /\max\{ M_{ \substack{ m,m',n,n'
\\ \mathbf q , \kk}}\}<M_c$, where $M_c$ is a cutoff that, if chosen to be $M_c \simeq 5\cdot 10^{-3}$  does not appreciably affect the final results. In principle by using an auxiliary localised basis set\cite{schwegler:9708,faber2014excited} one can obtain a further reduction of the matrix dimensions, but in the present work we did not explore this strategy.

\chapter{Induced field and response functions} \label{appA}

One of the objectives of atomistic simulations is the calculation of
the macroscopic dielectric function or of related response functions of dielectrics.
Within TD-DFT such goal is achieved via the calculations of the
microscopic density--density response function $\tchirr$, defined via the equation
\be
\delta n_{\sss \GG}(\qq,\w)=\tchirr_{\sss \GG\GG'} (\qq,\w)\ \delta v^{\text{ext}}_{\sss \GG'} (\qq,\w).
\label{eq:chi_red} 
\ee
Here $\GG$ are the reciprocal lattice vectors and $\w$
the frequency obtained from the Fourier transforms $\rr\ra\GG$ and $t\ra\w$.
In addition to  $\tchirr$, 
 the irreducible response function $\chirr$ and
 the auxiliary response function $\bchirr$ can be defined via
 \bea
 \delta n_{\sss \GG}(\qq,\w)&=&\chirr_{\sss \GG\GG'}(\qq,\w)\
  \delta v^{\text{tot}}_{\sss \GG'}(\qq,\w) \label{eq:chi_irr} \\
  \delta n_{\sss \GG}(\qq,\w)&=&\bchirr_{\sss \GG\GG'} (\qq,\w)
   [\delta v^{\text{ext}}_{\sss \GG'}(\qq,\w)+\delta \bar v^\text{H}_{\sss \GG'}(\qq,\w)]. \label{eq:chi_bar}
   \eea

   To linear order and at finite momentum (i.e. $\qq\neq\zero$),
   the longitudinal microscopic dielectric function can be derived from the response functions,
   \bea
   \epsilon^{-1}_{\sss \GG\GG'}(\qq,\w)=\delta_{\sss \GG,\GG'}
     + 4\pi \frac{\tchirr_{\sss \GG\GG'}(\qq,\w)}{|\qq+\GG||\qq+\GG'|}
     \label{eq:eps_M1_micro} , \\
     \epsilon_{\sss \GG\GG'}(\qq,\w)=\delta_{\sss \GG,\GG'}
       - 4\pi \frac{\chirr_{\sss \GG\GG'}(\qq,\w)}{|\qq+\GG||\qq+\GG'|}
       \label{eq:eps_micro}
       \text{.}
       \eea
       The longitudinal macroscopic dielectric function can then be obtained as
       $\epsilon_M(\qq,\w)=1/\epsilon^{-1}_{\sss \zero\zero}(\qq,\w)$.
       Absorption experiment however are described at $\qq=\zero$
       where the dielectric function 
       $\epsilon_M(\w)\equiv \epsilon_M(\zero,\w)$
       can be obtained only via a limiting process.
       They are defined as
       \bea
       \epsilon_M(\w)&=&\left[ 1 + 4\pi\lim_{\qq\ra 0} \frac{\tchirr_{\sss \zero\zero}(\qq,\w)}{|\qq|^2}\right]^{-1} \\
       \epsilon_M(\w)&=&\      1 - 4\pi\lim_{\qq\ra 0} \frac{\bchirr_{\sss \zero\zero}(\qq,\w)}{|\qq|^2}
       \text{.}
       \eea
       As we observed in the introduction this approach is at least problematic in real time simulation,
       where it is numerically more convenient to directly work at $\qq=0$ and thus the density--density
       response function cannot be used.

       Within DPFT the key quantity is the one which relates the macroscopic electric
       field $\Efield^{tot}$ or $\Efield^{ext}$ to the first order polarisation $\PPo 1$.
       \bea
       \PPo 1(\w)=\newtensor{\tilde{\chi}}(\w) \Efield^{ext}(\w) \label{eq:tsusc1} \\
       \PPo 1(\w)=\newtensor{       \chi }(\w) \Efield^{tot}(\w)  \label{eq:susc1}     
       \text{.}
       \eea
       $\newtensor{\chi}(\w)=\newtensor{\chi}^{(1)}(\w)$ is the (first--order) polarizabilty;
       $\newtensor{\tilde{\chi}}(\w)=\newtensor{\tilde{\chi}}^{(1)}(\w)$ is the quasi--polarizability.
       Since we obtain the polarizability dividing the Fourier transform of the time--dependent
       polarisation with the input electric field, we obtain either
       $\newtensor{\tilde{\chi}}(\w)$ or $\newtensor{\chi}(\w)$
       depending on whether we divide by $\Efield^{\text{ext}}$ or $\Efield^{\text{tot}}$.
       Notice that in this framework we have already made the distinction between macroscopic fields,
       described in terms of $\Efield^{\text{ext}}/\Efield^{\text{tot}}$, and microscopic ones,
       described in terms of $\bar v^{\text{tot}}/\bar v^{\text{tot}}$.
       $\newtensor{\tilde{\chi}}(\w)$ and $\newtensor{\chi}(\w)$
       are thus macroscopic functions.
       The longitudinal dielectric function can be obtained, to first order in the field, as
       \bea
       \epsilon_M(\w) &=& \left[ 1 + \tilde{\chi}_{ii}(\w)\right]^{-1}, \\
       \epsilon_M(\w) &=&\       1 -        \chi_{ii}(\w),
       \eea
       where $\tilde{\chi}_{ii}$ is any of the diagonal components of $\newtensor{\chi}$.

       More in general the $n$-order polarisation can be expressed as
       \begin{multline}\label{eq:PpwrfE}
               \PPo n(t)= \int dt_1\ ...\ dt_n\times \\ \susc n(t-t_1,\ ...\ ,t-t_n)\times  \\ 
                        \Efield^{\text{tot}}(t_1)\ ...\ \Efield^{\text{tot}}(t_n) ,
                \end{multline}
                where $\susc n$ is the $n$-order polarizability related to $n$-order nonlinear
                optical properties.
                Also here we could define the $\tsusc n$ as the response to the external field.
                The two can be related from the equation
                \be
                \tsusc n(\w)=\susc n(\w)(1-\susc 1)^n
                \ee
                As for the linear case we obtain either $\tsusc n(\w)$ ot $\susc n(\w)$
                depending for whether field we divide the polarisation $\Efield^{ext}$ or $\Efield^{tot}$.
                However, since usually $\susc n(\w)$ is the quantity considered in the literature the
                last choice is more convenient in nonlinear optics.

In practice, in real-time simulations we can choose to provide as input field  the total $\Efield^{tot}$ or the external one $\Efield^{ext}$.
In the first case we propagate the equation:
\be
i\hbar  \frac{d}{dt}| v_{m\kk} \rangle = \left( H^{\text{mb}}_{\kk} +i \Efield^{tot} \cdot \tilde \partial_\kk\right) |v_{m\kk} \rangle,
\ee
while in the second case we propagate the coupled Sch\"odinger plus Maxwell equations:
\bea
\left\{
\begin{array}{cc}
    i\hbar  \frac{d}{dt}| v_{m\kk} \rangle = \left( H^{\text{mb}}_{\kk} +i \Efield^{tot} \cdot \tilde \partial_\kk\right) |v_{m\kk} \rangle   \\
    \Efield^{tot} =  \Efield^{ext} + 4 \pi \PP.  
\end{array}
\right.
\eea
In this last case the total field is generated directly by the Maxwell equation. Since response functions are independent from the field, the two formulations provide the same response functions. However in numerical simulations, the first approach is preferable because we know analytically the total field. This allow us to have less numerical noise while extracting the $\susc n(\w)$ coefficients and we can probe precisely the frequencies we are interested in. 

\begin{figure}[h]
\centering
\includegraphics[width=0.7\textwidth]{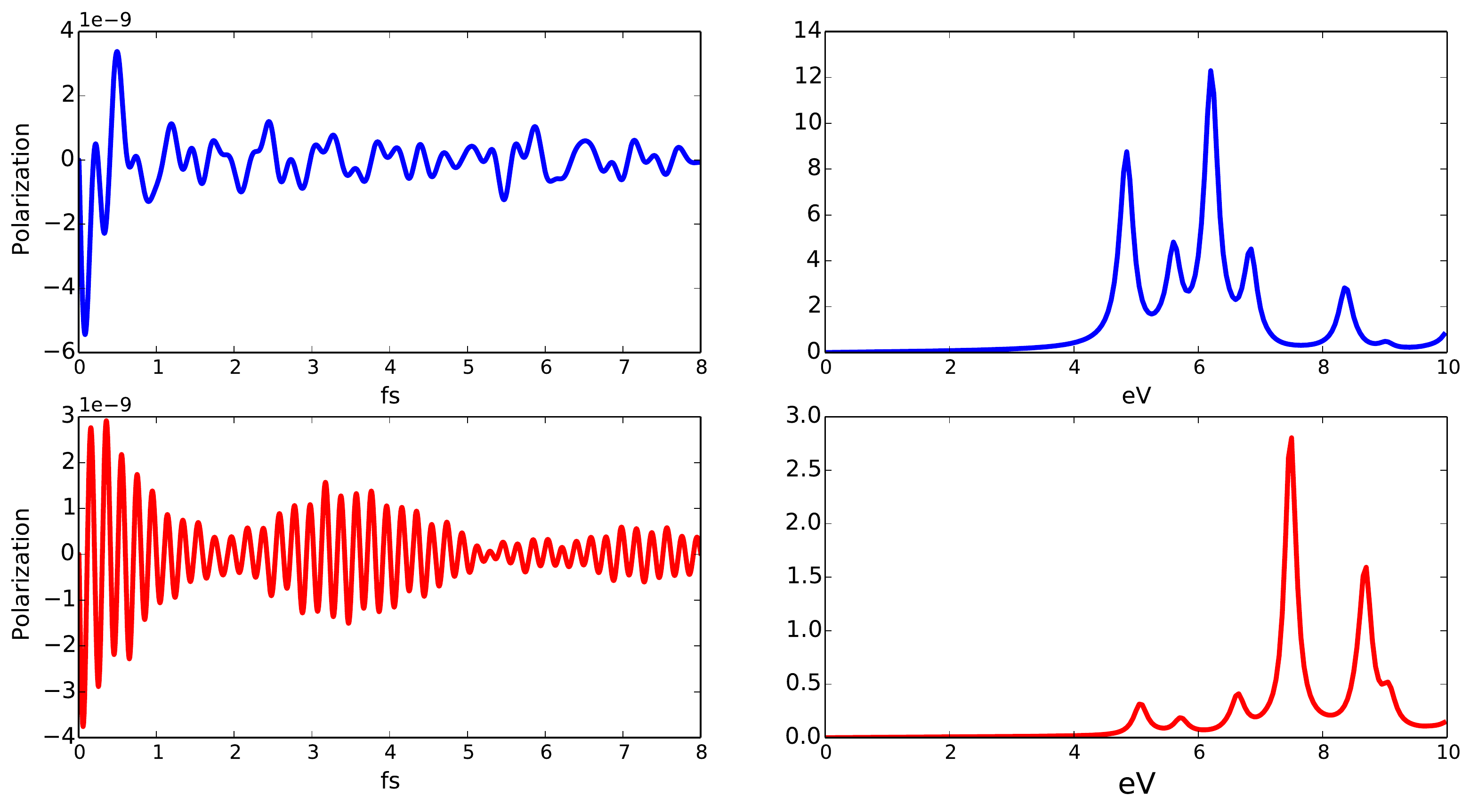}
\caption{\footnotesize{Polarisation, dielectric constant and EELs in hexagonal-BN at IPA. We excite the system with a delta function electric field at $\Efield^{inp} = \mathcal{E}_0 \delta(t)$. (top left) The real-time polarisation obtained using $\Efield^{tot} = \Efield^{inp}$, (bottom left) the polarisation obtained using $\Efield^{ext} = \Efield^{inp}$. (top right) The dielectric constant, (bottom right) the EELS.}\label{induced}}
\end{figure}  

Nevertheless if one is interested in quantities that depend from the field intensity, as for instance electro-absorption, saturation etc.. the correct formulation is the second one. In fact in a real material electrons cannot couple directly to the external field but they couple to the sum the external plus the field generated by the electron them-self. For a discussion on this point and a more rigorous treatment see Ref.~\cite{PhysRevB.85.045134}.\\
As example of these two formalisms let's consider the case of an insulator  subject to a delta function field,   $\Efield^{inp} = E_0 \delta(t)$, in independent particle approximation. If you use $\Efield^{inp}$ as the total one, the polarisation will oscillate according to the electron-hole frequencies, while if we couple the Sch\"odinger equation to the Maxwell one the polarisation will oscillate according to the plasmon frequencies, as depicted in Fig.~\ref{induced}. This can be easily understood from Eq.~\ref{eq:tsusc1} and \ref{eq:susc1}. In the first case the polarisation divided by the total field (that is a constant) is proportional to the EELs, while in the second case dividing the polarisation for the external one gives directly the dielectric constant. This explain the different behaviour of the polarisation in the two cases.  
\chapter{List of publications}
\begin{itemize}
    \item \emph{Optical properties of periodic systems within the current-current response framework: numerical pitfalls and solutions}\\
        D. Sangalli, J. A. Berger, C. Attaccalite, M. Gr\"uning and P. Romaniello, in preparation (2016) 
    \item \emph{Excitonic effects in third harmonic generation: the case of carbon nanotubes and nanoribbons}\\
        C. Attaccalite, E. Cannuccia and M. Gr\"uning, in preparation (2016) 
    \item \emph{Dielectrics in a time-dependent electric field: A real-time approach based on density-polarization functional theory}\\ M. Gr\"uning, D. Sangalli, and C. Attaccalite, Phys. Rev. B \textbf{94}, 035149 (2016)
    \item \emph{Performance of polarisation functionals for linear and nonlinear optical properties of bulk zinc chalcogenides ZnX (X = S, Se, and Te)}\\
        M. Gr\"uning and   C. Attaccalite, Phys. Chem. Chem. Phys., \textbf{18}, 21179 (2016) 
    \item \emph{Optical properties of Cu-chalcogenide photovoltaic absorbers from self-consistent GW and the Bethe-Salpeter equation} \\
        Sabine K\"orbel, David Kammerlander, Rafael Sarmiento-P\`erez, Claudio Attaccalite, Miguel A. L. Marques, and Silvana Botti, Phys. Rev. B \textbf{91}, 075134 (2015)
    \item \emph{Strong second harmonic generation in SiC, ZnO, GaN two-dimensional hexagonal crystals from first-principles many-body calculations} \\
        C. Attaccalite, A. Nguer, E. Cannuccia, M Gr\"uning, Phys. Chem. Chem. Phys. \textbf{17}, 9533 (2015)
    \item \emph{Second harmonic generation in h-BN and MoS 2 monolayers: Role of electron-hole interaction}\\ M. Gr\"uning, C. Attaccalite, Phys. Rev. B \textbf{89}, 081102 (2014)
    \item \emph{Nonlinear optics from an ab initio approach by means of the dynamical Berry phase: Application to second-and third-harmonic generation in semiconductors}\\ C. Attaccalite, M. Gr\"uning, Phys. Rev. B \textbf{88}, 235113 (2013)
    \item  \emph{Real-time approach to the optical properties of solids and nanostructures: Time-dependent Bethe-Salpeter equation}\\ C. Attaccalite, M. Gr\"uning, A. Marini, Phys. Rev. B \textbf{84}, 245110 (2011)
\end{itemize}

\clearemptydoublepage

\addcontentsline{toc}{chapter}{Bibliography}

{\footnotesize
\bibliography{slowlight.bib,nloptics.bib,tddft.bib,corrnlinear2d.bib,PDFT.bib,stst.bib,hdr.bib}
\bibliographystyle{alpha}}

\end{document}